\documentclass[english, reqno]{amsart}
\usepackage{hyperref}
\hypersetup{colorlinks=true,linkcolor=magenta,anchorcolor=green,citecolor=cyan,filecolor=black,menucolor=black,urlcolor=brown}

\usepackage{cite}
\usepackage[table]{xcolor}

\usepackage{mathtools}
\usepackage{enumitem}
\usepackage{mathrsfs}
\usepackage{amsthm}
\usepackage{amscd}
\usepackage{amstext}
\usepackage{amssymb}
\usepackage{graphicx}
\usepackage{setspace}
\usepackage{esint}
\usepackage{rotating}
\usepackage{bm}
\usepackage{nicefrac}
\usepackage{shuffle}
\usepackage{fancybox}
\usepackage{tikz}
\usetikzlibrary{calc,decorations.pathreplacing,decorations.markings,bending,arrows.meta}

\allowdisplaybreaks

\newtheorem{thm}{Theorem}[section]
 \newtheorem{cor}[thm]{Corollary}
\newtheorem{defn}[thm]{Definition}
 \newtheorem{lemma}[thm]{Lemma}

\makeatletter
\newcommand{\customlabel}[2]{%
   \protected@write \@auxout {}{\string \newlabel {#1}{{#2}{\thepage}{#2}{#1}{}} }%
   \hypertarget{#1}{#2}
}
\makeatother

\usepackage{amsfonts}
\usepackage{amssymb}
\usepackage{shuffle}
\usepackage[all]{xy}
\usepackage{array}
\usepackage{lmodern}
\usepackage[OT2,T1]{fontenc}

\usepackage{linearb}

\numberwithin{equation}{section}
\numberwithin{figure}{section}
\newtheorem{propos}[thm]{Proposition}

\def\L{\mathcal{L}}

\makeatother

\usepackage{babel}

\newcount\hh
\newcount\mm
\mm=\time
\hh=\time
\divide\hh by 60
\divide\mm by 60
\multiply\mm by 60
\mm=-\mm
\advance\mm by \time
\def\thistime{\number\hh:\ifnum\mm<10{}0\fi\number\mm}

\def\Li#1(#2){\textrm{Li}_{#1}\left(#2\right)}
\def\cLi_#1(#2){\mathcal{L}_{#1}\left(#2\right)}
\def\bLi_#1(#2){\mathbf{L}_{#1}\left(#2\right)}

\title[Single-valuedness of closed string amplitudes]{\bf 
Single-valued hyperlogarithms, correlation functions \\ and closed string amplitudes}

\author[P. Vanhove]{Pierre Vanhove}
 \address{
 Institut de Physique Th{\'e}orique\\
CEA, IPhT, F-91191 Gif-sur-Yvette, France\\
CNRS, URA 2306, F-91191 Gif-sur-Yvette, France}
\address{
National Research University Higher School of Economics, Russian Federation
}
\author[F. Zerbini]{Federico Zerbini}
 \address{
 Institut de Physique Th{\'e}orique\\
CEA, IPhT, F-91191 Gif-sur-Yvette, France\\
CNRS, URA 2306, F-91191 Gif-sur-Yvette, France}

\thanks{IPHT-t18/005}
\date{\today}

\begin{document}

\begin{abstract}
We give new proofs of a global and a local property of the integrals which compute closed string theory amplitudes at genus zero. Both kinds of properties are related to the newborn theory of single-valued periods, and our proofs provide an intuitive understanding of this relation. The global property, known in physics as the KLT formula, is a factorisation of the closed string integrals into products of pairs of open string integrals. We deduce it by identifying closed string integrals with special values of single-valued correlation functions in two dimensional conformal field theory, and by obtaining their conformal block decomposition. The local property is of number theoretical nature. We write the asymptotic expansion coefficients as multiple integrals over the complex plane of special functions known as single-valued hyperlogarithms. We develop a theory of integration of single-valued hyperlogarithms, and we use it to demonstrate that the asymptotic expansion coefficients belong to the ring of single-valued multiple zeta values.

\end{abstract}
\maketitle
\newpage
\tableofcontents
\newpage
\section{Introduction}
\label{sec:introduction}

In this work we study the integrals
\begin{equation}\label{e:IntroFormula}
  M_{N+3}(\pmb s,\pmb n,\pmb{\tilde n})= 
 \bigg(\frac{i}{2\pi}\bigg)^N\int_{\mathbb{C}^N}\prod_{0\leq i<j\leq N+1}
 |z_i-z_j|^{2 s_{ij}}
 (z_i-z_j)^{n_{ij}}  (\bar z_i-\bar z_j)^{\tilde n_{ij}} \prod_{i=1}^{N} dz_i\,d\bar z_i \,,
\end{equation}
with $z_0:=0$ and $z_{N+1}:=1$, for $N\in \mathbb{N}$ and $n_{ij},\tilde{n}_{ij}\in\mathbb{Z}$, as functions of the complex variables $s_{ij}$.

These integrals are building blocks of \emph{closed} string theory amplitudes on genus-zero surfaces, and are known to admit a meromorphic analytic continuation to $\pmb s\in\mathbb{C}^{N(N+3)/2}$ (see Section~\ref{sec:closedamp}).

More specifically, we are interested in a well-established
\emph{global} property, known in physics as the KLT formula (see below), and in a recently proven \emph{local} property, which concerns the number-theoretical nature of certain asymptotic expansion coefficients. These two properties are related, in two different ways, to the newborn theory of single-valued periods, as explained below. 

\emph{Periods} are special complex numbers, defined as the entries of the matrices~$P$ that represent Grothendieck's isomorphism $H^n_{\rm dR}\otimes_\mathbb{Q}\mathbb{C}\tilde\to H^n_{\rm B}\otimes_\mathbb{Q} \mathbb{C}$ between the de Rham and the Betti (relative) cohomology groups of an algebraic variety over~$\mathbb{Q}$. \emph{Single-valued periods} are defined as the entries of the matrices $P^{\rm sv}:=\bar P^{-1}P$, that represent an automorphism $H^n_{\rm dR}\otimes_\mathbb{Q}\mathbb{C}\tilde\to H^n_{\rm dR}\otimes_\mathbb{Q} \mathbb{C}$ of the de Rham cohomology induced by the action of complex conjugation on the complex points of the variety~\cite{Brown:2013gia}.

To explain the origin of the terminology ``single-valued'', consider a family of algebraic varieties, which depends algebraically on parameters living on a base space. Then the associated periods also depend on the parameters and, as functions of them, they satisfy special differential equations, known as Picard-Fuchs equations. In particular, they can be seen as \emph{multi-valued functions} on the base space, in the sense that they are local solutions of the Picard-Fuchs equations, whose analytic continuation along a path depends on the homotopy class of the path. The associated single-valued periods are, on the other hand, well-defined\footnote{In general, single-valued periods in families are well-defined Zariski locally (the bundle with connection underlying the family may be non-trivial), but in the cases that we consider they are well-defined on the whole base space.} on the base space, so they are \emph{single-valued functions} of the parameters, and single-valued periods are special values of these functions.

There is no canonical way in general to associate to a (motivic) period a single-valued period. Such a canonical map exists, however, for (motivic) periods of mixed Tate motives~\cite{Brown:2018omk}, which contain all periods appearing in this article; we call it a \emph{single-valued projection}\footnote{This is standard but perhaps unfortunate terminology in the field of string amplitudes. In particular, it should not be confused with the \emph{single-valued period map} of~\cite{Brown:2018omk, Brown:2019wna}, which maps de Rham motivic periods to single-valued periods. Our single-valued projection is the composition of that map with the \emph{de Rham projection} from motivic periods to de Rham periods, which exists in the catgory of mixed Tate motives but not in general.}. We remark that single-valued projections are a priori only defined on motivic periods, but assuming the period conjecture we can think of them also as maps on actual periods.

The introduction in mathematics of this single-valued formalism was mainly motivated by the surprising relations noticed in physics between the integrals~\eqref{e:IntroFormula} and the integrals
\begin{equation}\label{e:IntroFormula2}
A_{N+3}(\pmb s,\pmb n,\rho)= 
 \int_{0\leq x_{\rho(1)}\leq \cdots \leq x_{\rho(N)}\leq 1}\prod_{0\leq i<j\leq N+1}
 |x_i-x_j|^{s_{ij}}
 (x_i-x_j)^{n_{ij}} \prod_{i=1}^{N} dx_i\,,
\end{equation}
where $x_0:=0$, $x_{N+1}:=1$, and $\rho\in \mathfrak{S}_N$ is a permutation, which are the building blocks of \emph{open} string theory amplitudes at genus zero.

The first of these surprising relations is the KLT formula, discovered by (and named after) Kawai, Lewellen and Tye in~\cite{Kawai:1985xq}, which expresses closed string integrals~\eqref{e:IntroFormula} in terms of products of pairs of open string integrals~\eqref{e:IntroFormula2}. This identity is a global property of the integrals~\eqref{e:IntroFormula}, as it holds for any~$\pmb{s}$. It was noticed by Mizera that the KLT formula can be interpreted in terms of twisted\footnote{The Betti twisted cohomology is the cohomology with coefficients in the local system associated with the monodromies of a multi-valued ``twist function'', in this case of the form $\prod (x_i-x_j)^{s_{ij}}$. The de Rham twisted cohomology is the cohomology with coefficients in an algebraic vector bundle with a connection integrated by the twist function.} cohomology~\cite{Mizera:2016jhj, Mizera:2017cqs}. Recently, Brown and Dupont built on this observation to show that, in all cases relevant for closed string amplitudes, while the integrals~\eqref{e:IntroFormula2} are (twisted) periods\footnote{These are not periods (in families) in the classical sense.}, the integrals~\eqref{e:IntroFormula} are single-valued (twisted) periods, obtained from the former via the single-valued projection~\cite{Brown:2019wna}; the KLT formula is then a consequence of the fact that $P^{\rm sv}:=\bar P^{-1}P$. In other words, the KLT formula is intrinsically related to the single-valued nature of the closed string integrals~\eqref{e:IntroFormula}.

The second surprising relation between the closed string integrals~\eqref{e:IntroFormula} and the open string integrals~\eqref{e:IntroFormula2}, discovered experimentally, is of local nature. One of the most important goals in the field of string amplitudes is that of studying the asymptotic expansion at the pole $\pmb{s}=\textbf{0}$. Standard techniques allow to show~\cite{Broedel:2013aza} that the asymptotic expansion coefficients at $\pmb{s}=\textbf{0}$ of the integrals~\eqref{e:IntroFormula2} belong\footnote{This is proven only in the cases which are relevant in open string theory.} to the ring of \emph{multiple zeta values}, real numbers defined for $k_1,\ldots ,k_{r}\in \mathbb{N}$, $k_r\geq 2$, by the absolutely convergent nested series
\begin{equation}
\zeta(k_1,\ldots ,k_r):=\sum_{0<n_1<\cdots <n_r}\frac{1}{n_1^{k_1}\cdots n_r^{k_r}}.
\end{equation}

Extensive computations~\cite{Schlotterer:2012ny,Stieberger:2013wea,Stieberger:2014hba,Fan:2017uqy} led to conjecture that the analogous asymptotic expansion coefficients for the closed string integrals~\eqref{e:IntroFormula2} should be \emph{single-valued multiple zeta values}, i.e. single-valued periods~$\zeta^{\rm sv}(k_1,\ldots ,k_r)$ obtained as single-valued projections of multiple zeta values~\cite{Brown:2013gia}. Furthermore, it was conjectured in~\cite{Stieberger:2013wea} that the whole asymptotic expansions of closed string integrals could be obtained from those of open string integrals by applying term-by-term the single-valued projection. An argument proving this conjecture was recently sketched in~\cite{Schlotterer:2018zce}, and shortly afterwards a (different) full proof appeared in the already mentioned paper of Brown and Dupont~\cite{Brown:2019wna}. This result gives a second, different\footnote{Despite the fact that the proofs of the global and of the local results are contained in the same paper~\cite{Brown:2019wna}, they do not seem to be related in any simple way. The recent work~\cite{Brown:2019jng} starts exploring the possibility of relating them via a theory of \emph{hypergeometric motives}.} relation between the integrals~\eqref{e:IntroFormula} and the theory of single-valued periods.

The proofs of the single-valued projection from open to closed string integrals contained in~\cite{Brown:2019wna} and in~\cite{Schlotterer:2018zce} are rather abstract, and use some deep algebraic geometry. The main purpose of our paper is to give new more elementary analytic proofs of the KLT formula and of the appearence of single-valued multiple zeta values in the asymptotic expansion, which also provide a more heuristic explanation for the ``single-valued nature'' of the integrals~\eqref{e:IntroFormula}.

Our idea, in both cases, is to add a variable\footnote{The idea of adding an extra variable has been exploited also in other contexts related to string amplitudes computations, e.g. in~\cite{Broedel:2013aza} or~\cite{Mizera:2019gea} at genus zero or~\cite{Broedel:2019gba,DHoker:2015wxz} at genus one.} to obtain single-valued
functions (single-valued periods in families), and to study their
analytic properties. More specifically, in the global case our idea is
to understand the KLT formula as a specialisation of the conformal
block decomposition of a \emph{correlation function}, whose
single-valuedness is also required for physical reasons. In the local
case, our idea is to develop a theory of integration of
\emph{single-valued hyperlogarithms}, i.e. single-valued special
functions containing single-valued multiple zeta values as special
values, and to use it to algorithmically compute the coefficients of
the asymptotic expansion. Precise statements of our results are
contained in the next section.  

An important motivation for our work is that it seems suited for studying analogues of the integrals~\eqref{e:IntroFormula} on \emph{higher-genus surfaces}, i.e. higher orders in the perturbative expansion of closed string amplitudes. More precisely, the existence of global and local relations between open and closed string integrals, and their interpretation in terms of suitable single-valued periods and single-valued projections, is expected to hold in some form also at higher genus. Since a conformal block
decomposition must exist also for correlation functions on higher-genus surfaces~\cite{DiFrancesco:1997nk}, it would be interesting to use it to identify higher-genus analogues of the KLT formula. It would also be interesting to adapt our results on the integration of single-valued hyperlogarithms to compute (a degeneration of) the asymptotic expansion of genus-one closed string amplitudes in terms of single-valued multiple zeta values\footnote{The building
blocks of open and closed genus-one string amplitudes are functions of the modulus $\tau$ of the genus-one surface. One can consider a double expansion, first at $\pmb s=\pmb 0$, then at the degeneration point $\tau= i\infty$. In the open string case, the coefficients of this expansion are known to be multiple zeta values~\cite{Broedel:2014vla,Broedel:2015hia,Broedel:2017jdo}. In the closed string case, it is still a conjecture that the coefficients are single-valued multiple zeta values~\cite{Zerbini:2015rss,DHoker:2015wxz}, obtained via a single-valued projection from the open string coefficients~\cite{Brown:2017qwo,BrownNewClassII,Broedel:2018izr,Zerbini:2018sox}. Special cases of this conjecture were proven in~\cite{DHoker:2019xef,Zagier:2019eus,Vanhove:2020qtt}.}, along the lines of our article~\cite{Vanhove:2020qtt}. 

\subsection{Content}

\subsubsection{Closed string amplitudes and correlation functions}
Here we review our results of global nature. The starting point is that the integrals~\eqref{e:IntroFormula} can be obtained as special values (either at $\eta=0$ or at $\eta=1$, with the appropriate identification of the exponent parameters) of the correlation function
\begin{multline}\label{e:Gdefgeneralintro}
  \mathcal G_N\left({\pmb a\, \pmb b\, \pmb c\, \pmb d\atop \pmb{\tilde a}\,
      \pmb{\tilde  b}\, \pmb{\tilde c}\, \pmb{\tilde d}}\Big|\eta\right)\,:=\,\bigg(\frac{i}{2\pi}\bigg)^N\times\cr \int_{\mathbb C^N}  
\prod_{i=1}^{N}  z_i^{a_i}\bar   z_i^{\tilde a_i} (1-z_i)^{b_i}(1-\bar z_i)^{\tilde b_i} (\eta-z_i)^{c_i}  (\bar \eta-\bar z_i)^{\tilde c_i} \prod_{1\leq i<j\leq N}  (z_i-z_j)^{ d_{ij}}   (\bar z_i-\bar z_j)^{\tilde d_{ij}}  
  \prod_{i=1}^{N} dz_i\,d\bar z_i \,,
\end{multline}
with exponents $(\pmb a, \pmb b, \pmb c, \pmb d,\pmb{\tilde a},\pmb{\tilde  b}, \pmb{\tilde c}, \pmb{\tilde d})\in\mathbb{C}^{N(N+5)}$ such that $(\pmb a-\pmb{\tilde a}, \pmb b-\pmb{\tilde  b}, \pmb c-\pmb{\tilde c}, \pmb d-\pmb{\tilde d})\in\mathbb{Z}^{N(N+5)/2}$ (see Section~\ref{sec:closedampcorrfun}). This condition on the exponents insures that the integral~\eqref{e:Gdefgeneralintro} defines, in its region of absolute convergence, a single-valued real-analytic function of $\eta\in\mathbb{C}$. This is a physically motivated property of correlation functions in conformal field theory (see Section~\ref{sec:cft-correlators}).

Conformal field theory also predicts that the function~$\mathcal G_N$ has a \emph{conformal block decomposition}
\begin{equation}\label{e:mGgenintro}
 \mathcal G_N\left({\pmb a\, \pmb b\, \pmb c\, \pmb d\atop \pmb{\tilde a}\,
      \pmb{\tilde  b}\, \pmb{\tilde c}\, \pmb{\tilde d}}\Big|\eta\right)= \sum_{r,s=1}^{M}  G_{r,s}\!\left({\pmb a\, \pmb b\, \pmb c\, \pmb d\atop \pmb{\tilde a}\,
      \pmb{\tilde  b}\, \pmb{\tilde c}\, \pmb{\tilde d}}\right) \,I_r(\pmb
  a,\pmb b,\pmb c;\pmb d;\eta) \,I_s(\pmb
  {\tilde a},\pmb {\tilde b},\pmb {\tilde c};\pmb {\tilde d};{\bar \eta})\,,
\end{equation}
for some $M\in\mathbb{N}$, some coefficients~$G_{r,s}$ independent of~$\eta$ and some functions~$I_r$ which are holomorphic in~$\eta$, called \emph{conformal blocks}. This kind of formula is also called a \emph{holomorphic factorisation} of the function~$\mathcal G_N$ (see Section~\ref{sec:cft-correlators}).

In order to write down such a formula explicitly, and demonstrate it, we first study the functions which will play the role of conformal blocks. It turns out that they are contained in a class of generalised hypergeometric functions, systematically studied in the 1980s by Aomoto~\cite{Aomoto} and, independently, by Gel'fand~\cite{GelfandAlone}, given by the integrals
\begin{equation}\label{AGgendef1varIntro}
F_{\Delta}(\pmb a,\pmb b,\pmb c;\pmb d;\eta)\,:=\,\int_{\Delta}\prod_{i=1}^N\,|z_i|^{a_i}\,|z_i-1|^{b_i}\,|z_i-\eta|^{c_i}\,\prod_{1\leq i<j\leq N}|z_i-z_j|^{d_{ij}}\,\prod_{i=1}^N\,dz_i\,,
\end{equation}
where $\eta\in (0,1)$ and $\Delta$ is a connected component of $\{(z_1,\ldots ,z_N)\in \mathbb{R}^{N}\,|\,z_i\neq 0,1,\eta\, \text{ and } \,z_i\neq z_j\}$. We call the integrals~\eqref{AGgendef1varIntro} \emph{Aomoto-Gel'fand hypergeometric functions}. As functions of~$\eta$, they can be extended to holomorphic multi-valued functions on $\mathbb{C}\setminus\{0,1\}$.

Our first result is to identify a basis of these functions, suitable for our purposes (see Theorem~\ref{thmbasisint}).
\begin{thm}\label{thmbasisintintro}
Let $N\in\mathbb{N}$. The $(N+1)!$ Aomoto-Gel'fand hypergeometric
functions $I_{(\rho,\sigma)}(\pmb a,\pmb b,\pmb c;\pmb
d;\eta):=F_{\Delta_{(\rho,\sigma)}(\eta)}(\pmb a,\pmb b,\pmb c;\pmb
d;\eta)$ which are associated, for permutations $\rho\in\mathfrak{S}_r$,
$\sigma\in\mathfrak{S}_s$ with $r+s=N$, to the domains
\begin{equation}
  \Delta_{(\rho,\sigma)}(\eta):= \{0\leq z_{\sigma( 1)} \leq \cdots
  \leq z_{\sigma( s)}\leq \eta\leq 1\leq z_{\rho( 1)}\leq \cdots\leq z_{\rho( r)}\}\,,
\end{equation}
form a basis the vector space of ($N$-dimensional) integrals~\eqref{AGgendef1varIntro} over the field
\begin{equation}
\mathbb{F}:=\mathbb{Q}(e^{\pi ia_1},\ldots ,e^{\pi ia_N},e^{\pi ib_1},\ldots ,e^{\pi ib_N},e^{\pi ic_1},\ldots ,e^{\pi ic_N},e^{\pi id_{12}},\ldots ,e^{\pi id_{N-1\,N}}).
\end{equation}
\end{thm}
This theorem could be proven using results from the twisted cohomology literature. In particular, we rely on a result of Aomoto for the linear independence~\cite{Aomoto}, but we provide an alternative and more constructive proof that the functions $I_{(\rho,\sigma)}(\pmb a,\pmb b,\pmb c;\pmb
d;\eta)$ generate the whole space, using a contour deformation technique which is well-known in physics~\cite{Dotsenko:1984nm,Dotsenko:1984ad,BjerrumBohr:2009rd}, but less so in mathematics (see Proposition~\ref{prop:generatingfamily}).

The Aomoto-Gel'fand hypergeometric functions $I_{(\rho,\sigma)}(\pmb a,\pmb b,\pmb c;\pmb d;\eta)$ take the role of conformal blocks in our holomorphic factorisation of the correlation function~$\mathcal G_N$ (see Theorem~\ref{prop:holfac}):
\begin{thm}\label{prop:holfacintro}
There exist unique coefficients $G_{(\rho,\sigma),(\tilde\rho,\tilde\sigma)}(\pmb a, \pmb b, \pmb c;\pmb d)\in\mathbb{F}$ such that
\begin{equation}\label{eqintroholfac}
\mathcal G_{N}\left({\pmb a\, \pmb b\, \pmb c\, \pmb d\atop \pmb{\tilde a}\,
      \pmb{\tilde  b}\, \pmb{\tilde c}\, \pmb{\tilde d}}\Big|\eta\right)=\bigg(\frac i{2\pi}\bigg)^N\,\sum_{\substack{(\rho,\sigma),(\tilde\rho,\tilde\sigma)\in\mathfrak{S}_r\times\mathfrak{S}_s\\r,s\geq 0,\,\,r+s=N}}G_{(\rho,\sigma),(\tilde\rho,\tilde\sigma)}(\pmb a, \pmb b, \pmb c;\pmb d)\,I_{(\rho,\sigma)}(\pmb
  a,\pmb b,\pmb c;\pmb d;\eta) \,I_{(\tilde\rho,\tilde\sigma)}(\pmb
  {\tilde a},\pmb {\tilde b},\pmb {\tilde c};\pmb {\tilde d};{\bar \eta})\,.
\end{equation}
\end{thm}

The fact that~$\mathcal G_N$ is a single-valued function of~$\eta$ allows us to immediately deduce constraints on the form of the matrix of the coefficients $G_{(\rho,\sigma),(\tilde\rho,\tilde\sigma)}$ (see Proposition~\ref{prop:matrixformsv}). In fact, one may try to construct a holomorphic factorisation by imposing that the monodromies of the Aomoto-Gel'fand functions must disappear, along the lines of the classical paper of Dotsenko and Fateev~\cite{Dotsenko:1984nm}, where the existence (but not the uniqueness) of a holomorphic factorisation like~\eqref{eqintroholfac} is proven for special kinds of exponents.

Our proof is constructive, and it combines
Theorem~\ref{thmbasisintintro} with the contour deformation method
proposed in the original proof of the KLT formula. In particular, it
is not surprising that specialising Theorem~\ref{prop:holfacintro}
to~$\eta=1$ or to~$\eta=0$ one gets back the original KLT formula\footnote{More precisely, we introduce $J_{(\rho,\sigma)}(\pmb a,\pmb b,\pmb c;\pmb d;\eta):=I_{(\rho,\sigma)}(\pmb a,\pmb b,\pmb c;\pmb d;1-\eta)$, which can also be seen as Aomoto-Gel'fand functions~$F_{\Delta}$ for suitable domains~$\Delta$, and constitute an alternative basis (see Corollary~\ref{prop:changebasis}). This basis is then best suited, after specialising to~$\eta=1$, to a comparison with a version of the KLT formula stated in~\cite{Mizera:2016jhj}.}. In other words, the KLT relations are obtained by specialising a physically motivated formula which combines a chosen basis of holomorphic and anti-holomorphic multi-valued functions, which are twisted periods in families, into a (uniquely determined) single-valued function. This is a heuristic and physically motivated justification of the fact that the KLT formula is related to the theory of single-valued periods.
\subsubsection{Closed string amplitudes and single-valued hyperlogarithms}
Let us now turn our attention to local aspects of the integrals~\eqref{e:IntroFormula}. More precisely, we focus on the subfamily of integrals
\begin{equation}\label{SupAmpJintro}
M_{\rho,\sigma}(\pmb s)\,:=\bigg(\frac{i}{2\pi}\bigg)^N\,\int_{(\mathbb{P}^1_{\mathbb{C}})^N}\frac{\prod_{0\leq i<j\leq N+1}|z_j-z_i|^{2s_{ij}}\,\prod_{i=1}^N\,dz\,d\bar z}{z_{\rho(1)}\,\overline{z}_{\sigma(1)}(1-z_{\rho(N)})(1-\overline{z}_{\sigma(N)})\prod_{i=2}^N(z_{\rho(i)}-z_{\rho(i-1)})(\overline{z}_{\sigma(i)}-\overline{z}_{\sigma(i-1)})}\,,
\end{equation}
where $\rho, \sigma\in\mathfrak{S}_N$, which is obtained for special values of $\pmb n$, $\tilde{\pmb n}$ and which contains all relevant information for closed superstring theory, and we look at the asymptotic expansion at $\pmb s=\pmb 0$ of these integrals.
We then give a new\footnote{This statement was first proved (in a
  stronger form) in~\cite{Schlotterer:2018zce}
  and~\cite{Brown:2019wna} while we were writing our paper.} proof of the following (see Theorem~\ref{MainThmSec7}):
\begin{thm}\label{mainintroteo}
The asymptotic expansion coefficients of the integrals~\eqref{SupAmpJintro} are single-valued multiple zeta values.
\end{thm}

As explained in Section~\ref{sec:clos-string-part}, one may combine Theorem~\ref{prop:holfacintro} with folklore results on the asymptotic expansion of Aomoto-Gel'fand hypergeometric functions (see Section~\ref{ssec:AsympAG}) to deduce information on the asymptotic expansion coefficients of the integrals~\eqref{SupAmpJintro}. Unfortunately, this does not suffice to prove Theorem~\ref{mainintroteo}. Our proof takes instead a completely different route, and it relies on a technical result for the integration of a class of functions called single-valued hyperlogarithms.

\emph{Hyperlogarithms} are homotopy invariant iterated integrals of rational functions on the punctured complex plane. Once the starting point of the integration path is fixed, hyperlogarithms only depend on the position of the endpoint (holomorphically) and on the homotopy class of the path, and so they are multi-valued functions of the endpoint. Brown described a way to combine hyperlogarithms and their complex conjugates to remove their dependence on the homotopy class of the path, thus constructing a family of (real-analytic) functions defined on the whole punctured plane, called single-valued hyperlogarithms~\cite{BrownNote}. This construction can also be explained, a posteriori, with the theory of single-valued periods: hyperlogarithms are periods in families, and single-valued hyperlogarithms are their images under the single-valued projection.

In the case where the points removed from~$\mathbb{C}$ are~$0$ and~$1$, (single-valued) hyperlogarithms are called (single-valued) multiple polylogarithms, and their special values at $z=1$ are (single-valued) multiple zeta values. Even though all the integrals that we consider can eventually be expressed in terms of (multi-valued or single-valued) multiple polylogarithms and multiple zeta values, our analysis of the asymptotic expansion of the integrals~\eqref{SupAmpJintro} crucially relies on the general theory of hyperlogarithms. 

Our idea is to generalise at the same time results of Panzer on the
integration of (multi-valued) hyperlogarithms~\cite{Panzer:2015ida}, and
results of Schnetz on the integration of single-valued multiple
polylogarithms~\cite{Schnetz:2013hqa}. We demonstrate that (absolutely convergent) integrals of the kind
\begin{equation}\label{eqintrorand}
\int_{\mathbb{C}}\frac{f(z)\,dz\,d\bar z}{\prod_{r=1}^m(z-\sigma_{i_r})\prod_{s=1}^n(\overline{z}-\overline{\sigma}_{j_s})},
\end{equation}
for $\sigma_i\in\mathbb{C}$ and~$f$ a single-valued hyperlogarithm, can be written as single-valued hyperlogarithms in any of the variables~$\sigma_i$, with coefficients contained in a ring of special values of single-valued hyperlogarithms which depends on~$f$ (see Theorem~\ref{LemmaAppD} for a precise statement). This is the key result to prove Theorem~\ref{mainintroteo}, but it is also an interesting general result\footnote{Similar kinds of results were obtained with a more geometric approach in~\cite{DelDuca:2016lad} to study multi-Regge kinematics or in~\cite{Banks:2018rul} to study Kontsevich's deformation quantization.} which can be useful in other contexts, as already shown in~\cite{Vanhove:2020qtt}. 

The asymptotic expansion coefficients of~\eqref{SupAmpJintro} can be written\footnote{This is a rather delicate point, as one first needs to separate a polar part from an absolutely convergent part of the integrals~\eqref{SupAmpJintro}. We indicate an analytic procedure to do this, but we do not give the details in the general case. A different approach to separate and compute the polar part in the general case can be found in~\cite{Brown:2019wna}.} as multiple integrals over~$\mathbb{C}^N$ of single-valued hyperlogarithms. As integrals over any of the coordinates~$z_i$, they have the form~\eqref{eqintrorand}, hence they can be recursively integrated using our above-mentioned result, which eventually implies the statement of Theorem~\ref{mainintroteo}. It is possible (see Sections~\ref{Sectionk=1} and~\ref{Sectionk=2}) to implement our proof's method and explicitly compute the asymptotic expansion coefficients of~\eqref{SupAmpJintro}, without any knowledge of the analogous open string coefficients, contrary to all previous approaches. Moreover, our method exposes the fact that these coefficients are special values of single-valued functions, thus providing an intuitive explanation for their single-valued period nature.

\subsubsection{Organisation of the paper}

Sections~\ref{Sec:Hyperlogs} and~\ref{Sec3} contain the mathematical and physical background of the article, respectively. In Section~\ref{Sec:Hyperlogs}, we give an overview of some aspects of the theory of hyperlogarithms. In particular, we introduce multiple polylogarithms and multiple zeta values (both multi-valued and single-valued). A reader who is already familiar with these notions can skip Section~\ref{Sec:Hyperlogs} and still be able to read the paper until Section~\ref{sec:inthyp}, where we go back to the general theory of hyperlogarithms and discuss their integration. In Section~\ref{Sec3}, we give an overview of the physics involved in this paper. In particular, we introduce correlation functions, and genus-zero closed string amplitudes.

In Sections~\ref{sec:general-selberg} and~\ref{sec:factorisation} we focus on global aspects of the closed string amplitudes. In Section~\ref{sec:general-selberg}, which may also be read independently from the rest of the paper, we introduce Aomoto-Gel'fand functions and study some of their properties. The results obtained are used in Section~\ref{sec:factorisation}, where we prove the holomorphic factorisation of the correlation function~\eqref{e:Gdefgeneralintro} and deduce some of its consequences. In particular, in Section~\ref{sec:5.4} we derive the KLT formula.

In Sections~\ref{sec:inthyp} and~\ref{Sec:Final} we focus on local aspects of the closed string amplitudes. In Section~\ref{sec:inthyp}, which is the logical continuation of Section~\ref{Sec:Hyperlogs}, we recall some known results on the integration of hyperlogarithms, and we present new results for the integration of single-valued hyperlogarithms. These results are then used in Section~\ref{Sec:Final}, which is entirely devoted to prove Theorem~\ref{mainintroteo}. Before proving the general case in Section~\ref{SectionGen}, where due to notation complexity we omit some details, we give detailed proofs in Sections~\ref{Sectionk=1} and~\ref{Sectionk=2} of the $N=1$ and $N=2$ cases, respectively. This is also intended to help the reader interested in a practical algorithmic implementation of our method.

Finally, in Appendix~\ref{Sec:ConvReg} we discuss the convergence of the integrals considered.

\section{Hyperlogarithms}\label{Sec:Hyperlogs}

The material presented in this section is not original, except for the formulation of some results in slightly more general contexts. We prove only those statements whose proof is not an obvious adaptation of arguments from the literature, and we provide precise references for the omitted proofs.

\subsection{Definition and first properties}

Let~$M$ be a smooth manifold, let $\omega_1,\ldots,\omega_r$ be smooth complex-valued 1-forms on~$M$ and let $\gamma:[0,1]\rightarrow M$ be a parametrization of a piecewise smooth path. We can write $\gamma^*\omega_i=f_i(t)dt$ for some piecewise smooth function $f_i:[0,1]\rightarrow\mathbb{C}$, where $1\leq i\leq r$. The \emph{iterated integral} of $\omega_1,\ldots,\omega_r$ along $\gamma$ (which does not depend on the chosen parametrisation of $\gamma$) is
\begin{equation}\label{itint}
\int_\gamma \omega_1\cdots\omega_r:=\int_{1\geq t_1\geq\cdots\geq t_r\geq 0}f_1(t_1)\cdots f_r(t_r)\,dt_1\cdots dt_r.
\end{equation}
We say that~$r$ is the \emph{length} of the iterated integral. We will need the following properties:
\begin{itemize}
\item \emph{Composition of paths}.
\begin{equation}\label{pathconc}
\int_{\gamma_1\cdot\gamma_2} \omega_1\cdots\omega_r=\sum_{i=0}^r\int_{\gamma_1}\omega_1\cdots\omega_i\int_{\gamma_2}\omega_{i+1}\cdots\omega_r
\end{equation}
\item \emph{Shuffle product}. Denote by $\shuffle(r,s)$ the set of permutations $\rho$ of $\{1,\ldots ,r+s\}$ such that $\rho(1)<\rho(2)<\cdots <\rho(r)$ and $\rho(r+1)<\rho(r+2)<\cdots <\rho(r+s)$. Then
\begin{equation}\label{shuffleitint}
\int_\gamma\omega_1\cdots\omega_r\int_\gamma\omega_{r+1}\cdots\omega_{r+s}=\sum_{\rho\in\shuffle(r,s)}\int_\gamma\omega_{\rho^{-1}(1)}\cdots\omega_{\rho^{-1}(r+s)}.
\end{equation}
\end{itemize}

Let $X:=\{x_0,x_1,\ldots ,x_n\}$ be an alphabet of $n+1$ letters, and let $\Sigma:=\{\sigma_0,\sigma_1,\sigma_2\ldots ,\sigma_n\}\subset \mathbb{C}$ be a set of $n+1$ distinct complex numbers obtained as the image of an injective map $X\hookrightarrow \mathbb{C}$ (so~$\sigma_i$ is associated to the letter~$x_i$). We always assume that $\sigma_0=0$ and $\sigma_1=1$. We denote by~$X^*$ the free non-commutative monoid containing all possible words in the alphabet~$X$, including the empty word~$e$, and denote by $R\langle X \rangle$ the free $R$-module on~$X^*$, equipped with the (commutative) shuffle product $\shuffle$ which makes it into a ring. A string of~$r$ consecutive letters~$x_i$ will be denoted by~$x_i^r$, and the \emph{length} of a word $w=x_{i_1}\cdots x_{i_r}$ is $|w|=r$.

Define $D:=\mathbb{C}\setminus \Sigma$, then any iterated integral of the differential forms $dx/(x-\sigma_i)$ is homotopy invariant on $D$ (by holomorphicity). Let us fix a simply connected domain~$U$ obtained from~$D$ by cutting out closed half-lines $l(\sigma_i)$ starting at the points~$\sigma_i$ and not intersecting among themselves, and let us choose a branch of the logarithm on~$\mathbb{C}\setminus l(0)$. If  $w=x_{i_1}\cdots x_{i_r}$ and $i_r\neq 0$, we define for $z\in U$ the \emph{hyperlogarithm} associated with $w$ as the homotopy invariant iterated integral
\begin{equation}\label{defhypitint}
L_{w}(z)=\int_{[0,z]}\frac{dx}{x-\sigma_{i_1}}\cdots \frac{dx}{x-\sigma_{i_r}},
\end{equation} 
where $[0,z]$ is any path contained\footnote{The starting point~$0$ lies outside~$U$. This forces us to define separately the hyperlogarithms associated with words ending with~$x_0$, because the corresponding iterated integral would diverge. Alternatively, one can define all hyperlogarithms as iterated integrals making use of a regularization procedure.} in~$U$ starting at~$0$ and ending at~$z$. In order to define hyperlogarithms for any word $w\in X^*$, we set $L_{x_0^r}(z)=\log^r(z)/r!$, $L_e(z)=1$, and we require that $w\rightarrow L_w(z)$ respects the shuffle product (which by~(\ref{defhypitint}) and~(\ref{shuffleitint}) is already true for words not ending with~$x_0$). We say that~$w$ is the \emph{label} and~$z$ is the \emph{argument} of $L_w(z)$. 

All hyperlogarithms extend to holomorphic multi-valued functions on~$D$. For instance, for $i\neq 0$
\begin{equation}
L_{x_i^r}(z)\,=\,\frac{1}{r!}\log^r\bigg(1-\frac{z}{\sigma_i}\bigg).
\end{equation}

If $X=\{x_0,x_1\}$ and $\Sigma=\{0,1\}$ then hyperlogarithms are called \emph{multiple polylogarithms} (in one variable)~\cite{BrownSVMPL}, and they contain for instance all the classical polylogarithms:
\begin{equation}
L_{x_0^{n-1}x_1}(z)\,=\,-\sum_{k>0}\frac{z^k}{k^n}\,=\,-\text{Li}_n(z).
\end{equation}

Hyperlogarithms are characterised as follows:
\begin{thm}[Brown,~\cite{BrownNote}]\label{ThmBrownmvhyp}
The hyperlogarithms $\{L_w(z):w\in X^*\}$ constitute the unique family of holomorphic functions satisfying for $z\in U$
\begin{equation}\label{DiffHyperInt}
\frac{\partial}{\partial z}L_{x_iw}(z)\,=\,\frac{L_w(z)}{z-\sigma_i}\,,
\end{equation}
such that $L_e(z)=1$, $L_{x_0^r}(z)=\log^r(z)/r!$ for all $n\in\mathbb{N}$ and $L_w(z)\rightarrow 0$ as $z\rightarrow 0$ for any other word~$w$.
\end{thm}

\subsection{Rings of hyperlogarithms}\label{ssec:ringhyper}

For any ring\footnote{All results of~\cite{BrownNote} presented here were originally demonstrated only in the case $R=\mathbb{C}$. If their proof can be straightforwardly adapted to any ring~$R$, we will state the more general version and still attribute them to~\cite{BrownNote}.} $R\leq\mathbb{C}$ and any alphabet~$X$ we define the $R$-module $\mathcal{H}_{X,R}:=R\langle L_w(z)\,|\,w\in X^*\rangle$ (which implicitly depends also on the set $\Sigma$ associated to $X$). Because all hyperlogarithms satisfy shuffle-product identities, $\mathcal{H}_{X,R}$ is also a ring, contained in the ring of holomorphic functions on $U$.

We consider also the ring $\mathcal{A}_{X,R}:=\mathcal{O}_{\Sigma,R}\otimes_R \mathcal{H}_{X,R}$, where
\begin{equation}\label{RegFun}
\mathcal{O}_{\Sigma,R}:=R\bigg[z,\sigma_i,\frac{1}{z-\sigma_i},\frac{1}{\sigma_j-\sigma_i}:\sigma_i,\sigma_j\in \Sigma, \sigma_i\neq\sigma_j\bigg]\,.
\end{equation}

$\mathcal{A}_{X,R}$ is closed under holomorphic differentiation, and via $w\to L_w$ it is isomorphic, as a differential ring, to the \emph{universal ring of hyperlogarithms} $\mathcal{O}_{\Sigma,R}\otimes_R R\langle X\rangle$, equipped with a suitable derivation~$\partial$ \cite[Theorem 3.7]{BrownNote}. Using this, one can prove that every function $f(z)\in\mathcal{A}_{X,R}$ has a primitive in~$\mathcal{A}_{X,R}$, which is unique up to a constant, and that shuffle-product identities between hyperlogarithms are the only algebraic relations in~$\mathcal{A}_{X,R}$ \cite[Corollary 3.13]{BrownNote}. In particular, both~$\mathcal{A}_{X,R}$ and~$\mathcal{H}_{X,R}$ inherit a grading by the length of the labels, and~$\mathcal{H}_{X,R}$ is then isomorphic as a graded ring to~$R\langle X\rangle$ via $w\to L_w$.

\subsection{Special values of hyperlogarithms}\label{Ssec:specialhyper}
For any $w\in X^*$ and for any $\sigma_i\in \Sigma$ there exists an integer $K_i(w)\geq 0$ such that, in the intersection of a neighborhood of~$\sigma_i$ with the cut plane~$U$,
\begin{equation}\label{Exp1}
L_w(z)=\sum_{k=0}^{K_i(w)}\sum_{j\geq 0}c_{k,j}^{(i)}(w)\,(z-\sigma_i)^j\,\log^k(z-\sigma_i)\,,
\end{equation}
where $c_{j,k}^{(i)}(w)\in\mathbb{C}$. Similarly, for $z\in U$ close to~$\infty$ there are $K_\infty(w)\geq 0$ and $c_{k,j}^{(\infty)}(w)\in\mathbb{C}$ such that
\begin{equation}\label{Exp2}
L_w(z)=\sum_{k=0}^{K_\infty(w)}\sum_{j\geq 0}c_{k,j}^{(\infty)}(w)\,z^{-j}\,\log^k(z).
\end{equation}

For $\sigma_i\in \Sigma\cup\{\infty\}$ we define the \emph{regularized value} of $L_w(z)$ at $\sigma_i$ to be $L_w(\sigma_i):=c_{0,0}^{(i)}(w)$. In particular, $L_w(0)=\delta_{w,e}$. From now on, whenever we talk of special values of hyperlogarithms, we implicitly mean regularized values. For $i\neq 0$, we define the \emph{weight} $W(L_w(\sigma_i))$ of $L_w(\sigma_i)$ to be the length $|w|$ of the label~$w$.

For an alphabet $X=\{x_j\}_{0\leq j\leq n}$, a ring $R\leq \mathbb{C}$ and for every\footnote{We exclude the trivial case where $j=0$.} $1\leq j\leq n$ we define the $R$-module $\mathcal{S}_{X,R,j}=R\langle L_w(\sigma_j):w\in X^*\rangle$ (keeping again the dependence on the set $\Sigma$ associated to $X$ implicit). This is a subring of $\mathbb{C}$ because of the shuffle-product property~(\ref{shuffleitint}) of iterated integrals. Then we define $\mathcal{S}_{X,R}$ to be the smallest subring of $\mathbb{C}$ containing all $\mathcal{S}_{X,R,j}$. This ring is endowed with a filtration\footnote{Computer experiments, as well as the period conjecture, suggest that this filtration should be a grading in some cases, including that where $\mathcal{S}_{X,R}=\mathcal{S}_{\{x_0,x_1\},\mathbb{Q}}$ leading to multiple zeta values, but such conjectures seem out of reach.} induced by the weight.

If $X=\{x_0,x_1\}$ and $\Sigma=\{0,1\}$, the special values $L_w(1)$ are \emph{multiple zeta values}. These (real) numbers generate a $\mathbb{Q}$-algebra $\mathcal{Z}:=\mathcal{S}_{\{x_0,x_1\},\mathbb{Q},1}\equiv\mathcal{S}_{\{x_0,x_1\},\mathbb{Q}}$ and are usually defined as the nested series
\begin{equation}\label{mzvsumrep}
\zeta(k_1,\ldots ,k_r):=\sum_{0<n_1<\cdots <n_r}\frac{1}{n_1^{k_1}\cdots n_r^{k_r}},
\end{equation}
where $k_i\in\mathbb{N}$, $k_r\geq 2$. It is a simple exercise to verify that $\zeta(k_1,\ldots ,k_r)=L_{x_0^{k_r-1}x_1\cdots x_0^{k_1-1}x_1}(1)$. The weight of $\zeta(k_1,\ldots ,k_r)$ is therefore $|x_0^{k_r-1}x_1\cdots x_0^{k_1-1}x_1|=k_1+\cdots +k_r$.

\subsection{Generating series and monodromy}\label{sec:monodromyseries}

For a given alphabet~$X$ (and its associated set~$\Sigma$) consider the $\mathbb{C}\langle\langle X\rangle\rangle$-valued generating series
\begin{equation}
L_X(z)=\sum_{w\in X^*} L_w(z)\,w
\end{equation}
of all hyperlogarithms, where $\mathbb{C}\langle\langle X\rangle\rangle$ denotes the ring of formal series in $X^*$ with complex coefficients. Theorem~\ref{ThmBrownmvhyp} is then equivalent to saying that
\begin{equation}\label{KZeq}
\frac{\partial}{\partial z}L_X(z)=\sum_{i=0}^n\frac{x_i}{z-\sigma_i}L_X(z),
\end{equation}
and that $L_X$ is the only $\mathbb{C}\langle\langle X\rangle\rangle$-valued solution of~(\ref{KZeq}) on the simply connected domain $U$ such that
\begin{equation}
L_X(z)=f_0(z)\exp(x_0\log(z)),
\end{equation}
where $f_0(z)$ is a holomorphic function on $U$ with $\lim_{z\rightarrow 0}f_0(z)=1$~\cite{GonzalesLorca}. If $X=\{x_0,x_1\}$ and $\Sigma=\{0,1\}$, the generating series $L_{\{x_0,x_1\}}(1)$ of regularized special values at $z=1$ is known as the \emph{Drinfel'd associator}, and equation~(\ref{KZeq}) is known as the \emph{Knizhnik}-\emph{Zamolodchikov equation}. More generally, we will be interested in all (regularized) special values $L_X(\sigma_i)$.

Let now $C^\infty(U)$ denote the algebra of real-analytic functions on $U$, and let us fix $z_0\in U$. The fundamental group $\pi_1(D,z_0)$ of the punctured plane $D$ is the free group on generators $\gamma_0,\ldots ,\gamma_n$, where each $\gamma_i$ is a loop based at~$z_0$ and winding around~$\sigma_i$ once in the positive direction. For each $0\leq i\leq n$ we write $M_{\sigma_i}:C^\infty(U)\rightarrow C^\infty(U)$ for the monodromy operator given by analytic continuation of functions around~$\gamma_i$. 
\begin{propos}[Brown,~\cite{BrownNote}]\phantomsection\label{MonodromyThm}
For each $0\leq i\leq n$ we have
\begin{equation}\label{monodromy}
M_{\sigma_i}L_X(z)=L_X(z)(L_X(\sigma_i))^{-1}e^{2\pi ix_i}L_X(\sigma_i).
\end{equation}
\end{propos}

\subsection{Single-valued hyperlogarithms}
We define a map $\sim:X^* \rightarrow X^*$ by sending $w=x_{i_1}\cdots x_{i_r}$ to $\tilde{w}:=x_{i_r}\cdots x_{i_1}$, and we extend it by linearity to $\mathbb{C}\langle\langle X\rangle\rangle$. We call \emph{single-valued hyperlogarithms} the coefficients $\mathcal{L}_w(z)$ of words $w\in X^*$ in the generating series
\begin{equation}
\mathcal{L}_X(z):=L_X(z)\widetilde{\overline{L_{X^\prime}}}(z),
\end{equation}
where $X^\prime$ is an alphabet associated to the same set $\Sigma$ formed by letters $x_i'\in\mathbb{C}\langle\langle X\rangle\rangle$ which satisfy
\begin{equation}\label{AlphabetSV}
\widetilde{\overline{L_{X^\prime}}}(\sigma_i)x_i^\prime\widetilde{\overline{L_{X^\prime}}}(\sigma_i)^{-1}=L_X(\sigma_i)^{-1}x_iL_X(\sigma_i).
\end{equation}

This construction was first proposed by F. Brown in~\cite{BrownNote}, where eq.~(\ref{AlphabetSV}) was proven to admit a unique solution $x_i'\in\mathbb{C}\langle\langle X\rangle\rangle$. In particular, one has $x_0^\prime=x_0$ and, for $i\neq 0$, $x_i^\prime = x_i$ modulo words $w\in X^*$ with $|w|\geq 4$. Knowing that such a solution exists, Proposition~\ref{MonodromyThm} implies that, for all $0\leq i\leq n$,
\begin{align}
M_{\sigma_i}\mathcal{L}_X(z)&=L_X(z)L_X(\sigma_i)^{-1}e^{2\pi ix_i}L_X(\sigma_i)\widetilde{\overline{L_{X^\prime}}}(\sigma_i)e^{-2\pi ix^\prime_i}\widetilde{\overline{L_{X^\prime}}}(\sigma_i)^{-1}\widetilde{\overline{L_{X^\prime}}}(z) \notag \\
&=L_X(z)\widetilde{\overline{L_{X^\prime}}}(z)=\mathcal{L}_X(z),
\end{align}
which proves that $\mathcal{L}_X(z)$ is indeed well-defined, i.e. single-valued, on the whole punctured complex plane~$D$.

The simplest examples of single-valued hyperlogarithms are $\mathcal{L}_{x_0^{n}}(z)=\log^n|z|^2/n!$ and (for $i\neq 0$)
\begin{equation}
\mathcal{L}_{x_i^n}(z)=\frac{1}{n!}\log^n\bigg|1-\frac{z}{\sigma_i}\bigg|^2.
\end{equation}
They are indeed ``single-valued versions'' of the corresponding hyperlogarithms $L_{x_0^{n}}(z)$ and $L_{x_i^n}(z)$.

One can characterise single-valued hyperlogarithms as follows.
\begin{thm}[Brown,~\cite{BrownNote}]\phantomsection\label{BrownsvHyperlog}
The series $\mathcal{L}_X(z)$ is the unique real-analytic solution on $D$ to the differential equations
\begin{equation}
\frac{\partial}{\partial z}\mathcal{L}_X(z)=\sum_{i=0}^n\frac{x_i}{z-\sigma_i}\mathcal{L}_X(z)
\end{equation}
and
\begin{equation}\label{AntiHoloSV}
\frac{\partial}{\partial \overline{z}}\mathcal{L}_X(z)=\mathcal{L}_X(z)\sum_{i=0}^n\frac{x_i^\prime}{\overline{z}-\overline{\sigma}_i}
\end{equation}
such that $\mathcal{L}_X(z)\sim \exp(x_0\log|z|^2)$ as $z\rightarrow 0$.
\end{thm}

\subsection{Rings of single-valued hyperlogarithms}\label{sec:ringsv}
For any subring $R\leq \mathbb{C}$ and any alphabet~$X$ consider $\mathcal{H}_{X,R}\otimes_R\overline{\mathcal{H}_{X,R}}$ and $\mathcal{A}_{X,R}\otimes_R\overline{\mathcal{A}_{X,R}}$, which are subrings of $C^{\infty}(U)$. The only algebraic relations inside both rings are induced by shuffle identities, and so in particular all products $L_{w_1}(z)\,\overline{L_{w_2}(z)}$ are linearly independent over\footnote{$\overline{\mathcal{O}_{\Sigma,R}}$ is the ring generated over~$R$ by the complex conjugates of the generators of~$\mathcal{O}_{\Sigma,R}$ from~(\ref{RegFun}).} $\mathcal{O}_{\Sigma,\mathbb{C}}\otimes_\mathbb{C}\overline{\mathcal{O}_{\Sigma,\mathbb{C}}}$ \cite[Corollary~7.2]{BrownNote}.

We define the $R$-module $\mathcal{H}_{X,R}^{\rm sv}:=R\langle\mathcal{L}_w(z)\,|\,w\in X^* \rangle$, as well as $\mathcal{A}^{\rm sv}_{X,R}:=\mathcal{O}_{\Sigma,R}\otimes_R\overline{\mathcal{O}_{\Sigma,R}}\otimes_R \mathcal{H}_{X,R}^{\rm sv}$. They are both contained in $C^\infty(D)$. By construction, the restriction of~$\mathcal{H}^{\rm sv}_{X,R}$ and~$\mathcal{A}^{\rm sv}_{X,R}$ to the simply connected subdomain $U\subset D$ are submodules of $\mathcal{H}_{X,R}\otimes_R\overline{\mathcal{H}_{X,R}}$ and $\mathcal{A}_{X,R}\otimes_R\overline{\mathcal{A}_{X,R}}$, respectively. An important result of~\cite{BrownNote} (see Theorem 7.4) is that, even though single-valued hyperlogarithms are not defined as iterated integrals, they satisfy shuffle identities
\begin{equation}\label{LemmaShuffle}
\mathcal{L}_{w_1}(z)\mathcal{L}_{w_2}(z)=\sum_{w\in w_1\shuffle w_2}\mathcal{L}_{w}(z),
\end{equation}
so that~$\mathcal{H}^{\rm sv}_{X,R}$ and~$\mathcal{A}^{\rm sv}_{X,R}$ are rings and are isomorphic (as graded rings) to $R\langle X\rangle$ and $\mathcal{O}_{\Sigma,R}\otimes_R\overline{\mathcal{O}_{\Sigma,R}}\otimes_R R\langle X\rangle$, respectively, via the map $w\to \mathcal{L}_w$. More precisely, the ring~$\mathcal{A}^{\rm sv}_{X,R}$ is isomorphic as a differential graded ring (with respect to the holomorphic differentiation) to $\mathcal{O}_{\Sigma,R}\otimes_R\overline{\mathcal{O}_{\Sigma,R}}\otimes_R R\langle X\rangle$; it is a single-valued realisation of the universal ring of hyperlogarithms mentioned in Section~\ref{ssec:ringhyper}. The ``single-valued projection''\footnote{This is a single-valued projection from periods to single-valued periods, in the sense of the introduction.} ${\rm sv}:L_w\mapsto \mathcal{L}_w$ then induces isomorphisms $\mathcal{H}_{X,R}\simeq\mathcal{H}^{\rm sv}_{X,R}$ and $\mathcal{A}_{X,R}\otimes_R\overline{\mathcal{O}_{\Sigma,R}}\simeq \mathcal{A}^{\rm sv}_{X,R}$.

As in the multi-valued case, there are important consequences of these isomorphisms. First of all, the only algebraic relations inside both~$\mathcal{H}^{\rm sv}_{X,R}$ and~$\mathcal{A}^{\rm sv}_{X,R}$ are induced by shuffle identities. Moreover, importantly for us, every element in~$\mathcal{A}^{\rm sv}_{X,R}$ has a primitive in~$\mathcal{A}^{\rm sv}_{X,R}$ with respect to $\partial /\partial z$.

One may also consider the anti-holomorphic differentiation. Eq.~(\ref{AntiHoloSV}) implies that~$\mathcal{A}^{\rm sv}_{X,R}$ is closed under its action as soon as $R$ is a subring of $\mathbb{C}$ which contains the coefficients of every series $x_i'$, and in fact with these assumptions every element in~$\mathcal{A}^{\rm sv}_{X,R}$ has a primitive in~$\mathcal{A}^{\rm sv}_{X,R}$ with respect to $\partial /\partial \overline{z}$.

We conclude with a characterisation of~$\mathcal{H}^{\rm sv}_{X,R}$ and~$\mathcal{A}^{\rm sv}_{X,R}$.
\begin{propos}\label{CorollaryRefinement}
Let~$F$ be a single-valued function on~$D$ which is an $R$-linear combination (resp. $\mathcal{O}_{\Sigma,R}\otimes_R\overline{\mathcal{O}_{\Sigma,R}}$-linear combination) of products $L_{w_1}(z)\overline{L_{w_2}(z)}$. Then $F\in\mathcal{H}^{\rm sv}_{X,R}$ (resp. $F\in\mathcal{A}^{\rm sv}_{X,R}$).
\end{propos}
\begin{proof}
These statements were proven by Brown in~\cite{BrownNote} (see Theorem~8.1) for $R=\mathbb{C}$. We want to prove them for general subrings $R<\mathbb{C}$, and we only prove the first, as the proof of the second is completely similar. Suppose that $F(z)=\sum_{u_1,u_2}c_{u_1,u_2}L_{u_1}(z)\overline{L_{u_2}(z)}$ is single-valued and that $c_{u_1,u_2}\in R$. By Brown's result, we can also write $F(z)=\sum_{w}k_w\mathcal{L}_w(z)$ for some $k_w\in \mathbb{C}$. Because $x_i'=x_i+$ higher-length terms, for any word~$w$ we have $\mathcal{L}_w(z)=L_w(z)+(\mathcal{L}_w(z)-L_w(z))$, with $\mathcal{L}_w(z)-L_w(z)=\sum_{w_1,w_2}l_{w_1,w_2}L_{w_1}(z)\overline{L_{w_2}(z)}$ such that $|w_2|\geq 1$. We have mentioned at the beginning of this section that all products $L_{w_1}(z)\overline{L_{w_2}(z)}$ are linearly independent over $\mathcal{O}_{\Sigma,\mathbb{C}}\otimes_\mathbb{C}\overline{\mathcal{O}_{\Sigma,\mathbb{C}}}$. In particular, we can compare our two different expressions for $F(z)$ term by term. Comparing the holomorphic monomials, obtained in the second expression by discarding the terms $k_w(\mathcal{L}_w(z)-L_w(z))$, we find $k_w=c_{w,e}\in R$, as claimed.
\end{proof}

\subsection{Special values of single-valued hyperlogarithms}\label{sec:specialvalues}

Similarly to the holomorphic case, for each $\sigma_i\in \Sigma$ there exists an integer $K_i(w)\geq 0$ such that, in a neighborhood of~$\sigma_i$,
\begin{equation}\label{FinPtsExp}
\mathcal{L}_w(z)=\sum_{k=0}^{K_i(w)}\sum_{m=0}^{\infty}\sum_{n=0}^{\infty} c_{k,m,n}^{(i)}(w)\,(z-\sigma_i)^m\,(\overline{z}-\overline{\sigma}_i)^n\,\log^k|z-\sigma_i|^2\,,
\end{equation}
and also there exists $K_\infty(w)\geq 0$ such that, in a neighborhood of~$\infty$,
\begin{equation}\label{InfExp}
\mathcal{L}_w(z)=\sum_{k=0}^{K_\infty(w)}\sum_{m=0}^{\infty}\sum_{n=0}^{\infty} c_{k,m,n}^{(\infty)}(w)\,z^{-m}\,\overline{z}^{-n}\,\log^k|z|^2\,,
\end{equation}
where $c_{k,m,n}^{(i)}(w),c_{k,m,n}^{(\infty)}(w)\in\mathbb{C}$.

Just as we did in the holomorphic case, we can therefore define regularized values of single-valued hyperlogarithms at points $\sigma_i\in \Sigma\cup\{\infty\}$ as $\mathcal{L}_w(\sigma_i):=c_{0,0,0}^{(i)}(w)$, we can consider their regularized generating series~$\mathcal{L}_X(\sigma_i)$, and for $i\neq 0$ we can define their \emph{weight} $W(\mathcal{L}_w(\sigma_i)):=|w|$.
\begin{defn}[Single-valued multiple zeta values~\cite{Brown:2013gia}]
Let $X=\{x_0,x_1\}$ and $\Sigma =\{0,1\}$. We call single-valued multiple zeta values the (regularised) special values $\mathcal{L}_w(1)$.
\end{defn}

We denote by~$\mathcal{Z}^{\rm sv}$ the $\mathbb{Q}$-algebra generated by single-valued multiple zeta values. The restriction to $z=1$ of the morphism ${\rm sv}:L_w\mapsto \mathcal{L}_w$ from the previous section induces a surjective map\footnote{Assuming the period conjecture, this is a morphism of $\mathbb{Q}$-algebras~\cite{Brown:2013gia}, otherwise it must be regarded as a map of sets.} from~$\mathcal{Z}$ to~$\mathcal{Z}^{\rm sv}$. Using the notation $\zeta(k_1,\ldots ,k_r)$ for multiple zeta values from~\eqref{mzvsumrep}, and denoting by $\zeta^{\rm sv}(k_1,\ldots ,k_r)$ the image of $\zeta(k_1,\ldots ,k_r)$ in~$\mathcal{Z}^{\rm sv}$ via~${\rm sv}$, so that $\zeta^{\rm sv}(k_1,\ldots ,k_r)=\mathcal{L}_{x_0^{k_r-1}x_1\cdots x_0^{k_1-1}x_1}(1)$, one finds that $\zeta^{\rm sv}(2k)=0$ and that $\zeta^{\rm sv}(2k+1)=2\,\zeta(2k+1)$ for all $k\geq 1$. We refer to~\cite{Brown:2013gia} for more details on~$\mathcal{Z}^{\rm sv}$, including a conjectural description of its $\mathbb{Q}$-algebra structure.

It is important to remark the fact that~$\mathcal{Z}^{\rm sv}$ is contained in~$\mathcal{Z}$, as explained below. By definition, for any $X$ and any $w\in X^*$ the single-valued hyperlogarithm $\mathcal{L}_w(z)$ can be written in a unique way as a $\mathbb{C}$-linear combination of products $L_{w_1}(z)\overline{L_{w_2}(z)}$. Looking more carefully at the equations defining the alphabet~$X^\prime$, however, one can see that the coefficients of these linear combinations must in fact belong to $\mathcal{S}_{X,\mathbb{Q}}\otimes_\mathbb{Q}\overline{\mathcal{S}_{X,\mathbb{Q}}}\subset\mathbb{C}$. In particular, if $X=\{x_0,x_1\}$ and $\Sigma =\{0,1\}$, the coefficients belong to~$\mathcal{Z}$, and so $\mathcal{L}_w(1)\in\mathcal{Z}$. The next proposition refines this observation and provides an optimal statement about the ring containing these coefficients, generalising the analogous result for the alphabet $\{x_0,x_1\}$ demonstrated by Schnetz in~\cite{Schnetz:2013hqa}. In order to be able to state it, however, we need some notation.

For an alphabet $X$ and a ring $R\leq \mathbb{C}$, we define $\mathcal{S}^{\rm sv}_{X,R,j}=R\langle\mathcal{L}_w(\sigma_j):w\in X^*\rangle$ (by eq.~(\ref{LemmaShuffle}) this is a ring) and $\mathcal{S}^{\rm sv}_{X,R}$ to be the ring generated over $R$ by all $\mathcal{S}^{\rm sv}_{X,R,j}$ with $j\neq 0$. Therefore if $X=\{x_0,x_1\}$ we have $\mathcal{Z}^{\rm sv}=\mathcal{S}^{\rm sv}_{\{x_0,x_1\},\mathbb{Q},1}\equiv\mathcal{S}^{\rm sv}_{\{x_0,x_1\},\mathbb{Q}}$. The ring $\mathcal{S}^{\rm sv}_{X,R}$ is endowed with a filtration induced by the weight. In particular, we say that $c\in \mathcal{S}^{\rm sv}_{X,R}$ has homogeneous weight if it is an $R$-linear combination of monomials (given by products of special values), and each monomial has the same weight.
 
\begin{propos}\phantomsection\label{LemmaCoeffSV}
For all $w\in X^*$ one can write
\begin{equation}
\mathcal{L}_w(z)=\sum_{w_1,w_2\in X^*}c_{w_1,w_2}L_{w_1}(z)\overline{L_{w_2}(z)},
\end{equation}
where the coefficients $c_{w_1,w_2}$ belong to $\mathcal{S}^{\rm sv}_{X,\mathbb{Q}}$ and have homogeneous weight $W(c_{w_1,w_2})=|w|-|w_1|-|w_2|$.
\end{propos}
\begin{proof} 
First of all, by Theorem~\ref{BrownsvHyperlog} we have for all $0\leq j\leq n$
\begin{align}\label{Aux2}
\lim_{z \rightarrow\sigma_j}(\overline{z}-\overline{\sigma}_j)\frac{\partial}{\partial \overline{z}}\mathcal{L}_X(z)&=\lim_{z \rightarrow\sigma_j}(\overline{z}-\overline{\sigma}_j)\mathcal{L}_X(z)\Big(\frac{x_0^\prime}{\overline{z}}+\cdots +\frac{x_n^\prime}{\overline{z}-\overline{\sigma}_n}\Big)\notag \\
&=\mathcal{L}_X(\sigma_j)x_j^\prime\,,
\end{align}
which implies that
\begin{equation}\label{SideTwo}
\lim_{z \rightarrow\sigma_j}(\overline{z}-\overline{\sigma}_j)\frac{\partial}{\partial \overline{z}}\mathcal{L}_w(z)=(x_j^\prime|w)+\sum_{\substack{uv=w\\|v|<|w|}}\mathcal{L}_u(\sigma_j)(x_j^\prime|v)\,,
\end{equation}
where $(x_j^\prime|w)$ denotes the coefficient of the word $w$ in the series $x_j^\prime\in\mathbb{C}\langle\langle X\rangle\rangle$.

In particular, since for $j=0$ we know that $x_0^\prime=x_0$ and that $\mathcal{L}_w(0)=\delta_{w,e}$, we get
\begin{equation}\label{Aux1}
\lim_{z \rightarrow 0}\overline{z}\frac{\partial}{\partial \overline{z}}\mathcal{L}_w(z)=\delta_{w,x_0}.
\end{equation}
Moreover, we have the following:
\begin{lemma}\phantomsection\label{lemmaconcatsv}
For any $1\leq i\leq n$ consider the injection $\phi_i:X\hookrightarrow \mathbb{C}$ given by $x_j\mapsto \sigma_i-\sigma_j$ for $0\leq j\leq n$. Let $\mathcal{L}_{X,\phi_i}(z)$ be the generating series of single-valued hyperlogarithms associated to the alphabet $X$ and to the set of points $\phi_i(X)$, then
\begin{equation}\label{eqconc}
\mathcal{L}_X(\sigma_i-z)=\mathcal{L}_{X,\phi_i}(z)\mathcal{L}_X(\sigma_i)
\end{equation}
\end{lemma}
\begin{proof}[Proof of the lemma]
Formula~(\ref{pathconc}) for the composition of paths of iterated integrals implies that
\begin{equation}
\sum_{j\geq 0}\int_0^{\sigma_i-z}\underbrace{\omega_{X}(z)\cdots \omega_{X}(z)}_{j}=\sum_{j\geq 0}\int_{\sigma_i}^{\sigma_i-z}\underbrace{\omega_{X}(z)\cdots \omega_{X}(z)}_{j}\sum_{j\geq 0}\int_0^{\sigma_i}\underbrace{\omega_{X}(z)\cdots \omega_{X}(z)}_{j}
\end{equation}
where $\omega_{X}(z):=\sum_{i=0}^n\frac{x_i\,dz}{z-\sigma_i}$. The left-hand side of this formula is $L_X(\sigma_i-z)$, while the second term on the right-hand side is $L_X(\sigma_i)$. Substituting $z'=\sigma_i-z$ in the integrands of the first term on the right-hand side leads to the identity
\begin{equation}
L_X(\sigma_i-z)=L_{X,\phi_i}(z)L_X(\sigma_i),
\end{equation}
where $L_{X,\phi_i}(z)$ is the generating series of hyperlogarithms associated to the alphabet $X$ and to the set of points $\phi_i(X)$. This implies that
\begin{equation}\label{eqconc2}
\mathcal{L}_X(\sigma_i-z)=L_{X,\phi_i}(z)L_X(\sigma_i)\widetilde{\overline{L_{X^\prime}}}(\sigma_i)\widetilde{\overline{L_{X^\prime,\phi_i}}}(z).
\end{equation}
The right-hand sides of~(\ref{eqconc}) and~(\ref{eqconc2}) both satisfy the differential equation
\begin{equation}
\frac{\partial}{\partial z}F(z)\,=\,\sum_{j=0}^n\frac{x_j}{z-\tau_j}F(z),
\end{equation}
where $\tau_j:=\phi_i(x_j)=\sigma_i-\sigma_j$. Since their asymptotic behaviour at $z=0$ is the same, they must coincide.
\end{proof}
Combining this lemma and eq.~(\ref{Aux1}) we find that for any $w\in X^*$
\begin{equation}\label{SideOne}
\lim_{z \rightarrow\sigma_j}(\overline{z}-\overline{\sigma}_j)\frac{\partial}{\partial \overline{z}}\mathcal{L}_w(z)\,=\,\lim_{z \rightarrow 0}\overline{z}\frac{\partial}{\partial \overline{z}}\mathcal{L}_w(\sigma_j-z)\,\in\, \mathcal{S}^{\rm sv}_{X,\mathbb{Q},j}\,.
\end{equation}

Comparing eqs.~(\ref{SideTwo}) and~(\ref{SideOne}) and using induction on the length of the words, we conclude that the coefficients of $x_j^\prime$ belong to $\mathcal{S}^{\rm sv}_{X,\mathbb{Q},j}$ (recall that $x_j^\prime=x_j+$ higher-length terms). This immediately implies the first statement of the proposition. To demonstrate the second statement about the homogeneity of the weight, we observe that if we assign weight $-1$ to each letter $x_j$  of the alphabet $X$, we obviously have that $L_X(z)$ has weight zero. Therefore, by the definition of $\mathcal{L}_X(z)$, it is enough to show that each $x_j^\prime$ has homogeneous weight $-1$, i.e. that $(x_j^\prime|w)$ has weight $|w|-1$ for any word $w$. Once again, this follows by comparing eqs.~(\ref{SideTwo}) and~(\ref{SideOne}) and using induction.
\end{proof}

\section{Correlation functions and closed string amplitudes}\label{Sec3}

\subsection{Single-valuedness of CFT correlators}
\label{sec:cft-correlators}

Conformal field theories are local  two dimensional quantum
field theories.
Correlation functions are vacuum expectation values of the product of
dynamical composite fields (or vertex operators in string theory)
$V_i(\pmb x)$ with $\pmb x=(x,y)\in\mathbb R^2$
(see~\cite{Ginsparg:1988ui,DiFrancesco:1997nk} for some review and
introduction) 
\begin{equation}\label{e:GReal}
 \mathcal  G(\pmb x_1,\dots,\pmb x_n)= \left\langle \prod_{i=1}^n V_i(\pmb x_i)\right\rangle
                    ={1\over Z}\int   D V \, \prod_{i=1}^n
                      V_i(\pmb x_i) \, e^{-S}\,.
                    \end{equation}
                     $Z$ is the partition function. 
The action $S$ is a   function of the elementary fields and of the
two-dimensional metric. This metric is determined by the
geometry of the Riemann surface $\Sigma$ on which the theory is
considered. 

One axiom of conformal field theories is the requirement of single-valuedness of the
correlation functions as functions of the positions $\pmb x_i=(x_i,y_i)$  of the vertex
operators in the euclidean plane $\mathbb R^2$ (locality of the theory  depends on the  absence of
branch cuts)~\cite[Chap.~5]{DiFrancesco:1997nk}.
It means that a physical correlator should not have monodromies when
one varies the position of a given operator around the
position of other operators.

After complexification one can consider that the composite fields 
are functions $V_i(z,\tilde z)$ on $\mathbb C\times \mathbb C$.
The correlation function becomes a multi-valued function of the doubled coordinates
$\mathcal G(z_1,\dots,z_n,\tilde z_1,\dots, \tilde z_n)$ in $\mathbb
C^n\times \mathbb C^n$.
The euclidean real space is recovered when~$\tilde z$ is identified
with the complex conjugate of~$z$, so that $V_i(z,\bar z)= V_i(x,y)$ for $z=x+iy$ and $\bar
z=x-iy$ with $x,y\in\mathbb R$.
It is only on the real slice that the correlation functions are
single-valued.

This is particularly clear in the case of the two-point correlation
function  determined by the $SL(2,\mathbb C)\times SL(2,\mathbb C)$
Ward identities 
\begin{equation}\label{e:C2}
 \mathcal G(z_1,\tilde z_1,z_2,\tilde z_2)= \langle V_1(z_1,\tilde z_1)  V_2(z_2,\tilde z_2)  \rangle  =
{ \delta_{\Delta_1=\Delta_2}\delta_{\tilde\Delta_1=\tilde\Delta_2} N_{12} \over (z_1-z_2)^{2\Delta_1}(\tilde z_1-\tilde z_2)^{2\tilde\Delta_1}}\,.
\end{equation}
The two-point correlation function is non-vanishing only for fields $V_i(z,\tilde z)$
with the same conformal dimensions\footnote{The holomorphic and
  anti-holomorphic conformal
  dimensions are the exponents $\Delta$ and $\tilde\Delta$ under
a coordinate transformation $V(z,\tilde z)\to \left(\partial f(z)\over
\partial z\right)^\Delta \left(\partial \tilde f(\tilde z)\over
\partial\tilde z\right)^{\tilde \Delta} V(f(z),\tilde f(\tilde z))$.}
$\Delta_1=\Delta_2$ and $\tilde \Delta_1=\tilde\Delta_2$, as indicated by the Kronecker
delta functions, and it is therefore single-valued for $\tilde z=\bar z$, as required. $N_{12}$ is  a constant determined by the normalisation of the
fields.
Likewise, the three-point correlation function is fixed to be 
\begin{multline}\label{e:C3}
  \mathcal G(z_1,\tilde z_1,z_2,\tilde z_2,z_3,\tilde z_3)
  =\left\langle \prod_{i=1}^3 V_i(z_i,\tilde z_i) \right \rangle \cr =
{ C_{123} \over 
  (z_1-z_2)^{\Delta_1+\Delta_2-\Delta_3}
  (z_1-z_3)^{\Delta_1+\Delta_3-\Delta_2}
  (z_2-z_3)^{\Delta_2+\Delta_3-\Delta_1}}\cr
\times {1\over  (\tilde z_1-\tilde z_2)^{\tilde\Delta_1+\tilde\Delta_2-\tilde\Delta_3}
  (\tilde z_1-\tilde z_3)^{\tilde\Delta_1+\tilde\Delta_3-\tilde\Delta_2}
  (\tilde z_2-\tilde z_3)^{\tilde\Delta_2+\tilde\Delta_3-\tilde\Delta_1}}
  \,.
\end{multline}
In this case, there is no restriction on the individual conformal dimensions and
$C_{123}$ is a constant depending the type of fields.

The single-valuedness of the correlation functions~\eqref{e:C2}
and~\eqref{e:C3} on the real slice $\tilde
z=\bar z$ imposes that the difference of the conformal weights
$\Delta_i-\tilde\Delta_i$ (the spins) has to be integral. The
single-valued condition does not determine the value of 
the constants $N_{12}$ nor $C_{123}$.

The four-point correlation function is a non-trivial function of the
unique independent cross-ratio in two dimensions
\begin{equation}
  \eta= {(z_1-z_2)(z_3-z_4)\over (z_1-z_3)(z_2-z_4)}\,,
\end{equation}
and reads
\begin{equation}
 \mathcal G(z_1,\tilde z_1,\dots,z_4,\tilde z_4)=\left\langle \prod_{i=1}^4 V_i(z_i,\tilde z_i)  \right\rangle  =
  {G_{12|34}(\eta,\tilde\eta)\over
    \prod_{1\leq i<j\leq 4} (z_i-z_j)^{\Delta_i+\Delta_j-\Delta}
    (\tilde z_i-\tilde z_j)^{\tilde\Delta_i+\tilde\Delta_j-\tilde\Delta} }\,,
\end{equation}
with $\Delta=\frac13\sum_{i=1}^4\Delta_i$ and
$\tilde\Delta=\frac13\sum_{i=1}^4\tilde\Delta_i$.

Using the associativity of the operator
product expansion one can expand the four-point function on the
conformal blocks $\mathcal F_{1234}(k; \eta)$ and  $\widetilde{\mathcal
F}_{1234}(k; \tilde\eta)$ as 
\begin{equation}\label{e:Gdef}
    G_{12|34}(\eta,\tilde\eta) = \sum_{k,\tilde k}
    G_{12|34}^{k,\tilde k}  \mathcal F_{12|34}(k; \eta)
    \widetilde{\mathcal F}_{12|34}(\tilde k; \tilde\eta)\,.
  \end{equation}
The conformal blocks $\mathcal F_{1234}(k; \eta)$ are
holomorphic functions of $\eta$ and the conformal blocks $\widetilde{\mathcal
F}_{1234}(k; \tilde\eta)$ are holomorphic functions of $\tilde\eta$. We call this formula a \emph{holomorphic factorisation}. Keeping in mind that we are interested in the case $\tilde{\eta}=\bar{\eta}$, we call anti-holomorphic blocks the
functions of $\tilde\eta$ only. The $n$-point correlation function $\mathcal G(z_1,\tilde
z_1,\dots,z_n,\tilde z_n)$ can be inductively decomposed into a sum of holomorphically factorised contributions generalising the $n=4$ case. We refer to~\cite[\S9.3]{DiFrancesco:1997nk} for a more
extensive discussion.

These  four-point conformal blocks  are not single-valued when $\eta$
or $\tilde \eta$ move in the complex plane because they have monodromy around the points
$\eta =0$ and $\eta =1$. The coefficients $G_{12|34}^{k,\tilde k}$
need to be  such that the correlation function  $ \mathcal G_{12|34}(\eta,\bar\eta) $ evaluated on the real slice
$\tilde\eta=\bar\eta$ is free of
monodromies.

For the minimal models described by a Coulomb-gas, Dotsenko and Fateev
in~\cite{Dotsenko:1984nm,Dotsenko:1984ad} showed that the
single-valuedness condition of the four-point correlation function
determines the correlation functions, up to an overall constant.

We will determine the higher point correlations functions that evaluate to closed string building blocks at a special value, but with the important difference that we will determine the overall constant.  

\subsection{Closed string theory amplitudes building blocks}\label{sec:closedamp}
Let $N\in\mathbb{N}$. Any closed string tree-level amplitudes are finite  linear
combinations~\cite{DHoker:1988pdl,Green:1987sp,Polchinski:1998rq,Polchinski:1998rr}
\begin{equation}\label{e:Treedecomp}
  \mathcal M_{N+3}(  \pmb s,\pmb \epsilon)= \sum_r
  c_r(  \pmb s,\pmb \epsilon)\,  M_{N+3}( \pmb s, \pmb n^r, \pmb {\tilde n}^r)
\end{equation}
of the building blocks given by the following integrals
\begin{multline}\label{e:IntGeneric}
  M_{N+3}(\pmb s,\pmb n,\pmb{\tilde n})= \int_{\mathbb C^N} \prod_{i=1}^{N} |z_i|^{2 s_{0i}}
 |1-z_i|^{2 s_{i\,N+1}}  z_i^{n_{0i}}\bar z_i^{\tilde n_{0i}}  (1-z_i)^{n_{i\,N+1}}(1-\bar
 z_i)^{\tilde n_{i\,N+1}} \cr
 \times
 \prod_{1\leq i<j\leq N}
 |z_i-z_j|^{2 s_{ij}}
 (z_i-z_j)^{n_{ij}}  (\bar z_i-\bar z_j)^{\tilde n_{ij}} \prod_{i=1}^{N} d^2z_i \,,
\end{multline}
with the  integration measure $d^2z:= dz d\bar z/(-2\pi i)$.
The exponents~$n_{ij}$ and~$\tilde n_{ij}$, which depend on the theory, belong to~$\mathbb{Z}$ and form tuples $\pmb n:=(n_{ij})_{0\leq i<j\leq N+1}$ and
$\pmb {\tilde n}:=(\tilde n_{ij})_{0\leq i<j\leq N+1}$. The tuple of (Mandelstam) kinematic invariants
$\pmb s:=(s_{ij}:=\alpha' k_i\cdot k_j)_{0\leq i< j\leq N+1}$ is the (dimensionless) product of the inverse string tension~$\alpha'$ with a tuple of scalar products of the external momenta~$k_i$, which are~$N+3$ vectors of a $D$-dimensional\footnote{Tree-level amplitudes are defined for dimension $D\leq 26$ in bosonic string theory , and for $D\leq 10$ in superstring theory. } Minkowsky space-time~$\mathbb R^{1,D-1}$ with metric $(+-\cdots-)$, subject to the momentum 
conservation condition $k_0+\cdots +k_{N+2}=0$
and the condition $\alpha' k_i^2\in -4+4\mathbb N$.

As explained in Appendix~\ref{Sec:ConvReg}, for any fixed tuples $n_{ij},\tilde n_{ij}$ there exists a non-empty domain~$U$ contained in $\mathbb{C}^{N(N+3)/2}$ (which depends on $n_{ij},\tilde n_{ij}$) such that the integrals~\eqref{e:IntGeneric} converge absolutely for $\pmb s \in U$. In fact, as (holomorphic) functions of the Mandelstam variables $\pmb s \in U$, all closed string building blocks $M_{N+3}(\pmb s,\pmb n,\pmb{\tilde n})$ can be analytically continued\footnote{One rigorous direct way to do this is described in~\cite{Bocardo-Gaspar:2019pzk}. Alternatively, one can define the analytic continuation combining the KLT relations with the (well-known) analytic continuation of open string building blocks.} to define meromorphic function on $\mathbb{C}^{N(N+3)/2}$. For given values of the scalar products $k_i\cdot k_j$, with $0\leq i\leq j\leq N+1$, the small $\alpha'$-expansion, which is a Laurent series in~$\alpha'$, gives the higher-order string corrections to the field theory limit $\alpha'\rightarrow 0$. 

The coefficients $c_r(\pmb s,\pmb \epsilon)$ in~(\ref{e:Treedecomp}) depend on $\pmb s$ as above and on a polarisation vector $\pmb \epsilon=(\epsilon_i)_{0\leq i\leq N+1}$, with $\epsilon_i\in\mathbb{R}^{1,D-1}$ and $k_i\cdot\epsilon_i=0$. More precisely, they are rational functions in the
kinematic invariants~$s_{ij}$, in the product of the external momenta and the polarisation vectors $\sqrt{\alpha'} \,k_i\cdot
  \epsilon_j$ and in the product of the  components of the  polarisation vector $\epsilon_i\cdot
  \epsilon_j$ (for $0\leq i<j\leq N+1$).  The precise form of the coefficients $c_r(\pmb s,\pmb \epsilon)$ depends on the closed string  theory one considers---i.e. the bosonic string, or the type-II
  superstring or the heterotic string. 
 For the heterotic string with external gauge fields
  they would depend as well on the colour factors through a product of 
  traces (see e.g.~\cite[Vol~1, Chap.~6]{Green:1987sp}). 
Expressions for the four-point amplitudes can be found in
e.g~\cite{Green:1987sp,DHoker:1988pdl,Gross:1986mw}, higher-point  open
superstring amplitudes in~\cite{Mafra:2011nv}, while   open and closed
bosonic string and heterotic string  amplitudes can be found in~\cite{Azevedo:2018dgo}. 

The simplest tree-level bosonic closed string 
amplitude is that between $N+3$ tachyon fields which is given by one
building block~\cite{DHoker:1988pdl,Green:1987sp,Polchinski:1998rq,Polchinski:1998rr}
\begin{equation}\label{tachyonicInt}
\mathcal   M^{\rm tachyon}_{N+3}(\pmb s)=\int_{\mathbb{C}^N} \, \prod_{i=1}^{N} |z_i|^{2s_{0i}}
  |1-z_i|^{2s_{i\,N+1}}\prod_{1\leq i<j\leq N} |z_i-z_j|^{2s_{ij}}\,   \prod_{i=1}^{N} d^2z_i  
\end{equation}
with the on-shell condition $\alpha'k_i^2=-4$ for $0\leq i\leq
N+2$.

The building blocks in~\eqref{e:IntGeneric} arise from the correlation
functions between $N+3$ physical vertex operators $V_i(z) :=V_i(z,\bar z)$:
\begin{equation}\label{e:Gcft}
   M_{N+3}(\pmb s,\pmb n,\pmb{\tilde n})=\int_{\mathbb C^N} 
  \prod_{i=1}^{N} d^2z_i\,\left\langle V_{0}(0) \prod_{i=1}^{N} V_i(z_i) \,
  V_{N+1}(1) V_{N+2}(\infty) \right\rangle\,.
\end{equation}
We refer
to~\cite{DHoker:1988pdl,Green:1987sp,Polchinski:1998rq,Polchinski:1998rr}
for details about the form of the vertex operators in string
theory. The results of this work do not depend on the precise form of
the vertex operators, but only on the structure of the integrals given
in~\eqref{e:IntGeneric}. 

\subsection{Conformal correlators for closed string building blocks}\label{sec:closedampcorrfun}

For $\eta\in \mathbb C$ and $N\in\mathbb{N}$ we introduce the function 
\begin{multline}\label{e:Gdefgeneral}
  \mathcal G_N\left({\pmb a\, \pmb b\, \pmb c\, \pmb d\atop \pmb{\tilde a}\,
      \pmb{\tilde  b}\, \pmb{\tilde c}\, \pmb{\tilde d}}\Big|\eta\right):=\cr \int_{\mathbb C^N}  
\prod_{i=1}^{N}  z_i^{a_i}\bar   z_i^{\tilde a_i} (1-z_i)^{b_i}(1-\bar z_i)^{\tilde b_i} (\eta-z_i)^{c_i}  (\bar \eta-\bar z_i)^{\tilde c_i} \prod_{1\leq i<j\leq N}  (z_i-z_j)^{ d_{ij}}   (\bar z_i-\bar z_j)^{\tilde d_{ij}}  
  \prod_{i=1}^{N} d^2z_i \,,
\end{multline}
where the tuples of exponents $\pmb a, \pmb b, \pmb c, \pmb d,\pmb{\tilde a},\pmb{\tilde  b}, \pmb{\tilde c}, \pmb{\tilde d}$ are formed by complex numbers which satisfy
\begin{equation}
\begin{aligned}
  \label{e:spinG}
  a_i-\tilde a_i,\,  b_i-\tilde b_i,\,  c_i-\tilde c_i,&\in\mathbb Z \qquad &1\leq i\leq N,\cr
d_{ij}- \tilde d_{ij}&\in\mathbb Z\qquad &1\leq i<j\leq N\,.
\end{aligned}
\end{equation}
We will often omit these parameters from the notation and simply write $\mathcal G_N(\eta)$. By condition~(\ref{e:spinG}), the integrand of $\mathcal G_N(\eta)$ is single-valued, which implies that the integral over $\mathbb{C}^N$ makes sense and defines a real-analytic single-valued function of $\eta\in\mathbb{C}\setminus\{0,1\}$ for any values of the parameters such that the integral is absolutely convergent. The region of absolute convergence is a non-empty subset of $\mathbb{C}^{N(N+5)}$ explicitly worked out in Appendix~\ref{Sec:ConvReg}. As mentioned in Section~\ref{sec:cft-correlators}, single-valuedness in~$\eta$ is a fundamental physics requirement for correlation functions of conformal field theories\footnote{Conformal fields with non-integer spin, like the twist fields or spin
fields~\cite{Green:1987sp,Polchinski:1998rq,Polchinski:1998rr}, induce
branch cuts. We are not considering amplitudes with such fields, whose $\alpha'$-expansion would not be given by single-valued multiple
zeta values.}.

The function $\mathcal{G}_N(\eta)$
  evaluates at  $\eta=1$ (or $\eta=0$, with a different parameter identification) to the closed string amplitude building
  block $M_{N+3}(\pmb s,\pmb n,\pmb{\tilde n})$
  given by the integral in eq.~\eqref{e:IntGeneric},
if we set
\begin{equation}\begin{aligned}\label{e:paramA}
  a_i&:= s_{0i}+n_{0i},  \quad   b_i+c_i:= s_{i\, N+1}+n_{i\,N+1},  \quad 1\leq i\leq N,\\
 \tilde a_i&:= s_{0i}+\tilde n_{0i},  \quad   \tilde b_i+\tilde
 c_i:=  s_{i\,
   N+1}+\tilde n_{i\,N+1},  \quad 1\leq i\leq N,\\
d_{ij}&:=     s_{ij}+n_{ij},  \quad \tilde d_{ij}:= s_{ij}+\tilde n_{ij}, \qquad\qquad\  1\leq i<j\leq
                 N.
\end{aligned}
\end{equation}
We will explain how to make use of the generalisation~(\ref{e:Gdefgeneral}), and of the fact that it is a single-valued function of $\eta$, to deduce informations about the closed string partial amplitudes.

Following the standard rules for the correlation functions of
conformal field theory minimal models~\cite{DiFrancesco:1997nk} and
string theory
in~\cite{DHoker:1988pdl,Green:1987sp,Polchinski:1998rq,Polchinski:1998rr},
one can write the function $\mathcal G_N(\eta)$~\eqref{e:Gdefgeneral}, with the
integer spin condition~\eqref{e:spinG}, as  the integrated 
correlation function over the positions of  the $N+3$ vertex operators
$V_i(z_i)$ and one unintegrated auxiliary
vertex operator $U(\eta):=U(\eta,\bar \eta)$
\begin{equation}\label{e:Gcftcorr}
  \mathcal G_N(\eta)=\int_{\mathbb C^N} \prod_{i=1}^{N} d^2z_i
  \left\langle V_{0}(0) \prod_{i=1}^{N} V_i(z_i)
  V_{N+1}(1) V_{N+2}(\infty)  U(\eta)\right\rangle  
\end{equation}

It will not be necessary for our purposes (and it is not the case in
general) that the total amplitude $\mathcal M_{N+3}(\pmb s,\pmb \epsilon)$ is given by
the special value at $\eta=1$ of a single-valued correlation
function. In particular, in order to obtain qualitative results on the
coefficients of the $\alpha'$-expansion, it is enough that each
partial amplitude $M_{N+3}( \pmb s, \pmb n^r, \pmb {\tilde n}^r)$ arises this way
because the kinematic coefficients $c_r(\pmb s,\pmb \epsilon)$ in~\eqref{e:Treedecomp} are rational  functions with rational coefficients of the kinematic
invariants (see~\cite{Mafra:2011nv} for some expressions using the
pure spinor formalism, and~\cite{Azevedo:2018dgo} for bosonic open
and closed string and heterotic string amplitudes).  This is the
case, for instance, of the  single- and double-trace
contributions to the heterotic-string amplitude given in~\cite{Gross:1986mw}.
 
\section{Aomoto-Gel'fand hypergeometric functions}\label{sec:general-selberg}

Let $p\geq 2$, $N\geq 1$, let $z_{N+1}<\cdots <z_{N+p}$ be real variables, and let us consider generalised hypergeometric functions given by the integrals
\begin{equation}\label{AGgendef}
F_{\Delta}(\pmb\lambda;z_{N+1},\cdots ,z_{N+p})\,:=\,\int_{\Delta} \prod_{1\leq i<j\leq N+p}|z_i-z_j|^{\lambda_{ij}}\,\prod_{i=1}^N\,dz_i\,.
\end{equation}
with $\Delta$ a connected component of $\{(z_1,\ldots ,z_N)\in \mathbb{R}^{N}\,|\,z_i\neq z_j\text{ for } 1\leq i\leq N,1\leq j\leq N+p\}$, depending on some tuple $\pmb\lambda$ of complex parameters $\lambda_{ij}$ such that~\eqref{AGgendef} is absolutely convergent. This kind of functions, and natural generalisations thereof, was systematically studied by Aomoto (see eq.~(0.1) of~\cite{Aomoto}) and Gel'fand~\cite{GelfandAlone, Gelfand} at the end of the 1980s. More specifically, here we are interested in the case where $p=3$ and $(z_{N+1},z_{N+2},z_{N+3})=(0,\eta,1)$, given by\footnote{We have set $\lambda_{ij}=a_i$ for $j=N+1$, $\lambda_{ij}=b_i$ for $j=N+3$, $\lambda_{ij}=c_i$ for $j=N+2$, $\lambda_{ij}=d_{ij}$ otherwise.}
\begin{equation}\label{AGgendef1var}
F_{\Delta}(\pmb a,\pmb b,\pmb c;\pmb d;\eta)\,:=\,\int_{\Delta}\prod_{i=1}^N\,|z_i|^{a_i}\,|z_i-1|^{b_i}\,|z_i-\eta|^{c_i}\,\prod_{1\leq i<j\leq N}|z_i-z_j|^{d_{ij}}\,\prod_{i=1}^N\,dz_i\,.
\end{equation}
We will call the integrals~\eqref{AGgendef1var} \emph{Aomoto-Gel'fand hypergeometric functions}. For any fixed $\eta$ and $\Delta$, the region of absolute convergence in terms of the complex parameters $a_i,b_i,c_i,d_{ij}$ is non-empty (see Appendix~\ref{Sec:ConvReg}), and the integrals make sense also in the divergent case by analytic continuation\footnote{This can be done by taking the ``finite part of divergent integrals'' in the sense of Hadamard~\cite{Aomoto}.}, thus defining meromorphic functions over~$\mathbb{C}^{N(N+5)/2}$, whose polar set is a union of affine hyperplanes defined over $\mathbb{Z}$~\cite{Aomoto}. As functions of~$\eta$, which originally belongs to the interval $(0,1)$, $I_{(\rho,\sigma)}(\pmb a,\pmb b,\pmb c;\pmb d;\eta)$ can be analytically continued (by deforming the integration path) to define multi-valued functions on $\mathbb{P}^1_{\mathbb{C}}\setminus\{0,1,\infty\}$.

\subsection{Basis of Aomoto-Gel'fand hypergeometric functions}

Let us denote by~$X_N$ a set of indices $\{1,\ldots ,N\}$ of cardinality $N$. We consider for $r+s=N$ ($r,s\geq 0$, $N\geq 1$) all inclusions $\rho :X_r\hookrightarrow X_N$ and $\sigma: X_s\hookrightarrow X_N$ such that $X_N=\rho (X_r)\sqcup \sigma (X_s)$. We denote
by~$\mathfrak S_r$ the set of permutations of~$r$ elements. By abuse of notation, we sometimes consider~$\rho$ and~$\sigma$ also as elements of~$\mathfrak{S}_r$ and~$\mathfrak{S}_s$, respectively.
For $(\rho,\sigma)$ as above, we introduce a special family of Aomoto-Gel'fand hypergeometric functions
\begin{equation}\label{e:IsigmarhoDef}
I_{(\rho,\sigma)}(\pmb a,\pmb b,\pmb c;\pmb d;\eta):=F_{\Delta_{(\rho,\sigma)}(\eta)}(\pmb a,\pmb b,\pmb c;\pmb d;\eta)\,,
\end{equation}
integrated over
\begin{equation}
  \label{e:DeltaDef}
  \Delta_{(\rho,\sigma)}(\eta):= \{0\leq z_{\sigma( 1)} \leq \cdots
  \leq z_{\sigma( s)}\leq \eta\leq 1\leq z_{\rho( 1)}\leq \cdots\leq z_{\rho( r)}\}\,.
\end{equation}

Let us now fix $\eta\in \mathbb{P}^1_{\mathbb{C}}\setminus\{0,1,\infty\}$, and consider the parameters $\pmb a,\pmb b,\pmb c,\pmb d$ as complex variables over any complex domain where the analytic continuation of the Aomoto-Gel'fand hypergeometric functions~\eqref{AGgendef1var} does not have poles. Let us introduce the field
\begin{equation}\label{FieldCoeffs}
\mathbb{F}:=\mathbb{Q}(e^{\pi ia_1},\ldots ,e^{\pi ia_N},e^{\pi ib_1},\ldots ,e^{\pi ib_N},e^{\pi ic_1},\ldots ,e^{\pi ic_N},e^{\pi id_{12}},\ldots ,e^{\pi id_{N-1\,N}})
\end{equation}
and the (finite dimensional) vector spaces
\begin{equation}
H_{N,\eta}:=\mathbb{F}\langle F_\Delta(\pmb a,\pmb b,\pmb c;\pmb d;\eta)\,|\,{\rm dim}(\Delta)=N \rangle\,,
\end{equation}
spanned over~$\mathbb{F}$ by all Aomoto-Gel'fand functions associated with $N$-dimensional integrals.
\begin{propos}\label{prop:generatingfamily}
For any fixed $\eta\in \mathbb{P}^1_{\mathbb{C}}\setminus\{0,1,\infty\}$ the functions $I_{(\rho,\sigma)}(\pmb a,\pmb b,\pmb c;\pmb d;\eta)$ generate the vector space $H_{N,\eta}$ over $\mathbb{F}$.
\end{propos}
\begin{proof}
We use a contour deformation method which was proposed in~\cite{Dotsenko:1984nm} to study the special case where $a_1=\cdots =a_N$, $b_1=\cdots =b_N$, $c_1=\cdots =c_N$ and $d_{12}=\cdots =d_{N-1\,N}$ (corresponding to correlation functions in 2D statistical models). The same argument was later used in~\cite{BjerrumBohr:2009rd, Stieberger:2009hq,BjerrumBohr:2010hn} to study open string amplitudes, which can be seen as special values of Aomoto-Gel'fand hypergeometric functions at $\eta=1$. We only need to adapt this method to our slightly more general context, following~\cite{BjerrumBohr:2009rd,BjerrumBohr:2010hn}. For this reason, we will skip a few details, which can be found in the above-mentioned references.

We prove the statement for $\pmb a,\pmb b,\pmb c,\pmb d$ contained in the region of common convergence of the functions $F_\Delta(\pmb a,\pmb b,\pmb c;\pmb d;\eta)$ (which is specified in Appendix~\ref{Sec:ConvReg}). Since these integrals depend holomorphically on the parameters $\pmb a,\pmb b,\pmb c,\pmb d$, and the region of convergence is open and non-empty, the result remains true also for their analytic continuations. Moreover, we suppose for simplicity that $\eta\in [0,1]$, but our argument can be adapted to any $\eta\in \mathbb{P}^1_{\mathbb{C}}\setminus\{0,1,\infty\}$ by deforming the real line. 

Suppose now that we can prove the following two claims:
\begin{itemize}
\item \textbf{Claim~\customlabel{claim1}{1}:} Let $F_\Delta(\pmb a,\pmb b,\pmb c;\pmb d;\eta)$ be associated with an $N$-dimensional domain $\Delta$ which contains $m\leq N$ negative coordinates $z_{i_1}< \cdots < z_{i_m}< 0$. Then $F_\Delta(\pmb a,\pmb b,\pmb c;\pmb d;\eta)$ can be written as an $\mathbb{F}$-linear combination of Aomoto-Gel'fand hypergeometric functions $F_{\Delta'}(\pmb a,\pmb b,\pmb c;\pmb d;\eta)$ associated with $N$-dimensional domains $\Delta'$ with $m-1$ negative coordinates.
\item \textbf{Claim~\customlabel{claim2}{2}:} Let $F_\Delta(\pmb a,\pmb b,\pmb c;\pmb d;\eta)$ be associated with a non-negative domain $\Delta\subset\big(\mathbb{R}^+\big)^N$ such that $m\leq N$ coordinates $\eta< z_{i_1}< \cdots < z_{i_m}< 1$ belong to the interval $[\eta,1]$. Then $F_\Delta(\pmb a,\pmb b,\pmb c;\pmb d;\eta)$ can be written as an $\mathbb{F}$-linear combination of Aomoto-Gel'fand hypergeometric functions $F_{\Delta'}(\pmb a,\pmb b,\pmb c;\pmb d;\eta)$ associated with positive domains $\Delta\subset\big(\mathbb{R}^+\big)^N$ with $m-1$ coordinates contained in $[\eta,1]$.
\end{itemize}

Then, if we consider an Aomoto-Gel'fand function $F_\Delta(\pmb a,\pmb b,\pmb c;\pmb d;\eta)$, we could recursively use the first claim to write it as an $\mathbb{F}$-linear combination of Aomoto-Gel'fand functions with all points contained in $[0,\infty]$. Subsequently, we could recursively use the second claim to write $F_\Delta(\pmb a,\pmb b,\pmb c;\pmb d;\eta)$ in terms of Aomoto-Gel'fand functions of the kind $I_{(\rho,\sigma)}(\pmb a,\pmb b,\pmb c;\pmb d;\eta)$, with all points contained either in $[0,\eta]$ or in $[1,\infty]$, which therefore generate the whole space~$H_{N,\eta}$ over~$\mathbb{F}$. 

Hence we are left with proving the two claims, both of which rely on the following observation.
\begin{lemma}\label{lemmaproofbasis}
Fix a connected component $\Delta\subset\mathbb{R}^{N-1}$ of $\{(z_1,\ldots ,z_{N-1})\in (\mathbb{R}\setminus\{0,\eta,1\})^{N-1}\,|\,z_i\neq z_j\text{ for } 1\leq i,j\leq N\}$. Let $\Delta^{(1)},\ldots ,\Delta^{(N+3)}\subset\mathbb{R}^N$ denote the $N+3$ distinct connected components of $\{(z_1,\ldots ,z_N)\in\Delta\times \big(\mathbb{R}\setminus\{0,\eta,1\}\big)\,|\,z_N\neq z_i \text{ for } 1\leq i\leq N-1\}$. Then the $N+3$ associated Aomoto-Gel'fand functions $F_{\Delta^{(j)}}(\pmb a,\pmb b,\pmb c;\pmb d;\eta)$ are $\mathbb{F}$-linearly dependent.
\end{lemma}
\begin{proof}
We fix $(z_1,\ldots ,z_{N-1})\in\Delta$. Without loss of generality, we can suppose that $\Delta$ is such that $0< z_1< \cdots < z_{N-1}< \eta< 1$ (the proof works in the same way for any orderings). Let~$D$ be the simply connected complex domain obtained by removing from the lower\footnote{The proof would work (with obvious adaptations) also if we chose the upper half-plane. This freedom will be very important in the proof of Claim~\ref{claim2}.} half-plane $N+2$ non-intersecting half-lines starting at the points $0,z_1,\ldots ,z_{N-1},\eta,1$. 

The function
\begin{equation}\label{eq:intab2}
\prod_{i=1}^N\,z_i^{a_i}\,(1-z_i)^{b_i}\,(\eta-z_i)^{c_i}\,\prod_{1\leq i<j\leq N}(z_j-z_i)^{d_{ij}}
\end{equation}
is well-defined for~$z_N$ real and $z_{N-1}\leq z_N\leq \eta$, and we denote in the same way its (unique) holomorphic extension\footnote{Here we made a choice by fixing the branch of the integrand in the domain~$D$ associated to $z_{N-1}\leq z_N\leq \eta$, but we could have chosen any other branch.} to $z_N\in D$. Its integral over the closed contour $\{\text{Im}(z_N)=\varepsilon,|z_N|\leq \varepsilon^{-1}\}\cup\{|z_N|=\varepsilon^{-1}, \text{Im}(z_N)\geq \varepsilon\}$ depicted in Figure~\ref{figdem} vanishes for any $\varepsilon>0$ by Cauchy's theorem. Taking the limit $\varepsilon\rightarrow 0$, the integral over the semi-circle $\{|z_N|=\varepsilon^{-1}, \text{Im}(z_N)\geq \varepsilon\}$ also vanishes. This implies that
\begin{equation}\label{eq:intab}
\lim_{\varepsilon\rightarrow 0}\int_{\substack{\text{Im}(z_N)=\varepsilon\\|z_N|\leq \varepsilon^{-1}}}\,\prod_{i=1}^N\,z_i^{a_i}\,(1-z_i)^{b_i}\,(\eta-z_i)^{c_i}\,\prod_{1\leq i<j\leq N}(z_j-z_i)^{d_{ij}}\,dz_N\,=\,0.
\end{equation}
\begin{figure}
  \centering
    \begin{tikzpicture}[scale=1.5]

    \draw[ultra thick,->] (-1.8,0) -- (3.3,0);
    \draw[fill] (0,0) circle (0.04);
    \draw[fill] (0.5,0) circle (0.04);
    \draw[fill] (0.7,0) circle (0.03);
    \draw[fill] (0.8,0) circle (0.03);
    \draw[fill] (0.9,0) circle (0.03);
    \draw[fill] (1.1,0) circle (0.04);
    \draw[fill] (1.6,0) circle (0.04);
    \draw[fill] (2.1,0) circle (0.04);

    \node[below] at (0,0) {$0$};
    \node[below] at (0.5,0) {$z_1$};
    \node[below] at (1.1,0) {$z_{N-1}$};
    \node[below] at (1.6,0) {$\eta$};
    \node[below] at (2.1,0) {$1$};
    \node[right] at (1.3,2) {${\color{blue} z_N}$};
    \draw[ultra thick, color=blue] (-1.5,.2) -- (3,.2);
    \draw[ultra thick, color=blue,postaction={decorate},decoration={
    markings,
    mark=at position 0.5 with {\arrow{>}}}] (3,.2)  arc (0:180:2.25cm);
    
  \end{tikzpicture}
  \caption{Integration contour of Lemma~\ref{lemmaproofbasis}.}
  \label{figdem}
\end{figure}
On the other hand, the integration domain of~\eqref{eq:intab} can be divided into the $N+3$ possible relative positions of the real part of the variable $z_N$ with respect to the points $0,z_1,\ldots ,z_{N-1},\eta,1$. Doing this, it is possible to (explicitly) determine phases $\varphi_1,\ldots ,\varphi_{N+3}\in\mathbb{F}$ such that 
\begin{equation}\label{eq:intab3}
\lim_{\varepsilon\rightarrow 0}\int_{\substack{\text{Im}(z_N)=\varepsilon\\|z_N|\leq \varepsilon^{-1}}}\,\prod_{i=1}^N\,z_i^{a_i}\,(1-z_i)^{b_i}\,(\eta-z_i)^{c_i}\,\prod_{1\leq i<j\leq N}(z_j-z_i)^{d_{ij}}\,dz_N\,=\,\sum_{j=1}^{N+3}\varphi_j\,F_{\Delta^{(j)}}(\pmb a,\pmb b,\pmb c;\pmb d;\eta).
\end{equation}
For instance, in the region $z_{N-1}<\text{Re}(z_{N})<\eta$ the corresponding phase~$\varphi_{N+1}$ is simply equal to~$1$, because of our definition of~\eqref{eq:intab2}, while in the next region $\eta<\text{Re}(z_{N})<1$ we have $\varphi_{N+2}=e^{-i\pi c}$.

The statement follows by combining~\eqref{eq:intab} and~\eqref{eq:intab3}.
\end{proof}

$\bullet$ \emph{Proof of Claim~\ref{claim1}}. 
We can suppose that the negative coordinates in~$\Delta$ are $z_1,\ldots ,z_m$. For each $1\leq i\leq m$, let us fix a permutation $\alpha_i\in\mathfrak{S}_{m-1}$ of the indices $\{1,\ldots ,i-1,i+1,\ldots m\}$. We define~$\Delta^{(i,\alpha_i)}$ to be the $N-1$-dimensional domains of Aomoto-Gel'fand functions with $m-1$ negative coordinates $z_{\alpha_i(1)}<\cdots <z_{\alpha_i(i-1)}<z_{\alpha_i(i+1)}<\cdots <z_{\alpha_i(m)}<0$ and the same $N-m$ positive ordered coordinates of the original  domain~$\Delta$. The number of such domains is~$m!$. 

We apply Lemma~\ref{lemmaproofbasis} to each~$\Delta^{(i,\alpha_i)}$, denoting by~$z_i$ the extra-variable integrated over the contour of Figure~\ref{figdem}. We obtain~$m!$ different relations of $\mathbb{F}$-linear dependence between Aomoto-Gel'fand functions associated with two kinds of $N$-dimensional domains: those where the inserted variable~$z_i$ is~$>0$, and those where it is~$<0$. The former kind of domains has~$m-1$ negative coordinates, the latter has~$m$ negative coordinates. In particular, all of the possible~$m!$ Aomoto-Gel'fand functions associated with domains with negative $z_1,\ldots ,z_m$ appear in these relations. 

Since the~$m!$ relations obtained are $\mathbb{F}$-linearly independent, we can solve the system and express any Aomoto-Gel'fand function associated with an $N$-dimensional domain with~$m$ negative coordinates, and in particular that associated with our initial domain $\Delta$, in terms of Aomoto-Gel'fand functions associated with $N$-dimensional domains~$\Delta'$ with $m-1$ negative coordinates, as claimed.

$\bullet$ \emph{Proof of Claim~\ref{claim2}}. Here the idea is the same as above: we want to find enough equations to express all of the possible domains with~$m$ coordinates in the interval $[\eta,1]$ in terms of ``better'' domains~$\Delta'$, with $m-1$ coordinates in the interval $[\eta,1]$ and still no negative coordinates. We cannot, however, repeat the exact same steps as above, because when we integrate a variable~$z_i$ over the contour described by Lemma~\ref{lemmaproofbasis}, we obtain also an Aomoto-Gel'fand function associated with the domain where $z_i< 0$, which is not one of the desired domains~$\Delta'$. 

This issue is circumvented by integrating~$z_i$ not only on the contour of Figure~\ref{figdem} (while picking the branch cuts in the lower half-plane), but also on the same kind of contour in the lower half-plane (while picking the branch cuts in the upper half-plane). In this way, for any~$\Delta^{(i,\alpha_i)}$ we obtain a pair of independent equations with coefficients in~$\mathbb{F}$, which can be combined to cancel the contribution from the domain with $z_i< 0$ (see Section~\ref{sec:S4pt} for more details). At this point, we are in the same situation as that of the proof of Claim~\ref{claim1}, and we conclude by the same argument.
\end{proof}

We remark that the relations constructed in the proof among $F_\Delta(\pmb a,\pmb b,\pmb c;\pmb d;\eta)$, viewed as functions of the parameters $\pmb a,\pmb b,\pmb c,\pmb d$ for fixed~$\eta\in\mathbb{P}^1_{\mathbb{C}}\setminus\{0,1,\infty\}$, do not depend on the position of~$\eta$; they can therefore be viewed also as relations among functions of~$\eta$.

For fixed $r\geq0$ and
  $s=N-r\geq0$ there are~$r!$ permutations~$\rho$ and~$s!$
  permutations~$\sigma$, for a total of $r!\times s!\times {N\choose r}=N!$
  distinct ordered integrals, hence since $0\leq r\leq N$ the total number\footnote{Another way to obtain this counting is
  to realise that the sets
  $\alpha=\{\sigma(1),\dots,\sigma(s),N+1,\rho(1),\dots,\rho(r)\}$ run over all permutations of $\{1,\dots,N+1\}$ as the permutations $\sigma$ and $\rho$ vary.} of distinct generators $I_{(\rho,\sigma)}(\pmb a,\pmb b,\pmb c;\pmb d;\eta)$ of $H_{N,\eta}$ is $N!\times (N+1)=(N+1)!$.   It follows\footnote{In order to apply this result, one needs to interpret the vector spaces $H_{N,\eta}$ in terms of twisted homology. This would also allow to give an alternative proof Proposition~\ref{prop:generatingfamily}, by adapting the construction in~\cite{Aomoto} of an explicit (different) basis of the twisted homology to our context. A nice dictionary between the homological approach and the more elementary methods presented here is provided in~\cite{Casali:2019ihm}.} from a theorem of Aomoto (see Theorem~1 of~\cite{Aomoto}) that the dimension of $H_{N,\eta}$ is precisely $(N+1)!$. Therefore, combining this counting with the above proposition, we obtain the following:
\begin{thm}[Basis of integrals]\label{thmbasisint}
For any fixed $\eta\in \mathbb{P}^1_{\mathbb{C}}\setminus\{0,1,\infty\}$ the functions $I_{(\rho,\sigma)}(\pmb a,\pmb b,\pmb c;\pmb d;\eta)$ are a basis of $H_{N,\eta}$.
\end{thm}
   
We also consider a different special family of Aomoto-Gel'fand hypergeometric function, related to our basis by a change of variables:
\begin{equation}\label{varchangeJI}
J_{(\rho,\sigma)}(\pmb a,\pmb b,\pmb c;\pmb d;\eta):= I_{(\rho,\sigma)}(\pmb b,\pmb a,\pmb c;\pmb d;1-\eta)\,.
\end{equation}
By the change of variables $z_i\mapsto 1-z_i$, with $1\leq i\leq N$, in the integrand of~$J_{(\rho,\sigma)}$, we can also write
\begin{equation}\label{e:IsigmarhoDefdual}
J_{(\rho
  ,\sigma)}(\pmb a,\pmb b,\pmb c;\pmb
                d;\eta)=F_{\tilde\Delta_{(\rho,\sigma)}(\eta):}(\pmb
                a,\pmb b,\pmb c;\pmb d;\eta)\,,
\end{equation}
integrated over the domain
\begin{equation}
  \label{e:DeltaDualDef}
  \tilde\Delta_{(\rho,\sigma)}(\eta):= \{z_{\rho(r)}\leq \cdots\leq
  z_{\rho(1)} \leq 0\leq \eta\leq z_{\sigma(s)} \leq \cdots
  \leq z_{\sigma(1)}\leq 1 \}\,.
\end{equation}

By~\eqref{varchangeJI} and Theorem~\ref{thmbasisint}, for any fixed~$\eta$ the functions $J_{(\rho,\sigma)}(\pmb a,\pmb b,\pmb c;\pmb d;\eta)$ are a basis of the vector space~$H_{N,1-\eta}$. Moreover, eq.~\eqref{e:IsigmarhoDefdual} implies that they can also be seen as (linearly independent) elements of the space~$H_{N,\eta}$, and therefore constitute an alternative basis of $H_{N,\eta}$. We want to study this change of basis. Let us introduce the vector notation 
\begin{align}\label{e:vectnot}
  \vec I_N(  \pmb a,\pmb b, \pmb c;\pmb d;\eta)&:=\big(I_{(\rho
  ,\sigma)}(\pmb a,\pmb b,\pmb c;\pmb
                d;\eta) \big)_{(\rho,\sigma)\in \mathfrak S_r\times
                \mathfrak S_s, r+s=N} \notag\\
 \vec J_N(  \pmb a,\pmb b, \pmb c;\pmb d;\eta)&:=\big(J_{(\rho
  ,\sigma)}(\pmb a,\pmb b,\pmb c;\pmb
                d;\eta) \big)_{(\rho,\sigma)\in \mathfrak S_r\times
                \mathfrak S_s, r+s=N}
\end{align}
such that the first $N!$ rows of the vectors are the integrals $I_{(\rho,\emptyset)}$ and $J_{(\rho,\emptyset)}$, respectively,  obtained by setting $s=0$, and the last $N!$ rows  of the vectors are the integrals  $I_{(\emptyset,\sigma)}$ and $J_{(\emptyset,\sigma)}$, respectively, obtained by setting  $r=0$. Then one has the following consequence of the previous theorem:

\begin{cor}[Change of basis]\phantomsection\label{prop:changebasis}
\noindent The vectors $\vec I_N(  \pmb a,\pmb b, \pmb c;\pmb d;\eta)$ and $\vec J_N(  \pmb a,\pmb b, \pmb c;\pmb d;\eta)$ are related by
\begin{equation}\label{e:ItotildeIa}
\vec  I_N(\pmb a,\pmb b,\pmb c;\pmb d;\eta)=
 R_N(\pmb  a,\pmb b,\pmb c;\pmb d)\, \vec J_N(  \pmb a,\pmb b, \pmb c;\pmb d;\eta)\,,
\end{equation}
where $R_N(\pmb  a,\pmb b,\pmb c;\pmb d)$ is an invertible $(N+1)!\times (N+1)!$ matrix with coefficients in~$\mathbb{F}$. In particular, $R_N(\pmb  a,\pmb b,\pmb c;\pmb d) $ is invariant under integer shifts of its parameters, and its inverse is 
\begin{equation}\label{RNinv}
    R_N(\pmb  a,\pmb b,\pmb c;\pmb d) ^{-1}=R_N(\pmb  b,\pmb a,\pmb c;\pmb d) \,.
\end{equation}
\end{cor}
\begin{proof}
We only need to prove~(\ref{RNinv}), which follows from
\begin{equation}
\vec  I_N(\pmb a,\pmb b,\pmb c;\pmb d;\eta)=
R_N(\pmb  a,\pmb b,\pmb c;\pmb d)\, \vec   I_N(\pmb b,\pmb a,\pmb c;\pmb d;1-\eta)=R_N(\pmb  a,\pmb b,\pmb c;\pmb d)\, R_N(\pmb b,\pmb  a,\pmb c;\pmb d)\, \vec  I_N(\pmb a,\pmb b,\pmb c;\pmb d;\eta)\,,
\end{equation}
and the fact that the integrals $I_{(\rho,\sigma)}(\pmb a,\pmb b,\pmb c;\pmb d;\eta)$ are linearly independent over $\mathbb{F}$.
\end{proof}

\subsection{The $N=1$ matrix $R_1(a,b,c)$}\label{sec:S4pt}

We derive an explicit expression for the matrix $R_N(\pmb a,\pmb b,\pmb c;\pmb d)$ when $N=1$ by specialising to this case the method of the proof of Proposition~\ref{prop:generatingfamily}. Since $N=1$, the tuple $\pmb d$ is empty and we omit it from the notation. Let us assume that the integrals considered are all absolutely convergent\footnote{This is the case in the non-empty open region where $\text{Re}(a),\text{Re}(b),\text{Re}(c),\text{Re}(-2-a-b-c)>-1$, and the relations obtained extend by analytic continuation.}. The vectors of Aomoto-Gel'fand integrals for $N=1$ read
\begin{align}
\vec I_1(a,b,c;\eta)&=
\begin{pmatrix}
{\displaystyle  I_{(Id,\emptyset)}(a,b,c;\eta))} \cr
{\displaystyle I_{(\emptyset,Id)}(a,b,c;\eta)}
\end{pmatrix}=\begin{pmatrix}
{\displaystyle  \int_1^{+\infty}
  z^{a} (z-1)^{b} (z-\eta)^{c}dz} \cr
{\displaystyle \int_0^\eta 
  z^{ a} (1-z)^{ b} ( \eta-z)^{ c}dz}
\end{pmatrix},\cr
\vec J_1(a,b,c;\eta)&=
\begin{pmatrix}
{\displaystyle  J_{(Id,\emptyset)}(a,b,c;\eta))} \cr
{\displaystyle J_{(\emptyset,Id)}(a,b,c;\eta)}
\end{pmatrix}=
\begin{pmatrix}
 {\displaystyle \int_{-\infty}^0 
   (-z)^{a} (1-z)^{b} (\eta-z)^{c}dz}\cr
 {\displaystyle\int_\eta^1 
  z^{ a} (1-z)^{ b} (z- \eta)^{ c}dz}
\end{pmatrix}.
\end{align}
Consider now the function $z^a(1-z)^b(\eta-z)^c$, first defined on $0<z<\eta<1$ and then extended by analytic continuation to the complex plane, with branch cuts contained in the complex upper (resp. lower) half-plane. We integrate it along the contour contained in the lower (resp. upper) half-plane depicted in Figure~\ref{fig:ContourDeformation}.
\begin{figure}[ht]
  \centering
    \begin{tikzpicture}[scale=1.5]

    \draw[ultra thick,->] (-.8,0) -- (2.3,0);
    \draw[fill] (0.5,0) circle (0.04);
    \draw[fill] (1,0) circle (0.04);
    \draw[fill] (1.5,0) circle (0.04);

    \node[below] at (.5,0) {0};
    \node[below] at (1,0) {$\eta$};
    \node[below] at (1.5,0) {1};
    \node[right] at (1,1) {${\color{blue} z}$};
     \node[right] at (1,-1) {${\color{red} z}$};
    
    \draw[ultra thick, color=blue] (0.7,.2)  arc (0:180:.2cm);
    \draw[ultra thick, color=blue] (1.2,.2)  arc (0:180:.2cm);
    \draw[ultra thick, color=blue] (1.7,.2)  arc (0:180:.2cm);

    \draw[ultra thick, color=blue] (-0.5,.2) -- (.3,.2);
    \draw[ultra thick, color=blue] (0.7,.2) -- (.8,.2);
     \draw[ultra thick, color=blue] (1.2,.2) -- (1.3,.2);
  \draw[ultra thick, color=blue] (1.7,.2) -- (2,.2);

    \draw[ultra thick, color=blue,postaction={decorate},decoration={
    markings,
    mark=at position 0.5 with {\arrow{>}}}] (2,.2)  arc (0:180:1.25cm);

    \draw[ultra thick, color=red] (.3,-.2)  arc (180:360:.2cm);
    \draw[ultra thick, color=red] (.8,-.2)  arc (180:360:.2cm);
    \draw[ultra thick, color=red] (1.3,-.2)  arc (180:360:.2cm);

     \draw[ultra thick, color=red] (-.5,-.2) -- (.3,-.2);
     \draw[ultra thick, color=red] (0.7,-.2) -- (0.8,-.2);
    \draw[ultra thick, color=red] (1.2,-.2) -- (1.3,-.2);
     \draw[ultra thick, color=red] (1.7,-.2) -- (2,-.2);

     \draw[ultra thick, color=red,postaction={decorate},decoration={
    markings,
    mark=at position 0.5 with {\arrow{<}}}] (-.5,-.2)  arc (180:360:1.25cm);
  \end{tikzpicture}
  \caption{Two contours leading to eqs.~\ref{eq:twocontours}.}
  \label{fig:ContourDeformation}
\end{figure}

Since there are no poles contained inside the integration contour, and the integrals over the half-circles tend to zero as the radius tends to infinity, we deduce the two linear
relations (corresponding to the two different contours)
\begin{align}\label{eq:twocontours}
  e^{-i\pi(b+c)}I_{(Id,\emptyset)}(a,b,c;\eta)+ e^{i\pi a}J_{(Id,\emptyset)}(a,b,c;\eta)+ e^{-i\pi c}
J_{(\emptyset,Id)}(a,b,c;\eta)+I_{(\emptyset,Id)}(a,b,c;\eta) &=0,\cr
e^{i\pi(b+c)} I_{(Id,\emptyset)}(a,b,c;\eta)+e^{-i\pi a}J_{(Id,\emptyset)}(a,b,c;\eta)+ e^{i\pi c}J_{(\emptyset,Id)}(a,b,c;\eta)+I_{(\emptyset,Id)}(a,b,c;\eta)&=0\,.
\end{align}
These relations, which remain valid for $\eta\in\mathbb{C}\setminus\{0,1\}$, can be written in matrix form as
\begin{equation}\label{e:MonJtoI}
   {  \vec I_1}(a,b,c;\eta)
={1\over \sin(\pi(b+c))}
  \begin{pmatrix}
    \sin(\pi a)& -\sin(\pi c)\cr
  - \sin(\pi(a+b+c))& -\sin(\pi b)
  \end{pmatrix}  {  \vec J_1}(a,b,c; \eta)
\end{equation}
or as
\begin{equation}\label{e:MonItoJ}
   {  \vec J_1}(a,b,c;\eta)
={1\over \sin(\pi(a+c))}
  \begin{pmatrix}
    \sin(\pi b)& -\sin(\pi c)\cr
  - \sin(\pi(a+b+c))& -\sin(\pi a)
  \end{pmatrix}  {  \vec I_1}(a,b,c; \eta)\,.
\end{equation}
Therefore we have that
\begin{equation}\label{eq:R1matr}
    R_1(a,b,c) ={1\over \sin(\pi(b+c))}
  \begin{pmatrix}
    \sin(\pi a)& -\sin(\pi c)\cr
  - \sin(\pi(a+b+c))& -\sin(\pi b)
  \end{pmatrix} 
\end{equation}
and, as expected,
\begin{equation}
    R_1(a,b,c) ^{-1} ={1\over \sin(\pi(a+c))}
  \begin{pmatrix}
    \sin(\pi b)& -\sin(\pi c)\cr
  - \sin(\pi(a+b+c))& -\sin(\pi a)
  \end{pmatrix} =   R_1(b,a,c) \,.
\end{equation}
Of course there are problems with defining or inverting $R_1$ if $a+c\in\mathbb{Z}$ or $b+c\in\mathbb{Z}$, but we can discard these values since they correspond to singularity divisors of ${  \vec I_1}$ and ${  \vec J_1}$.

\subsection{Monodromies of the Aomoto-Gel'fand hypergeometric functions}
\label{sec:monodromies}

We consider closed loops $\gamma_0$ and $\gamma_1$ around the points $\eta=0$ and $\eta=1$. We want to study the monodromy transformations of the Aomoto-Gel'fand hypergeometric
functions $I_{(\rho,\sigma)}(\pmb a,\pmb b,\pmb c;\pmb d;\eta)$ when $\eta$ moves around these loops (this only depends on the homotopy class of the loops, and does not depend on the starting point). We will then apply these results to our analysis of the correlation functions $\mathcal{G}_N$ in Section~\ref{ssec:factorisation}.


\subsubsection{The monodromy around $\eta=0$}
\label{sec:monodromy-around-z=0}

  The monodromy matrix around $\eta=0$ of the functions
  in~\eqref{e:IsigmarhoDef} is 
  obtained by performing the change of variable 
  \begin{align}
z_{\rho(m)}&=\hphantom{\eta} \zeta_{\rho(m)} \qquad  \textrm{for}\qquad 1\leq m\leq r \,,\cr
z_{\sigma(n)}&= \eta \zeta_{\sigma(n)}\qquad \textrm{for}\qquad
1\leq n\leq s \,,       
  \end{align}
which gives (setting $d_{ij}:=d_{ji}$ if $j<i$) 
 \begin{multline}\label{e:IsigmarhoDefBis}
 I_{(\rho   ,\sigma)}(\pmb a,\pmb b,\pmb c;\pmb d;\eta)=\prod_{n=1}^s\eta^{1+a_{\sigma(n)}+c_{\sigma(n)}}\prod_{1\leq k<l\leq s} \eta^{d_{ \sigma(k)\sigma(l)}}\int_{\Delta^*_{(\rho,\sigma)}}\prod_{j=1}^N|\zeta_j|^{a_j}\cr
\times \prod_{m=1}^r\prod_{n=1}^s|\zeta_{\rho(m)}-1|^{b_{\rho(m)}}|\zeta_{\rho(m)}-\eta|^{c_{\rho(m)}}|\eta\zeta_{\sigma(n)}-1|^{b_{\sigma(n)}}|\zeta_{\sigma(n)}-1|^{c_{\sigma(n)}} |\zeta_{\rho(m)}-\eta\zeta_{\sigma(n)}|^{d_{\rho(m)\sigma(n)}}\cr
\times\prod_{1\leq h<i\leq r}|\zeta_{\rho(h)}-\zeta_{\rho(i)}|^{d_{\rho(h)\rho(i)}}\prod_{1\leq k<l\leq s}|\zeta_{\sigma(k)}-\zeta_{\sigma(l)}|^{d_{\sigma(k)\sigma(l)}} \prod_{j=1}^Nd\zeta_j\,,
\end{multline}
integrated over the domain
\begin{equation}
  \label{e:DeltaBis}
  \Delta^*_{(\rho,\sigma)}:=  \{0\leq  \zeta_{\sigma(1)}\leq \cdots\leq
  \zeta_{\sigma(s)}\leq 1\leq \zeta_{\rho(1)} \leq \cdots
  \leq \zeta_{\rho(r)} \}\,.
\end{equation}
Only the powers of~$\eta$ in front of the integral give monodromies when~$\eta$ makes a loop around~$0$. 
This shows that, if we write
\begin{equation}\label{e:g0}
   \vec  I_N(\pmb a,\pmb b,\pmb c;\pmb d;\eta)
     \mathrel{\mathop{\to}\limits^{\gamma_0}}  g_0(\pmb a,\pmb c;\pmb d) \,\vec I_N(\pmb a,\pmb b,\pmb
   c;\pmb d;\eta),
\end{equation} 
then the monodromy matrix~$g_0$ is a diagonal matrix independent of $\pmb b$.

When $s=0$ and $\sigma$ is the empty permutation then all the points $z_i$ of the integration domain~\eqref{e:DeltaDef} belong to $(1,\infty)$ and the integrals have trivial monodromies around $\eta=0$,
i.e.
\begin{equation}
  I_{(\rho,\emptyset)}  (\pmb a,\pmb b,\pmb c;\pmb
                d;\eta) \mathrel{\mathop{\to}\limits^{\gamma_0}}   I_{(\rho,\emptyset)}  (\pmb a,\pmb b,\pmb c;\pmb
                d;\eta) \,.
\end{equation}

When $s=N$ and $\rho$ is the empty permutation then all the points $z_i$ of the integration domain~\eqref{e:DeltaDef} belong to $(0,1)$ and the integrals have the same  monodromies around $\eta=0$, i.e.
\begin{equation}\label{e:goaction}
  I_{(\emptyset,\sigma)}  (\pmb a,\pmb b,\pmb c;\pmb
                d;\eta) \mathrel{\mathop{\to}\limits^{\gamma_0}}
                \prod_{m=1}^N e^{2\pi i (a_m+c_m)}\prod_{1\leq m<n\leq
                  N} e^{2\pi i d_{mn}} I_{(\emptyset,\sigma)}  (\pmb a,\pmb b,\pmb c;\pmb
                d;\eta) \,.
              \end{equation}
Note that the arguments of the exponential phase factors can differ from $a_m,c_m,d_{mn}$ by arbitrary integers.

Therefore the monodromy matrix around $\eta=0$ has the diagonal form
\begin{equation}
  \label{e:g0matrix}
  g_0(\pmb a,\pmb c;\pmb d)=
  \begin{pmatrix}
   \mathbb I & 0 &0 \cr
    0 & \delta_0&0\cr
0 & 0 & \prod_{m=1}^N e^{2\pi i (a_m+c_m)}\prod_{1\leq m<n\leq
                  N} e^{2\pi i d_{mn}} \mathbb I
  \end{pmatrix}
\end{equation}
where $\mathbb I$ is the size $N!$ identity matrix and $\delta_0$ is a
diagonal matrix of size $(N-1) N!$ whose diagonal elements arise from the phases of the prefactor in~\eqref{e:IsigmarhoDefBis}.


\subsubsection{The monodromy around $\eta=1$}
\label{sec:monodromy-around-z=1}

The vectors $ {\vec I} (\pmb a,\pmb b,\pmb c;\pmb d;\eta)$ do not have
diagonal monodromy matrix around $\eta=1$. If instead we look at the
integrals $ {\vec J} (\pmb a,\pmb b,\pmb c;\pmb d;\eta)={\vec I} (\pmb b,\pmb
a,\pmb c;\pmb d;1-\eta)$ and at the transformation
\begin{equation}\label{e:g1}
\vec  J_N(\pmb a,\pmb b,\pmb c;\pmb d;\eta)
 \mathrel{\mathop{\to}\limits^{\gamma_1}}  g_{1}(\pmb b,\pmb c;\pmb d) \,\vec J_N(\pmb a,\pmb b,\pmb c;\pmb d;\eta),
\end{equation}
we find that the monodromy matrix $g_1$ is diagonal and independent of $\pmb a$, given by 
\begin{equation}
  \label{e:g1matrix}
  g_1(\pmb b,\pmb c;\pmb d)=
  \begin{pmatrix}
  \prod_{m=1}^N e^{2\pi i (b_m+c_m)}\prod_{1\leq m<n\leq
                  N} e^{2\pi i d_{mn}}  \mathbb I & 0 &0 \cr
    0 & \delta_1&0\cr
0 & 0 & \mathbb I
  \end{pmatrix}
\end{equation}
where $\mathbb I$ is the size $N!$ identity matrix and $\delta_1$ is a
diagonal matrix of size $(N-1) N!$.
\subsection{Asymptotic expansion of Aomoto-Gel'fand functions}\label{ssec:AsympAG}

We have already mentioned that the integrals~(\ref{e:IsigmarhoDef}), considered as functions of the $N(N+5)/2$ parameters $\pmb a,\pmb b,\pmb c,\pmb d$, can be analytically continued to define meromorphic functions on $\mathbb{C}^{N(N+5)/2}$, whose poles can only occur along divisors given by certain affine hyperplanes defined over $\mathbb{Z}$. 

With string theory applications in mind, it is of interest to consider the asymptotic expansion of Aomoto-Gel'fand functions at points with integer coordinates. Such points always belong to some polar divisor. It is believed but probably not rigorously demonstrated in such generality that, in the intersection of a small neighborhood of a singular point $(\pmb a_0,\pmb b_0,\pmb c_0,\pmb d_0)\in \mathbb{Z}^{N(N+5)/2}$ with the complement of the polar divisors, any Aomoto-Gel'fand function $F_{\Delta}(\pmb a,\pmb b,\pmb c;\pmb d;\eta)$ can be written as 
\begin{equation}\label{asympAGhyp}
F_{\Delta}(\pmb a,\pmb b,\pmb c;\pmb d;\eta)=\frac{F^{hol}(\pmb a,\pmb b,\pmb c;\pmb d;\eta)}{P(\pmb a,\pmb b,\pmb c;\pmb d)},
\end{equation}
for some homogeneous polynomial $P$ with integer coefficients and some holomorphic function $F^{hol}$ whose Taylor coefficients at $(\pmb a_0,\pmb b_0,\pmb c_0,\pmb d_0)$ belong to the algebra $\mathcal{H}_{\{0,1\},\mathcal{Z}[2\pi i]}$ of $\mathcal{Z}[2\pi i]$-linear combination of multiple polylogarithms (see Section~\ref{ssec:ringhyper}). A proof for certain special points (at the boundary of the convergence region) follows, for instance, by combining the methods of~\cite{Brown:2019wna} with Brown's work on periods of moduli spaces of genus zero curves~\cite{Brown:2009qja}. A proof of the general case would require a careful study of the analytic continuation.

Note that it is known that the coefficients of genus-zero amplitudes in open superstring theory, obtained by setting $\eta=0$ or $\eta=1$ and considering special integer points, belong to $\mathcal{Z}$~\cite{Broedel:2013aza}. The previous  statement about Aomoto-Gel'fand functions would only imply that they belong to $\mathcal{Z}[2\pi i]$. We do not know whether the coefficients of the asymptotic expansion of the functions~$F_{\Delta}(\pmb a,\pmb b,\pmb c;\pmb d;\eta)$ should belong to the smaller ring~$\mathcal{H}_{\{0,1\},\mathcal{Z}}$ (thus implying the optimal result for string theory amplitudes).

We also remark that the polynomials $P$ give massless field theory poles, and can be considered as well-understood in physics~\cite{Green:1987sp, Polchinski:1998rq, Polchinski:1998rr}.

\section{Holomorphic factorisation} \label{sec:factorisation}

The starting point of this section is a formula to write the correlation function $\mathcal G_N(\eta)$ defined by~\eqref{e:Gdefgeneral} as a bilinear product in our basis of (holomorphic) Aomoto-Gel'fand hypergeometric functions~\eqref{e:IsigmarhoDef} or~\eqref{e:IsigmarhoDefdual} and their complex conjugates. This yields an explicit conformal block decomposition, that~we call the \emph{holomorphic factorisation} of~$\mathcal G_N(\eta)$, which generalises known formulas from the literature (see e.g.~\cite{Dotsenko:1984nm,Dotsenko:1984ad}).

Our proof partly relies on the original method introduced in~\cite{Kawai:1985xq} to construct holomorphic factorisations of closed string building blocks in terms of open string building blocks, now known as KLT relations. It is of course necessary to adapt the method to our context with the extra-variable~$\eta$, which can be set to~$1$ (or to~$0$) to get the KLT relations back (see Section~\ref{ssec:KLTback}). Some more details, omitted in the general case, can be found in Section~\ref{sec:n=1-case} for the $N=1$-case, or for instance in~\cite{BjerrumBohr:2010hn} for the general (string amplitude) case.
\begin{thm}\phantomsection\label{prop:holfac} \emph{(Holomorphic factorisation)}. 
\begin{itemize}
\item[(i)] There exist square matrices $G_N(\pmb a,\pmb b,\pmb c;\pmb d)$ and $\hat G_N(\pmb a,\pmb b,\pmb c; \pmb d)$ of size
$(N+1)!$ with coefficients in the field $\mathbb{F}$ such that
  \begin{equation}\label{e:GholoFac}
      \mathcal G_{N}\left({\pmb a\, \pmb b\, \pmb c\, \pmb d\atop \pmb{\tilde a}\,
      \pmb{\tilde  b}\, \pmb{\tilde c}\, \pmb{\tilde d}}\Big|\eta\right)=\bigg(\frac i{2\pi}\bigg)^N\,\vec  I_N(\tilde{\pmb a},\tilde{\pmb b},\tilde{\pmb c};\tilde{\pmb d};\bar \eta)^T  \, G_N\left(\pmb a, \pmb b, \pmb c; \pmb d\right)  \, \vec I_N(\pmb a,\pmb
b,\pmb c;\pmb d;\eta) \,
\end{equation}
and
\begin{equation}\label{e:Gholobis}
      \mathcal G_{N}\left({\pmb a\, \pmb b\, \pmb c\, \pmb d\atop \pmb{\tilde a}\,
      \pmb{\tilde b}\, \pmb{\tilde c}\, \pmb{\tilde d}}\Big|\eta\right)=\bigg(\frac i{2\pi}\bigg)^N\,\vec J_N(\tilde{\pmb a},\tilde{\pmb b},\tilde{\pmb c};\tilde{\pmb d};\bar \eta)^T \,  \hat G_N\left(\pmb a, \pmb b, \pmb c; \pmb d\right)
\, \vec  J_N(\pmb a,\pmb
b,\pmb c;\pmb d;\eta) \,.
\end{equation}
\item[(ii)] The matrices $G_N(\pmb a,\pmb b,\pmb c;\pmb d)$ and $\hat G_N(\pmb a,\pmb b,\pmb c;\pmb d)$ such that~\eqref{e:GholoFac} and~\eqref{e:Gholobis} hold are unique.
\end{itemize}
\end{thm}

\begin{proof} $(i)$  For each $1\leq r\leq N$ we set $z_r=x_r+iy_r$ with $x_r, y_r \in
   \mathbb R$. The main idea is to consider the real variables $y_r$ as complex variables integrated over the real line, and to rotate these integration paths to the straight lines $y_r\to (i-2\epsilon)y_r$ contained in $\mathbb{C}$, with $\epsilon\in\mathbb{R}^+$, which tend to be purely imaginary as $\epsilon\rightarrow 0$. Because the poles of the integrand with respect to the variable~$y_r$ are located on the purely imaginary axis, and the integrand behaves nicely as $y_r\rightarrow \infty$, our original integral is invariant under this path deformation for any positive~$\epsilon$.
   
The complex variable $z_r=x_r+iy_r$ is thus deformed to $z_r= x_r-y_r+2i\epsilon y_r$, and so $\bar z_r= x_r-y_r-2i\epsilon y_r$. Setting $v_r^\pm :=x_r\pm y_r\in\mathbb{R}$ and $\delta_r:= v_r^+-v_r^-\in\mathbb{R}$, so that $z_r=v_r^-+i\epsilon\delta_r$ and  $\bar z_r= v_r^+-i\epsilon \delta_r$, the integral in~\eqref{e:Gdefgeneral} is therefore equal, for any positive~$\epsilon$, to\footnote{The prefactor comes from the change of variables.}
\begin{multline}\label{e:Gnepsdef}
\mathcal G^\epsilon_{N}(\eta):=\bigg(\frac i{2\pi}\bigg)^N\int_{\mathbb R^{2N}} \prod_{r=1}^N (v_r^-+i\epsilon \delta_r)^{a_r}
    (v_r^+-i\epsilon \delta_r)^{\tilde a_r}
    (v_r^--1+i\epsilon \delta_r)^{b_r}
     (v_r^+-1-i\epsilon \delta_r)^{\tilde b_r} \cr
    \times  (v_r^--\eta+i\epsilon \delta_r)^{c_r}
      (v_r^+-\bar \eta-i\epsilon \delta_r)^{\tilde c_r}\prod_{1\leq r<s\leq N} (v_r^--v_s^-+i\epsilon
    (\delta_r-\delta_s) )^{d_{rs}}(v_r^+-v_s^+-i\epsilon
    (\delta_r-\delta_s))^{\tilde d_{rs}} \prod_{r=1}^N dv_r^+
    dv_r^- 
   \,.
 \end{multline}
We are interested in the limit of this integral as $\epsilon\rightarrow 0$, which leads to the sought for factorisation. One must be careful when taking this limit and pay attention to the branch cuts of the integrand, in order for $\lim_{\epsilon}\mathcal G^\epsilon_{N}(\eta)$ to be well-defined and equal to $\mathcal G_N(\eta)$. This leads to the formula
\begin{multline}
\mathcal G_{N}(\eta)=\bigg(\frac i{2\pi}\bigg)^N
    \int_{\mathbb{R}^{2N}}\prod_{r=1}^N |v_r^-|^{a_r}
    |v_r^+|^{\tilde a_r}
    |v_r^--1|^{b_r}
     |v_r^+-1|^{\tilde b_r}|v_r^--\eta|^{c_r}
 |v_r^+-\bar \eta|^{\tilde c_r}\cr
    \prod_{1\leq r<s\leq N} |v_r^--v_s^-|^{d_{rs}}|v_r^+-v_s^+|^{\tilde d_{rs}} \,\varphi(\pmb a,\pmb b,\pmb c,\pmb d;\pmb v^+,\pmb v^-) \,\prod_{r=1}^N dv_r^+
    dv_r^-\,,
\end{multline}
where~$\varphi$ takes different values in $\mathbb{F}$ depending on the positions of each $v_r^+$ relative to its singular points $0,1,\eta,v_1^+,\ldots ,v_N^+$, and on the positions of each $v_r^-$ relative to its singular points $0,1,\eta,v_1^-,\ldots ,v_N^-$ (compare with~\cite[eq. (3.21)]{Kawai:1985xq}). It is very important to remark that these phases do not depend on the relative positions of the $v_r^+$ and $v_r^-$ variables, as one may expect from the factors $v_r^+\pm v_r^-$ appearing in the imaginary parts of the integral~\eqref{e:Gnepsdef}, due to the complete symmetry of the integrand with respect to $v_r^+\leftrightarrow v_r^-$ modulo integer shifts of the exponents. The consequence is that
\begin{multline}\label{e:Gepsfac}
\mathcal G_{N}(\eta)=\bigg(\frac i{2\pi}\bigg)^N\sum_{\alpha,\beta\in\mathfrak S_N}
  c_{\alpha,\beta}(\pmb a,\pmb b,\pmb c,\pmb d)
  \int_{\Delta_\alpha(\pmb v^+)\times \Delta_\beta(\pmb v^-)}  
\prod_{r=1}^N |v_r^-|^{a_r}
    |v_r^+|^{\tilde a_r}|v_r^--1|^{b_r}
     |v_r^+-1|^{\tilde b_r} \cr
    \prod_{r=1}^N |v_r^--\eta|^{c_r}
      |v_r^+-\bar \eta|^{\tilde c_r}\prod_{1\leq r<s\leq N} |v_r^--v_s^- |^{d_{rs}}|v_r^+-v_s^+|^{\tilde d_{rs}}  \prod_{r=1}^N dv_r^+
    dv_r^- 
   \,,
\end{multline}
where the domains~$\Delta_\alpha$ (resp.~$\Delta_\beta$) range over all the possible simplexes for which the $v_r^+$-part (resp. $v_r^-$-part) is well-defined, and the coefficients $c_{\alpha,\beta}(\pmb a,\pmb b,\pmb c,\pmb d)$ belong to $\mathbb{F}$. Each of the integrals in the sum is therefore the product of two Aomoto-Gel'fand hypergeometric functions, one in~$\eta$ and one in~$\bar \eta$. The statement follows from Theorem~\ref{thmbasisint}.

\medskip
$(ii)$  Suppose that eq.~\eqref{e:GholoFac} holds for two matrices $G_N$ and $G'_N$ with coefficients in $\mathbb{F}$. This implies that, for any $\eta$,
\begin{equation}\label{eqcontrad}
\vec  I_N(\tilde{\pmb a},\tilde{\pmb b},\tilde{\pmb c};\tilde{\pmb d};\bar \eta)^T  \, \big(G_N\left(\pmb a, \pmb b, \pmb c; \pmb d\right)\,-\,G'_N\left(\pmb a, \pmb b, \pmb c; \pmb d\right)\big)  \, \vec I_N(\pmb a,\pmb
b,\pmb c;\pmb d;\eta)\,=\,0\,.
\end{equation}

Suppose by contradiction that $G_N\neq G'_N$. We can assume that $(\tilde{\pmb a},\tilde{\pmb b},\tilde{\pmb c},\tilde{\pmb d})=(\pmb a, \pmb b, \pmb c; \pmb d)$, because integer shifts leave $G_N$ and $G'_N$ invariant. 

As explained in Section~\ref{ssec:AsympAG}, the coefficients of the asymptotic expansion of Aomoto-Gel'fand functions at some\footnote{Conjecturally, this should be true for all such points.} singular points with integer coordinates are known to belong to $\mathcal{H}_{\{0,1\},\mathcal{Z}}$, while those of the entries of $G_N - G'_N$ obviously belong to $\mathbb{Q}[2\pi i]$. Combining these expansions with eq.~\eqref{eqcontrad} yields non trivial $\mathcal{Z}[2\pi i]$-linear relations between products $L_{w_1}(z)\overline{L_{w_2}(z)}$, but this contradicts the fact, mentioned at the beginning of Section~\ref{ssec:ringhyper}, that such products are linearly independent over $\mathcal{O}_{\Sigma,\mathbb{C}}\otimes_\mathbb{C}\overline{\mathcal{O}_{\Sigma,\mathbb{C}}}$.
\end{proof}

Our construction applies as well to the holomorphic factorisation of
the four point correlation function of the minimal models considered
in~\cite{Dotsenko:1984nm}, where the authors
used an approach based on differential equations satisfied by the conformal blocks and on the fact that correlation functions are single-valued to construct a solution. They could only
determine the solution up to an overall scale, and the
argument that we have used fixes this scale.

\begin{cor}
The intertwining matrices $G_N$ and $\hat G_N$ in the holomorphic factorisation
in~\eqref{e:GholoFac} and~\eqref{e:Gholobis} 
are related by 
  \begin{equation}\label{e:GGhat}
  G_N(\pmb a, \pmb b, \pmb c; \pmb d)=R_N(\pmb a,\pmb b,\pmb
  c;\pmb d)^T\, \hat G_N(\pmb a, \pmb b, \pmb c; \pmb d)
  \, R_N(\pmb a,\pmb b,\pmb c;\pmb d)\,.
\end{equation}
where $R_N(\pmb a,\pmb b,\pmb c;\pmb d)$ is the matrix of change of basis from Corollary~\ref{prop:changebasis}.
\end{cor}

\begin{proof} This follows from the holomorphic factorisations~\eqref{e:GholoFac} and~\eqref{e:Gholobis} by combining the unicity of the factorisation with Corollary~\ref{prop:changebasis}.
\end{proof}
\subsection{Single-valuedness and holomorphic factorisation} \label{ssec:factorisation}

We derive constraints on the matrices $G_N$ and $\hat
G_N$ in the holomorphic factorisation of Theorem~\ref{prop:holfac} from the absence of monodromies around $\eta=0$ and
$\eta=1$ of the function $\mathcal G_N(\eta)$. 

\begin{propos}\label{prop:matrixformsv}
The intertwining matrices $G_N$  and $\hat G_N$ in the holomorphic
factorisation have the block diagonal form
\begin{equation}\label{e:GN}
  G_N=
  \begin{pmatrix}
    G_N^{(1)}&0&0\cr
    0&G_N^{(2)}&0\cr
    0&0&G_N^{(3)}
  \end{pmatrix}\,, \qquad\qquad 
  \hat  G_N=
  \begin{pmatrix}
   \hat  G_N^{(1)}&0&0\cr
    0&\hat G_N^{(2)}&0\cr
    0&0&\hat G_N^{(3)}
  \end{pmatrix}\,,
\end{equation}
where $G_N^{(i)}$ and $\hat G_N^{(i)}$ for $i=1,3$  are square-matrices of size
$N!$ while  $G_N^{(2)}$ and $\hat G_N^{(2)}$ are  diagonal matrices of
size $(N-1) N!$.
\end{propos}

\begin{proof}
The absence of monodromies of $\mathcal G_N(\eta)$ 
around $\eta=0$ implies that
\begin{multline}
 \vec I_N(\tilde{\pmb a},\tilde{\pmb
b},\tilde{\pmb c};\tilde{\pmb d};\bar \eta)^T \,  G_N(\pmb a, \pmb b, \pmb c; \pmb d)\,  \vec  I_N(\pmb a,\pmb
b,\pmb c;\pmb d;\eta)\\
=\vec I_N(\tilde{\pmb a},\tilde{\pmb
b},\tilde{\pmb c};\tilde{\pmb d};\bar \eta)^T \,
\left(  g_0(\tilde{\pmb a},\tilde{\pmb c};\tilde{\pmb d})^T\, G_N(\pmb a, \pmb b, \pmb c; \pmb d) \,g_0(\pmb a, \pmb c; \pmb d)\right)\, \vec  I_N(\pmb a,\pmb
b,\pmb c;\pmb d;\eta) \,,
\end{multline}
where $g_0$ is the monodromy matrix from~(\ref{e:g0}). The form of the entries of $g_0$ and equation~(\ref{e:spinG}) imply that $g_0(\tilde{\pmb a},\tilde{\pmb c};\tilde{\pmb d})=g_0(\pmb a, \pmb c; \pmb d)$. The unicity of the holomorphic factorisation implies that 
\begin{equation}
  G_N= g_0^T\, G_N \,g_0\,.  
\end{equation}
 Decomposing~$G_N$ into block matrices $(G_N)_{ij}$ with $1\leq i,j\leq 3$, and using the  block diagonal form for~$g_0$  given in
Section~\ref{sec:monodromy-around-z=0}, we have the following set of
equations 
\begin{align}
  (G_N)_{12}(1-\delta_0)&=0\cr
(G_N)_{13} \bigg(1-\prod_{m=1}^N e^{2\pi i (a_m+c_m)}\prod_{1\leq m<n\leq
                      N}e^{2\pi i d_{mn}} \mathbb I\bigg)&=0\cr
(1-\delta_0)(G_N)_{21}&=0\cr
 \bigg(1-\prod_{m=1}^N e^{2\pi i (a_m+c_m)}\prod_{1\leq m<n\leq
                      N}e^{2\pi i d_{mn}} \delta_0\bigg)(G_N)_{23}&=0\cr
 \bigg(1-\prod_{m=1}^N e^{-2\pi i (a_m+c_m)}\prod_{1\leq m<n\leq
                      N}e^{-2\pi i d_{mn}} \mathbb I\bigg)(G_N)_{31}&=0\cr
\bigg(1-\prod_{m=1}^N e^{-2\pi i (a_m+c_m)}\prod_{1\leq m<n\leq
                      N}e^{-2\pi i d_{mn}} \delta_0\bigg)(G_N)_{32}&=0\,.
\end{align}
The matrix $\delta_0$ is a diagonal matrix with diagonal elements
given by the phases factors in~\eqref{e:goaction}.  Since 
 the complex parameters $a_m$, $b_m$, $c_m$, $d_{mn}$ belong to a non-empty open domain, one 
concludes that~$G_N$ has the
block diagonal form
\begin{equation}\label{e:GNproof}
  G_N=
  \begin{pmatrix}
    G_N^{(1)}&0&0\cr
    0&G_N^{(2)}&0\cr
    0&0&G_N^{(3)}
  \end{pmatrix}\,,
\end{equation}
where $G_N^{(i)}:=(G_N)_{ii}$ for $i=1,3$  are square-matrices of size
$N!$ and  $G_N^{(2)}:=(G_N)_{22}$ is a  diagonal matrix of
size $(N-1)N!$. 

Similarly, the absence of monodromies around $\eta=1$ implies that the
matrix $\hat G_N$ in the holomorphic factorisation~\eqref{e:Gholobis}
satisfies the condition
\begin{equation}
 \hat G_N= g_1^T \,\hat G_N\, g_1\,.  
\end{equation}
Since $g_1$ has the  block diagonal form given in
Section~\ref{sec:monodromy-around-z=1} then $\hat G_N$ has the
block diagonal form
\begin{equation}\label{e:GhatNproof}
 \hat  G_N=
  \begin{pmatrix}
   \hat  G_N^{(1)}&0&0\cr
    0&\hat G_N^{(2)}&0\cr
    0&0&\hat G_N^{(3)}
  \end{pmatrix}\,,
\end{equation}
where $\hat G_N^{(i)}$ for $i=1,3$  are square-matrices of size
$N!$ and  $\hat G_N^{(2)}$ is a  diagonal matrix of
size $(N-1) N!$.
\end{proof}

\subsection{The $N=1$ case}
\label{sec:n=1-case}

We want to perform the holomorphic factorisation of the integral
  \begin{equation}\label{N=1correlator}
    \mathcal G_1 \left({ a\,  b\, c\atop {\tilde a}\,
      {\tilde  b}\, {\tilde c}}\Big|\eta\right)=\int_{\mathbb C}   z^{a}
   \bar z^{\tilde a}
     (z-1)^{b}
     (\bar z-1)^{\tilde b}
      (z-\eta)^{c}
      (\bar z-\bar \eta)^{\tilde c}d^2z \,,
    \end{equation}
with $a,\tilde a, b,\tilde b,c, \tilde c$ such that~(\ref{N=1correlator}) is absolutely convergent\footnote{This region is specified by the conditions $\text{Re}(a+\tilde{a}),\text{Re}(b+\tilde{b}),\text{Re}(c+\tilde{c})\geq -2$ and $\text{Re}(a+\tilde{a}+b+\tilde{b}+c+\tilde{c})\leq -2$.} and such that the integer spin  conditions $a-\tilde a\in\mathbb Z$, $b-\tilde b\in\mathbb Z$, $c-\tilde c\in\mathbb Z$ are satisfied.

As in the proof of the general case, we deform the integration path of the imaginary part so that $\mathcal{G}_1(\eta)$ is equal, for any positive $\epsilon$, to
    \begin{equation}\label{epsilonform1}
   \mathcal G_1^\epsilon(\eta)=\frac i{2\pi}\,\int_{\mathbb R^{2}}   (v^-+i\epsilon \delta)^{a}
    (v^+-i\epsilon \delta)^{\tilde a}
     (v^--1+i\epsilon \delta)^{b}
(v^+-1-i\epsilon \delta)^{\tilde b}
      (v^--\eta+i\epsilon \delta)^{c}
      (v^+-\bar \eta-i\epsilon \delta)^{\tilde c}dv^+
    dv^- \,,
    \end{equation}
with $\delta:=v^+-v^-$. Taking correctly the limit of this expression leads to the formula
\begin{equation}\label{prereductform}
   \mathcal G_1(\eta)=\frac i{2\pi}\,\int_{\mathbb R^{2}} |v^-|^{a} |v^+|^{\tilde a}
     |v^--1|^{b}
|v^+-1|^{\tilde b}
      |v^--\eta|^{c}
      |v^+-\bar \eta|^{\tilde c}\varphi(a,b,c;v^+,v^-)dv^+
    dv^- \,,
    \end{equation}
where $\varphi(a,b,c;v^+,v^-)\in \mathbb{F}$, and it takes different values depending on whether $v^+$ and $v^-$ belong to $(-\infty,0)$, $(0,\eta)$, $(\eta,1)$ or $(1,\infty)$. For instance, $\varphi=1$ if $v^+$ and $v^-$ both belong to $(-\infty,0)$ or $(0,\eta)$ or $(\eta,1)$ or $(1,\infty)$, but if $v^+$ (or $v^-$) belongs to $(-\infty,0)$ and $v^-$ (or $v^+$) belongs to $(1,\infty$) we get $\varphi=e^{i\pi (a+b+c)}$. We invite the reader to do the instructive exercise of determining the phase $\varphi$ in each possible domain. 

Since our purpose is to explicitly obtain the holomorphic factorisation, we now need to write each Aomoto-Gel'fand function in terms of our basis $\vec I_1(a,b,c;\eta)$ or $\vec J_1(a,b,c;\eta)$. To do so, we first follow a method of~\cite{Kawai:1985xq} to simplify the computation, and then we use the explicit change of basis worked out in Section~\ref{sec:S4pt}. We therefore take a step back from~\eqref{prereductform}, and we consider~\eqref{epsilonform1}. We fix $v^+$ in one of the four possible domains $(-\infty,0)$, $(0,\eta)$, $(\eta,1)$ or $(1,\infty)$, and consider the integration path of the variable $v^-+i\epsilon\delta$ for each of these cases, as in Figure~\ref{fig:v1m} below.

    \begin{figure}[ht]
      \centering
 \begin{tikzpicture}[scale=1.25]

\draw[ultra thick,->] (-2,0) -- (3,0);

\draw[fill] (-1,0) circle (0.04);
  \draw[fill] (0,0) circle (0.04);
    \draw[fill] (1,0) circle (0.04);
    \draw[fill] (2,0) circle (0.04);

\node[above] at (-.8,0) {$v^+$};
    \node[below] at (0,0) {0};
    \node[below] at (1,0) {$\eta$};
    \node[below] at (2,0) {1};
    \node[above] at (-2,.5) {$C^-$};

   \draw[ultra thick,postaction={decorate},decoration={
    markings,
    mark=at position 0.5 with {\arrowreversed{>}}}] (-1,0) to[out=135, in=45,distance=.5cm] (-2,.5);
    \draw[ultra thick, postaction={decorate},decoration={
    markings,
    mark=at position 0.5 with {\arrow{>}}}] (-1,0) to[out=-45, in=180,distance=1cm] (0,-.5);

 
\draw[ultra thick,->] (4,0) -- (9,0);

    \draw[fill] (5,0) circle (0.04);
    \draw[fill] (6,0) circle (0.04);
    \draw[fill] (7,0) circle (0.04);
     \draw[fill] (8,0) circle (0.04);

    \node[below] at (5,0) {0};
    \node[below] at (6,0) {$\eta$};
    \node[below] at (7,0) {1};
        \node[above] at (8.1,0) {$v^+$};

    \draw[ultra thick,postaction={decorate},decoration={
    markings,
    mark=at position 0.5 with {\arrowreversed{>}}}] (8,0) to[out=135, in=45,distance=.5cm] (7,.5);
    \draw[ultra thick, postaction={decorate},decoration={
    markings,
    mark=at position 0.5 with {\arrow{>}}}] (8,0) to[out=-45, in=180,distance=1cm] (9,-.6);

\node[below]  at (1,-1) {$(a)$};
  \node[below] at (6,-1) {$(b)$};

   \node[above] at (7,.5) {$C^-$};

 \end{tikzpicture}
   \begin{tikzpicture}[scale=1.25]

\draw[ultra thick,->] (-2,0) -- (3,0);

\draw[fill] (-1,0) circle (0.04);
  \draw[fill] (0,0) circle (0.04);
    \draw[fill] (1,0) circle (0.04);
    \draw[fill] (2,0) circle (0.04);

\node[below] at (-1,0) {0}; 
    \node[above] at (0.1,0) {$v^+$};
    \node[below] at (1,0) {$\eta$};
    \node[below] at (2,0) {1};
    \node[above] at (-1,.5) {$C^-$};

    \draw[ultra thick,postaction={decorate},decoration={
    markings,
    mark=at position 0.5 with {\arrowreversed{>}}}] (0,0) to[out=135, in=45,distance=.5cm] (-1,.5);
    \draw[ultra thick, postaction={decorate},decoration={
    markings,
    mark=at position 0.5 with {\arrow{>}}}] (0,0) to[out=-45, in=180,distance=1cm] (1,-.5);

 
\draw[ultra thick,->] (4,0) -- (9,0);

    \draw[fill] (5,0) circle (0.04);
    \draw[fill] (6,0) circle (0.04);
    \draw[fill] (7,0) circle (0.04);
     \draw[fill] (8,0) circle (0.04);

    \node[below] at (5,0) {0};
    \node[below] at (6,0) {$\eta$};
    \node[above] at (7.1,0) {$v^+$};
        \node[below] at (8,0) {1};

    \draw[ultra thick,postaction={decorate},decoration={
    markings,
    mark=at position 0.5 with {\arrowreversed{>}}}] (7,0) to[out=135, in=45,distance=.5cm] (6,.5);
    \draw[ultra thick, postaction={decorate},decoration={
    markings,
    mark=at position 0.5 with {\arrow{>}}}] (7,0) to[out=-45, in=180,distance=1cm] (8,-.5);

\node[below]  at (1,-1) {$(c)$};
  \node[below] at (6,-1) {$(d)$};

   \node[above] at (6,.5) {$C^-$};

  \end{tikzpicture}
      \caption{The contour of integration $C^-$ for the variable $v^-+i\epsilon\delta$ depending on the
      position of $v^+$ on the real axis.}
      \label{fig:v1m}
    \end{figure}

    The situation in the four different cases is the following:
    \begin{itemize}[label=$\blacktriangleright$]
    \item If $v^+\in (-\infty,0)$ then the integration contour for $v^-+i\epsilon\delta$
      is given in Figure~\ref{fig:v1m}(a). Since the only poles in the variable $v^-+i\epsilon\delta$ are at $0,1,\eta$, we obtain that the integral vanishes.
     \item If $v^+\in (1,+\infty)$ then the integration contour for $v^-+i\epsilon\delta$
      is given in Figure~\ref{fig:v1m}(b). Since the only poles in the variable $v^-+i\epsilon\delta$ are at $0,1,\eta$, we obtain that the integral vanishes.
       \item If $v^+\in (0,\eta)$ then the integration contour for $v^-+i\epsilon\delta$
      is given in Figure~\ref{fig:v1m}(c). By deforming the   part of
      the contour which is in the lower half-plane to the left to get a loop around~$0$ based at~$-\infty$,
      one picks the contribution from $v^-=0$. Hence we get
      \begin{equation}
     \lim_{\epsilon\to0}   \int_{C^-}      (v^-+i\epsilon \delta)^{a}
        (v^- -1+i\epsilon \delta)^{b} (v^- -\eta+i\epsilon \delta)^{c}dv^-   
=        2i\sin(\pi a)  \int_{-\infty}^0  (-v^-)^{a}
         (1-v^-)^{b}
         (\eta-v^-)^{c}dv^- .
       \end{equation}
\item If $v^+\in (\eta,1)$ then the integration contour for $v^-+i\epsilon\delta$
      is given in Figure~\ref{fig:v1m}(d). By deforming the   part of
      the contour which is in the upper half-plane to the right to get a loop around~$1$ based at~$-\infty$,
      one picks the contribution from $v^-=1$. Hence we get
      \begin{equation}
     \lim_{\epsilon\to0}   \int_{C^-}        (v^-+i\epsilon \delta)^{a}
        (v^- -1+i\epsilon \delta)^{b} (v^- -\eta+i\epsilon \delta)^{c} dv^- 
=        2i\sin(\pi b)  \int_\eta^1   (-v^-)^{a}
         (1-v^-)^{b}
         (\eta-v^-)^{c}dv^-.
       \end{equation}
    \end{itemize}
Therefore we obtain that
\begin{multline}\label{e:Ghol1}
  \mathcal G_1\left({ a\,  b\, c\atop {\tilde a}\,
      {\tilde b}\, {\tilde c}}\big|\eta\right)=  -\frac{\sin(\pi a)}{\pi} \int_0^\eta
  (v^+)^{\tilde a} (1-v^+)^{\tilde b} (\bar \eta-v^+)^{\tilde c}dv^+ \, \int_{-\infty}^0 
  (-v^-)^{a} (1-v^-)^{b} (\eta-v^-)^{c}dv^-\cr
  -  \frac{\sin(\pi b)}{\pi} \int_\eta^1 
  (v^+)^{\tilde a} (1-v^+)^{\tilde b} (v^+-\bar \eta)^{\tilde c}dv^+ \,\int_1^{+\infty}
  (v^-)^{a} (v^--1)^{b} (v^--\eta)^{c}dv^-\,.
\end{multline}
Following the notations and results of Section~\ref{sec:S4pt}, we have
\begin{equation}
\vec
    I_1 (a,b,c;\eta)=\begin{pmatrix}
{\displaystyle  I_{(Id,\emptyset)}(a,b,c;\eta)} \cr
{\displaystyle I_{(\emptyset,Id)}(a,b,c;\eta)}
\end{pmatrix}=
\begin{pmatrix}\label{e:I1def}
{\displaystyle \int_1^{+\infty}
  z^{a} (z-1)^{b} (z-\eta)^{c}dz }\cr
{\displaystyle\int_0^\eta 
  z^{ a} (1-z)^{ b} ( \eta-z)^{ c}dz}
\end{pmatrix}
\end{equation}
and
\begin{equation}
\vec
    J_1 (a,b,c;\eta)=\begin{pmatrix}
{\displaystyle  J_{(Id,\emptyset)}(a,b,c;\eta)} \cr
{\displaystyle J_{(\emptyset,Id)}(a,b,c;\eta)}
\end{pmatrix}=
\begin{pmatrix}\label{e:J1def}
{\displaystyle \int_{-\infty}^0 
  (-z)^{a} (1-z)^{b} (\eta-z)^{c}dz}\cr 
{\displaystyle \int_\eta^1 
  z^{ a} (1-z)^{ b} (z- \eta)^{ c}dz}
\end{pmatrix}\,.
\end{equation}
The  linear relations~\eqref{e:MonItoJ} give
\begin{align}
  J_{(Id,\emptyset)}(a,b,c;\eta)&= {\sin(\pi b)I_{(Id,\emptyset)}(a,b,c;\eta)
  -\sin(\pi c) I_{(\emptyset,Id)}(a,b,c;\eta)\over \sin(\pi (a+c))}, \cr
 J_{(\emptyset,Id)}(\tilde a,\tilde b,\tilde c;\bar\eta)&= {-\sin(\pi ( a+ b+ c)) I_{(Id,\emptyset)}(\tilde a,\tilde b,\tilde c;\bar\eta)
  -\sin(\pi a) I_{(\emptyset,Id)}(\tilde a,\tilde b,\tilde c;\bar\eta)\over \sin(\pi (a+ c))}\,.
\end{align}
We can therefore rewrite~\eqref{e:Ghol1} as 
\begin{equation}\label{e:G1resultI}
  \mathcal G_1\left({ a\,  b\, c\atop {\tilde a}\,
      {\tilde b}\, {\tilde c}}\big|\eta\right) =\frac{i}{2\pi}\,\vec I_1 (\tilde a,\tilde b,\tilde c;\bar \eta)^T
\,  G_1(a,b,c)\,
\vec I_1(a,b,c;\eta)\,.
\end{equation}
The off-diagonal elements of $G_1$ are $(G_1)_{12}=0$ and
\begin{equation}
  (G_1)_{21}=2i\sin(\pi b)\left({\sin(\pi a)\over \sin(\pi
      (a+c))}- {\sin(\pi a)\over \sin(\pi
      (a+c))}\right)  =0\,,
\end{equation}
as required by the absence of monodromies for
$\mathcal G_1\left({ a\,  b\, c\atop {\tilde a}\,
      {\tilde b}\, {\tilde c}}\big|\eta\right)$.
Hence we have 
\begin{equation}
 G_1(a,b,c)
={-2i\over \sin(\pi(a+c))} \begin{pmatrix}
\sin(\pi(a+b+c))  \sin(\pi
   b)  &0\cr
  0  &\sin(\pi a)\sin(\pi c)
 \end{pmatrix}\,.
\end{equation}

Similarly, by expressing the holomorphic factorisation using the other
basis of Aomoto-Gel'fand hypergeometric functions
\begin{equation}\label{e:holresultJ}
  \mathcal G_1\left({ a\,  b\, c\atop {\tilde a}\,
      {\tilde b}\, {\tilde c}}\big|\eta\right) =
   \frac i{2\pi}\, \vec  J_1(\tilde a,\tilde b,\tilde c;\bar \eta)^T
\, \hat G_1(a,b,c)\,
\vec J_1 (a,b,c;\eta),
\end{equation}
we get
\begin{equation}\label{e:G1resultJ}
 \hat G_1(a,b,c)=
{-2i\over\sin(\pi(b+c))} \begin{pmatrix}
 \sin(\pi(a+b+c))  \sin(\pi a) &0\cr
  0  &\sin(\pi b)\sin(\pi c)
 \end{pmatrix}\,.
\end{equation}

The relation between this expression  and the  holomorphic factorisation
of closed string theory amplitudes 
  in~\cite{Kawai:1985xq,BjerrumBohr:2010hn} is discussed in Section~\ref{sec:svM4}.
The arbitrariness in choosing to close the contour of integration to
the left or to the right is taken care by the linear relations
derived earlier between the various integrals appeared,
as discussed in details in~\cite{BjerrumBohr:2010hn}.

\subsection{The $\alpha'$-expansion of the correlation function $\mathcal{G}_N(\eta)$}
\label{sec:clos-string-part}

The meromorphic continuation in the parameters $(\pmb a,\pmb b,\pmb c,\pmb d)\in\mathbb{C}^{N(N+5)/2}$ of the Aomoto-Gel'fand hypergeometric functions can be combined with the holomorphic factorisation to analytically continue the correlation function $\mathcal G_N(\eta)$ outside the region of convergence of its defining integral. In particular, with the identification of the parameters in~\eqref{e:paramA}, we can consider the
small $\alpha'$-expansion (in the sense of Section~\ref{ssec:AsympAG}) of $\mathcal
G_N(\eta)$ for fixed values of the integers $n_{ij}$ and $\tilde{n}_{ij}$.

We have mentioned in Section~\ref{ssec:AsympAG} that the coefficients of the expansion of the Aomoto-Gel'fand functions should be $\mathcal{Z}[2\pi i]$-linear combinations of
multiple polylogarithms. Moreover, since the entries of the matrix~$G_N$ belong to the field~$\mathbb{F}$, the coefficients of their expansion belong to $\mathbb{Q}[2\pi i]$. Therefore, by the holomorphic factorisation, the asymptotic expansion coefficients of $\mathcal G_{N}$ should be $\mathcal{Z}[(2\pi i)^{\pm 1}]$-linear combinations\footnote{The negative powers of $\pi$ come from the prefactor $(i/2\pi)^N$. In fact, following more closely~\cite{BjerrumBohr:2010hn}, it is possible to prove that the coefficients of the $\alpha'$-expansion of $G_N$ belong to $(2\pi i)^N\mathbb{Q}[\pi]$, which implies that we can get rid of all negative powers of $\pi$, but we have preferred to skip these technical details.}
of products of holomorphic and anti-holomorphic multiple polylogarithms, which are single-valued as functions of~$\eta$. By Proposition~\ref{CorollaryRefinement} (with $R=\mathcal{Z}[(2\pi i)^{\pm 1}]$) we conclude that these
coefficients should then belong to the ring $\mathcal{H}^{\rm sv}_{\{0,1\},\mathcal{Z}[(2\pi i)^{\pm 1}]}$ of $\mathcal{Z}[(2\pi i)^{\pm 1}]$-linear combinations of single-valued multiple polylogarithms (see Section~\ref{sec:ringsv}).

We believe that these coefficients should even belong to $\mathcal{H}^{\rm sv}_{\{0,1\},\mathcal{Z}^{\rm sv}}$, and a proof of this fact, at least for special points at the boundary of the convergence region, should follow from the methods used in Section~\ref{Sec:Final} in the case of string amplitudes.

\subsection{Back to closed string amplitudes}\label{sec:5.4}

We now specialise some of the previous results on
$\mathcal{G}_N(\eta)$ to the value $\eta=1$. With the parameter identification from eq.~(\ref{e:paramA}), these special values correspond to the partial amplitudes $M_{N+3}(\pmb s, \pmb{n},\tilde{\pmb{n}})$, which are the building blocks of the closed string amplitude $\mathcal M_{N+3}(\pmb s,\pmb{\epsilon})$ (see eq.~\eqref{e:Treedecomp}). We could have also specialised $\mathcal{G}_N(\eta)$ to $\eta=0$, using a different parameter identification. Our choice of setting $\eta=1$ will lead us to use the holomorphic factorisation~\eqref{e:Gholobis} rather than~\eqref{e:GholoFac} (which would be more convenient if we set $\eta=0$ instead).

First of all, we remark that the considerations from the previous section, even if rigorously proven, would not imply that the coefficients of the $\alpha'$-expansion of $M_{N+3}(\pmb s, \pmb{n},\tilde{\pmb{n}})$ belong to the ring~$\mathcal{Z}^{\rm sv}$ of single-valued multiple zeta values (specializing to $\eta=1$ we would land in the larger space $\mathcal{Z}[(2\pi i)^{\pm 1}]$). As announced in the introduction, we will give a proof of the single-valued nature of these coefficients (see Section~\ref{Sec:Final}), but for this we will have to take a different route. We therefore turn to other aspects of the holomorphic factorisation.

\subsubsection{Relation with the KLT representation of closed string amplitudes}\label{ssec:KLTback}

Setting $\eta=1$ in the expression for the  holomorphic
factorisation in~\eqref{e:Gholobis} leads to a fairly simple expression,
because at $\eta=1$ only the first $N!$ rows $\vec \jmath_N$ of the vector of
Aomoto-Gel'fand hypergeometric functions $\vec J_N$
in~\eqref{e:vectnot} are non-vanishing:
\begin{equation}
   \vec J_N(\pmb a,\pmb b,\pmb c;\pmb d;1) =
    \begin{pmatrix}
    \vec \jmath_N(\pmb a,\pmb b,\pmb c;\pmb d)   \cr 0
    \end{pmatrix},
\end{equation}
and $
\vec\jmath_N(\pmb a,\pmb b,\pmb c;\pmb d)=(J_{(\rho,\emptyset)}(\pmb
a,\pmb b,\pmb c;\pmb d;1))_{\rho\in\mathfrak S_N}$
with
\begin{multline}\label{e:smallj}
  J_{(\rho,\emptyset)}(\pmb
a,\pmb b,\pmb c;\pmb d;1)=   \int_{z_{\rho(N)}\leq\cdots \leq
    z_{\rho(1)}\leq 0} \,\prod_{1\leq m<n\leq N }^N
  (z_{\rho(m)}-z_{\rho(n)})^{d_{\rho(m)\rho(n)}} \cr
\times  \prod_{m=1}^{N} (-z_{\rho(m)})^{a_{\rho(m)}} (1-z_{\rho(m)})^{b_{\rho(m)}+c_{\rho(m)}}\prod_{i=1}^N dz_i\,.
\end{multline}
Hence, using the parameter identification~(\ref{e:paramA}), the partial closed string amplitudes read
\begin{equation}\label{e:MnResult}
   M_{N+3}(\pmb s,\pmb n,\pmb {\tilde n})= \mathcal G_{N}\left({\pmb a\, \pmb b\, \pmb c\, \pmb d\atop \pmb{\tilde a}\,
      \pmb{\tilde  b}\, \pmb{\tilde c}\, \pmb{\tilde d}}\Big|1\right)=\bigg(\frac i{2\pi}\bigg)^N\, \vec \jmath_N(\tilde{\pmb
   a},\tilde{\pmb b},\tilde{\pmb c};\pmb d)^T  \, \hat G_N^{(1)}(\pmb a,\pmb b,\pmb c;\pmb d) \, \vec \jmath_N(\pmb
   a,\pmb b,\pmb c;\pmb d).
\end{equation}
In other words, although the end result for the closed string building block only
depends on $\hat G_N^{(1)}$, it was necessary to have the full matrix
$\hat G_N$ to be able to cancel the monodromies when $\eta$ varies.

The integrals in~\eqref{e:smallj} are colour-ordered open string amplitudes. The expression for the closed string amplitudes  in~\eqref{e:MnResult}, which is equivalent to the original KLT relations~\cite{Kawai:1985xq}, is the same appeared 
in~\cite[eq.~(4.3)]{Mizera:2016jhj} or~\cite[Example~3.1]{Mizera:2017cqs}, involving the inverse of the momentum
kernel from~\cite{BjerrumBohr:2009rd,Stieberger:2009hq}.  
 
\subsubsection{The single-valued expansion of the four-point amplitude}
\label{sec:svM4}

We illustrate the previous results on the $N=1$ case corresponding to
the four-point amplitude 

  \begin{equation}
      M_4(\pmb s,\pmb n,\pmb {\tilde n})=\int_{\mathbb C}
     | z|^{2s_{01}} z^{n_{01}}\bar z^{\tilde n_{01}} |1-z|^{2s_{12}}
     (1-z)^{n_{12}}
     (1-\bar z)^{\tilde n_{12}}
 \,d^2z\,.
    \end{equation}
We can write
\begin{equation}
      M_4(\pmb s,\pmb n,\pmb {\tilde n})=\mathcal G_1\left({a\, b\,
        c\atop \tilde a\,\tilde b\,\tilde c}\big| 1\right),
\end{equation}
with the identification of the parameters 
$  a=s_{01}+n_{01}$, $\tilde a=s_{01}+\tilde n_{01}$,
$b+c=s_{12}+n_{12}$ and $\tilde b+\tilde c=s_{12}+\tilde
n_{12}$ and the assumption that the Mandelstam variables $s_{01},s_{02}, s_{12}\notin\mathbb{Z}$ satisfy the momentum conservation condition $s_{01}+s_{02}+s_{12}=0$.

With such identifications, the holomorphic decomposition in~\eqref{e:holresultJ} evaluated
at $\eta=1$ gives
\begin{multline}\label{e:Gholo4pt}
    M_4(\pmb s,\pmb n,\pmb {\tilde n})\,=\,-\,{\sin(\pi s_{01})
      \sin(\pi s_{02})\over  \pi\sin(\pi s_{12})} \times\\
   \times \left(\int_{-\infty}^0                                                   (-z)^{ s_{01}+ n_{01}}  (1-z)^{ s_{12}+   n_{12}}\,dz\right)
      \left(\int_{-\infty}^0
                                                     (-z)^{ s_{01}+
                                                       \tilde n_{01}}
                                       (1-z)^{ s_{12}+
                                        \tilde n_{12}}\,dz\right) \,.
\end{multline}
By specialising the change of basis equation~\eqref{e:MonJtoI} to $\eta=1$ one gets\footnote{Alternatively, one can substitute $z\rightarrow 1/(1-z)$, use Euler's formula for the beta integral in terms of $\Gamma$-functions and Euler's reflection formula $\Gamma(x)\Gamma(1-x)=\tfrac \pi{\sin(\pi x)}$}
\begin{equation}\label{equaz}
\sin(\pi s_{02}) \int_{-\infty}^0
                                                     (-z)^{ s_{01}+ n_{01}}
                                       (1-z)^{ s_{12}+
                                         n_{12}}\,dz =\sin(\pi
s_{12}) \int_0^1
                                                     z^{ s_{01}+ n_{01}}
                                       (1-z)^{ s_{12}+ n_{12}}\,dz \,.
\end{equation}
Plugging this relation in~\eqref{e:Gholo4pt} (or specialising~\eqref{e:Ghol1} to $\eta=1$) we recover, up to a normalisation prefactor, the original KLT expression~\cite[eq.~(3.11)]{Kawai:1985xq}
\begin{align}
  M_4(\pmb s,\pmb n,\pmb {\tilde n})&= \,-\, \frac 1\pi\sin(\pi s_{01})\left(\int_{-\infty}^0(-z)^{ s_{01}+ n_{01}}(1-z)^{ s_{12}+n_{12}}\,dz\right) \left(\int_0^1 z^{ s_{01}+\tilde n_{01}}
(1-z)^{ s_{12}+ \tilde n_{12}}\,dz\right) \cr
&= \,-\,\frac 1\pi\sin(\pi s_{01})  \left(\int_{-\infty}^0
(-z)^{ s_{01}+\tilde n_{01}}(1-z)^{ s_{12}+\tilde   n_{12}}\,dz\right) \left(\int_0^1z^{ s_{01}+n_{01}}(1-z)^{ s_{12}+  n_{12}}\,dz\right)\,.
\end{align}

This illustrates how in this case the momentum kernel~$\mathcal S:= \sin(\pi s_{01})$ from~\cite{BjerrumBohr:2010hn} arises (up to a normalisation factor) as the product of
the holomorphic factorisation matrix~$\hat G_N^{(1)}$ times a change of basis matrix between colour-ordered amplitudes.

The $\Gamma$-function representations of the integrals appearing in eq.~(\ref{equaz}) are
\begin{align}
  \int_0^1
                                                     z^{ s_{01}+ n_{01}}
                                       (1-z)^{ s_{12}+ n_{12}}\,dz   &={\Gamma(1+
                          s_{01}+n_{01})\Gamma(1+ s_{12}+n_{12})\over\Gamma(2+s_{01}+s_{12}+n_{01}+n_{12})}\cr
  \int_{-\infty}^0
                                                     (-z)^{ s_{01}+ n_{01}}
                                       (1-z)^{ s_{12}+
                                         n_{12}}\,dz&=           {\Gamma(1+ s_{01}
                               +\tilde n_{01})\Gamma(-1- (s_{01}+s_{12})-\tilde n_{01}-\tilde n_{12})\over\Gamma(-s_{12}-\tilde n_{12})}\,.          
\end{align}

By momentum conservation and Euler's reflection formula the four-point partial amplitude reads 
\begin{equation}
  M_4(\pmb s,\pmb n,\pmb {\tilde n})= (-1)^{\bar
    n_{12}}
  {\Gamma(1 +  s_{01} + n_{01})\over \Gamma(- s_{01} - \tilde n_{01})}
   {\Gamma(1 + s_{12} + n_{12})\over \Gamma(-
          s_{12} - \tilde n_{12})}{\Gamma(-1
          +s_{02} - \tilde n_{01} - \tilde n_{12})\over \Gamma(2 -
          s_{02} + n_{01} + n_{12})} \,.
\end{equation}
Using repeatedly that
$\Gamma(1+x)=x\Gamma(x)$, one can find a rational function $Q(s_{01},s_{12},s_{02})$  with integer coefficients (depending on the integer parameters $n_{01}, n_{12}, \tilde n_{01}, \tilde n_{12}$) such that
\begin{equation}
M_4(\pmb s,\pmb n,\pmb {\tilde n})= Q(s_{01},s_{12},s_{02})
{\Gamma(1 +  s_{01})\over \Gamma(1- s_{01})}{\Gamma(1 + s_{12})\over \Gamma(1-s_{12})}{\Gamma(1+s_{02})\over \Gamma(1 -s_{02})} \,.
\end{equation}
Euler's formula
\begin{equation}
  \Gamma(1+x)= e^{-\gamma x} \exp\left(\sum_{m\geq2} {\zeta
      (m)\over m} (-x)^m\right),
\end{equation}
where $\gamma$ is the Euler-Mascheroni constant, implies that
\begin{equation}
  {\Gamma(1+x+ n)\over \Gamma(1-x- n)}=
  e^{-2\gamma (x+n)}
  \exp\left(-2\sum_{m=1}^\infty {\zeta(2m+1)\over 2m+1} (x+n)^{2m+1}\right)  \,.
\end{equation}
Therefore, by momentum conservation, the four-point partial amplitude takes the form
\begin{equation}
    M_4(\pmb s,\pmb n,\pmb {\tilde n})\,=\,Q(s_{01},s_{12},s_{02})\exp\left(-2\sum_{m=1}^\infty {\zeta(2m+1)\over 2m+1} \Big( s_{01}^{2m+1}+s_{12}^{2m+1}+s_{02}^{2m+1}\Big)\right) \,.
\end{equation}
The argument of the exponential factor only involves odd Riemann
zeta values, i.e. single-valued multiple zeta values (we recall that $\zeta^{\rm sv}(2n)=0$ and
$\zeta^{\rm sv}(2n+1)=2\zeta(2n+1)$). This shows that $ M_4(\pmb
s,\pmb n,\pmb {\tilde n})$ is the product of a rational function in
$s_{01}$, $s_{12}$, $s_{02}$
times a function whose Taylor expansion only involves single-valued multiple zeta
values. The rest of the paper is devoted to prove the same statement for any $N$.

\section{Integration of single-valued hyperlogarithms}\label{sec:inthyp}

We recall that in Section~\ref{Sec:Hyperlogs}, starting from an abstract alphabet $X$ and its associated set of points $\Sigma=\{0,1,\sigma_2,\ldots ,\sigma_n\}\subset\mathbb{C}$, we have defined two families of functions: hyperlogarithms $L_w(z)$, which are multi-valued on $D:=\mathbb{C}\setminus \Sigma$, and their single-valued analogues $\mathcal{L}_w(z)$. Both families are indexed by labels $w\in X^*$, i.e. words in the abstract non-commutative letters~$x_i\in X$. Throughout this and the next section, in order to lighten the notation, we will denote by $X$  both the alphabet and the set of complex points, and by $\sigma_i$ both a point and a formal letter. The meaning will always be clear from the context. The goal of this section is to prove a qualitative result on integrals of single-valued hyperlogarithms over $\mathbb{P}^1_\mathbb{C}$ (Theorem~\ref{LemmaAppD}) which plays a crucial role in Section~\ref{Sec:Final}.

\subsection{Label dependence in the multi-valued case}

For every $2\leq i\leq n$ we consider the alphabets $X_i:=X\setminus \{\sigma_i\}$. We also recall from Section~~\ref{Ssec:specialhyper} that $\mathcal{S}_{X,R}$ denotes the ring of regularised special values of hyperlogarithms over a ring $R$ with respect to an alphabet $X$. The next proposition, due to Panzer\footnote{This is a key ingredient of the Maple procedure \texttt{HyperInt} developed by Panzer~\cite{Panzer:2014caa} which allows to compute multiple integrals of hyperlogarithms.}, clarifies the behaviour of (special values of) hyperlogarithms as functions of a letter $\sigma_i$ in their label. We include a sketch of the proof, where we explain the main idea but omit the technical details\footnote{Special values of hyperlogarithms are defined in the divergent cases by a regularisation procedure. Here we will not verify that the steps of the proof are compatible with such regularisation. The omitted details can be found in~\cite{Panzer:2015ida}.}.
\begin{propos}[Panzer~\cite{Panzer:2015ida}]\phantomsection\label{PanzerThm}
For any $w\in X^*$, any $0\leq j\leq n$ and any $2\leq i\leq n$
\begin{equation}
L_w(\sigma_j)=\sum_{u} c_uL_u(\sigma_i),
\end{equation}
where $c_u\in \mathcal{S}_{X_i,\mathbb{Q}[2\pi i]}$ and the sum runs over a finite number of words $u\in X_i^*$.
\end{propos}
\begin{proof} 
The statement is proven by induction on the length~$|w|$ of the word $w$. If $|w|=1$ then the only interesting case is $w=\sigma_i$, where one needs to rewrite $\log(1-\sigma_j/\sigma_i)$ in terms of $L_0(\sigma_i)$ and $L_{\sigma_j}(\sigma_i)$. This introduces an integer multiple of~$\pi i$ which depends on the chosen branch of the logarithm. To prove the general case, we need the following:
\begin{lemma}\phantomsection\label{CorHyper}
For any $w\in X^*$, any $z\in\mathbb{C}$ and any $2\leq i\leq n$
\begin{equation}\label{eqstrasenznum}
\frac{\partial}{\partial \sigma_i}L_w(z)=\sum_{\substack{\tau \in X_{i}\cup \{z\}\\ v\in X^*}}\frac{\lambda_{\tau, v}}{\sigma_i-\tau}L_v(z),
\end{equation}
where $\lambda_{\tau, v}\in\mathbb{Q}$, $|v|<|w|$ and the sum is finite.
\end{lemma}
\begin{proof}[Proof of the lemma] 
If we denote $\sigma_{i_0}:=z$, $\sigma_{i_{r+1}}:=0$, $d\log(0):=0$, a simple computation shows that the total differential of any $L_{\sigma_{i_1}\cdots \sigma_{i_r}}(z)$ is given (in terms of regularised values) by
\begin{equation}\label{TotalDiff}
dL_w(z)=\sum_{k=1}^n L_{\sigma_{i_1}\cdots \sigma_{i_{k-1}}\sigma_{i_{k+1}}\cdots \sigma_{i_r}}(z)\,d\log\bigg(\frac{\sigma_{i_k}-\sigma_{i_{k-1}}}{\sigma_{i_{k+1}}-\sigma_{i_k}}\bigg).
\end{equation}
Comparing the two sides of eq.~(\ref{TotalDiff}) we get~(\ref{eqstrasenznum}).
\end{proof}
By the lemma and the inductive hypothesis,
\begin{equation}
\frac{\partial}{\partial \sigma_i}L_w(\sigma_j)=\sum_{\substack{\tau \in X_{i}\\ u\in X_{i}^*}}\frac{c_u}{\sigma_i-\tau}L_u(\sigma_i),
\end{equation}
where $c_u\in \mathcal{S}_{X_i,\mathbb{Q}[2\pi i]}$. Hence we have
\begin{equation}
L_w(\sigma_j)=\sum_{\substack{\tau \in X_{i}\\ u\in X_i^*}}c_uL_{\tau u}(\sigma_i) + c,
\end{equation}
and $c=L_w(\sigma_j)\big|_{\sigma_i=0}\in \mathcal{S}_{X_i,\mathbb{Q}[2\pi i]}$.
\end{proof}

\subsection{Label dependence in the single-valued case}
We have seen that (special values of) hyperlogarithms, as functions of the letters in the label, are again hyperlogarithms. We want to prove a similar result for single-valued hyperlogarithms. The first necessary step is to prove the following:
\begin{lemma}\phantomsection\label{LemmaSV}
For each $w\in X^*$, each $2\leq i\leq n$ and each $z\in\mathbb{C}$ the single-valued hyperlogarithm $\mathcal{L}_w(z)$ is single-valued also as a function of $\sigma_i\in\mathbb{C}$.
\end{lemma}
\begin{proof}
We need to show that $\sigma_i\rightarrow\mathcal{L}_w(z)$ is a well-defined function for all $w\in X^*$ and all $\sigma_i,z\in\mathbb{C}$. Recall that, by Theorem~\ref{BrownsvHyperlog}, $\mathcal{L}_w(z)$ is a uniquely determined and well-defined function of $z\in D$ for any $\sigma_i\in\mathbb{C}$, which we have extended to a function of $z\in\mathbb{C}$. This precisely means that $\mathcal{L}_w(z)$ takes a unique value for any $z$ and any $w$, so $\sigma_i\rightarrow\mathcal{L}_w(z)$ is well-defined.
\end{proof}

We are now ready to prove the single-valued analogue of Proposition~\ref{PanzerThm}, which will play an important role in the computation of multiple integrals of single-valued hyperlogarithms. We recall from Section~\ref{sec:specialvalues} that $\mathcal{S}^{\rm sv}_{X,R}$ denotes the ring of regularised special values of single-valued hyperlogarithms over the ring $R$ with respect to the alphabet $X$.
\begin{thm}\phantomsection\label{ThmLabelArgumentSV}
For any $w\in X^*$, any $2\leq i\leq n$ and any $0\leq j\leq n$
\begin{equation}
\mathcal{L}_w(\sigma_j)=\sum_u c_u\mathcal{L}_u(\sigma_i),
\end{equation}
where $c_u\in\mathcal{S}^{\rm sv}_{X_i,\mathbb{Q}}$ and the sum runs over a finite number of words $u\in X_i^*$.
\end{thm}
\begin{proof} 
If $|w|=1$ we have $\mathcal{L}_{\sigma_k}(\sigma_j)=\log|1-\sigma_j/\sigma_k|^2$. Therefore if $k\neq i$ then $\mathcal{L}_{\sigma_k}(\sigma_j)\in \mathcal{S}^{\rm sv}_{X_i,\mathbb{Q}}$ and if $k=i$ then $\mathcal{L}_{\sigma_i}(\sigma_j)=\mathcal{L}_{\sigma_j}(\sigma_i)-\mathcal{L}_{0}(\sigma_i)$. In order to proceed by induction on $|w|$, we first need the following single-valued analogue of Lemma~\ref{CorHyper}:
\begin{lemma}\phantomsection\label{CrucialLemmaSV}
For any $w\in X^*$, any $z\in\mathbb{C}$ and $2\leq i\leq n$, we have
\begin{equation}
\frac{\partial}{\partial\sigma_i}\mathcal{L}_w(z)=\sum_{\substack{\tau\in X_i\cup \{z\}\\v\in X^*}}\frac{\lambda_{\tau ,v}}{\sigma_i-\tau}\mathcal{L}_v(z),
\end{equation}
where the sum is finite and all $\lambda_{\tau, v}\in\mathcal{S}^{\rm sv}_{X,\mathbb{Q}}$ have homogenous weight\footnote{The notion of weight was introduced in Section~\ref{sec:specialvalues}.} $W(\lambda_{\tau, v})=|w|-|v|-1$.
\end{lemma}
\begin{proof}[Proof of the lemma]
We will use induction on the length $|w|$. If $|w|=1$ the statement clearly holds. For $|w|\geq 2$ let 
\begin{equation}
\mathcal{L}_w(z)=\sum_{w_1,w_2}c_{w_1,w_2}L_{w_1}(z)\overline{L_{w_2}(z)}.
\end{equation}
By Proposition~\ref{LemmaCoeffSV}, $c_{w_1,w_2}\in \mathcal{S}^{\rm sv}_{X,\mathbb{Q}}$ and $W(c_{w_1,w_2})=|w|-|w_1|-|w_2|$. Since some $c_{w_1,w_2}$ may depend on $\sigma_i$ and $\frac{\partial}{\partial\sigma_i}\overline{L_w(z)}=0$ for all $w$, we have
\begin{equation}
\frac{\partial}{\partial\sigma_i}\mathcal{L}_w(z)=\sum_{w_1,w_2}\Big(\frac{\partial}{\partial\sigma_i}c_{w_1,w_2}\Big)L_{w_1}(z)\overline{L_{w_2}(z)}+\sum_{w_1,w_2}c_{w_1,w_2}\Big(\frac{\partial}{\partial\sigma_i}L_{w_1}(z)\Big)\overline{L_{w_2}(z)}.
\end{equation}
It is trivial to see from the definition of single-valued hyperlogarithms that $|w_1|+|w_2|>0$. Therefore we can use the inductive hypothesis on the special values of single-valued hyperlogarithms at points of $X$ which constitute the coefficients $c_{w_1,w_2}$, obtaining that
\begin{equation}
\frac{\partial}{\partial\sigma_i}c_{w_1,w_2}=\sum_{\tau\in X_i}\frac{\mu_{\tau,w_1,w_2}}{\sigma_i-\tau},
\end{equation}
where each $\mu_{\tau,w_1,w_2}\in \mathcal{S}^{\rm sv}_{X,\mathbb{Q}}$ has homogeneous weight given by $W(\mu_{\tau,w_1,w_2})=W(c_{w_1,w_2})-1$. 

Moreover, by Lemma~\ref{CorHyper} we have
\begin{equation}
\frac{\partial}{\partial\sigma_i}L_{w_1}(z)=\sum_{\substack{\tau\in X_i\cup \{z\}\\v\in X^*}}\frac{\lambda_{\tau ,v}}{\sigma_i-\tau}L_v(z)
\end{equation}
with $\lambda_{\tau, v}\in \mathbb{Z}$ and $|v|=|w|-1$. Hence we are left with 
\begin{equation}
\frac{\partial}{\partial\sigma_i}\mathcal{L}_w(z)=\sum_{\tau\in X_i\cup \{z\}}\frac{f_\tau(z)}{\sigma_i-\tau},
\end{equation}
where each $f_\tau(z)$ is a single-valued $\mathcal{S}^{\rm sv}_{X,\mathbb{Q}}$-linear combination of products $L_{v_1}(z)\overline{L_{v_2}(z)}$ of homogeneous weight $|w|-1$. By applying Proposition~\ref{CorollaryRefinement} to the ring $R=\mathcal{S}^{\rm sv}_{X,\mathbb{Q}}$, we conclude that each $f_\tau(z)$ is a $\mathcal{S}^{\rm sv}_{X,\mathbb{Q}}$-linear combination of single-valued hyperlogarithms $\sum_{v}\lambda_{\tau ,v}\mathcal{L}_v(z)$ and $W(\lambda_{\tau ,v})+|v|=|w|+1$.
\end{proof}
Let us now continue our inductive argument. By Lemma~\ref{CrucialLemmaSV} we have
\begin{equation}
\frac{\partial}{\partial\sigma_i}\mathcal{L}_w(\sigma_j)=\sum_{\substack{\tau\in X_i\\v\in X^*}}\frac{\lambda_{\tau ,v}}{\sigma_i-\tau}\mathcal{L}_v(\sigma_j),
\end{equation}
with $\lambda_{\tau, v}\in\mathcal{S}^{\rm sv}_{X,\mathbb{Q}}$ of  homogenous weight $W(\lambda_{\tau, v})=|w|-|v|-1$.
We can therefore apply the inductive hypothesis both on the coefficients $\lambda_{\tau, v}$ and on the hyperlogarithms $\mathcal{L}_v(\sigma_j)$, obtaining (after performing enough shuffle products)
\begin{equation}
\frac{\partial}{\partial\sigma_i}\mathcal{L}_w(\sigma_j)=\sum_{\substack{\tau\in X_i\\u\in X_i^*}}\frac{\mu_{\tau, u}}{\sigma_i-\tau}\mathcal{L}_u(\sigma_i)
\end{equation}
with $\mu_{\tau, u}\in \mathcal{S}^{\rm sv}_{X_i,\mathbb{Q}}$, which implies by Theorem~\ref{BrownsvHyperlog} that
\begin{equation}
\mathcal{L}_w(\sigma_j)=\sum_{\substack{\tau\in X_i\\u\in X_i^*}}\mu_{\tau, u}\mathcal{L}_{\tau u}(\sigma_i)+f(\overline{\sigma_i})
\end{equation}
for some function $f$ annihilated by $\partial /\partial \sigma_i$. Because of Proposition~\ref{PanzerThm}, the function $f(\overline{\sigma_i})$ must belong to the ring of anti-holomorphic hyperlogarithms $\overline{\mathcal{H}_{X,\mathbb{C}}}$. By Lemma~\ref{LemmaSV} we know that $f$ is a single-valued function of $\sigma_i$, but by Proposition~\ref{MonodromyThm} every non-constant element of $\overline{\mathcal{H}_{X,\mathbb{C}}}$ has non-trivial monodromy, hence $f$ must be constant. Since $|\tau u|\geq 1$ we have $\mathcal{L}_{\tau u}(0)=0$. Therefore we conclude that $f=\mathcal{L}_w(\sigma_j)\big|_{\sigma_i=0}$, which belongs to $\mathcal{S}^{\rm sv}_{X_i,\mathbb{Q}}$.
\end{proof}

\subsection{Integrals of single-valued hyperlogarithms over $\mathbb{P}^1_{\mathbb{C}}$}

Let us consider a function $f(z)$ belonging to the ring $\mathcal{A}^{\rm sv}_{X,\mathbb{C}}$ introduced in Section~\ref{sec:ringsv}. We define the \emph{holomorphic and anti-holomorphic residues} of $f$ at a point $\sigma_i\in X$ as $\textup{Res}_{z=\sigma_i}f(z):=c_{0,-1,0}^{(i)}$, $\overline{\textup{Res}}_{z=\sigma_i}f(z):=c_{0,0,-1}^{(i)}$, respectively, referring to the coefficients $c_{k,m,n}^{(i)}$ in the asymptotic expansions obtained from eqs.~(\ref{FinPtsExp}) and~(\ref{InfExp}) by multiplication with elements in $\mathcal{O}_{\Sigma,\mathbb{C}}\otimes \overline{\mathcal{O}_{\Sigma,\mathbb{C}}}$. Similarly, at $\infty$ we set $\textup{Res}_{z=\infty}f(z):=c_{0,-1,0}^{(\infty)}$ and $\overline{\textup{Res}}_{z=\infty}f(z):=c_{0,0,-1}^{(\infty)}$. The following residue formula was first stated by Schnetz in~\cite{Schnetz:2013hqa} for $X=\{0,1\}$. We include the proof, because the same argument will occur again in Section~\ref{Sec:Final}. 
\begin{propos} (Schnetz,~\cite{Schnetz:2013hqa})\phantomsection\label{ResidueThm}
Suppose that $f(z)\in\mathcal{A}^{\rm sv}_{X,\mathbb{C}}$ and that $\int_{\mathbb{P}^1_{\mathbb{C}}}f(z)d^2z<\infty$, where we set $d^2z:=dzd\overline{z}/(-2\pi i)$. Then
\begin{eqnarray}\label{resFormula}
\int_{\mathbb{P}^1_{\mathbb{C}}}f(z)d^2z&=&\textup{Res}_{z=\infty}G(z)-\sum_{i=0}^n\textup{Res}_{z=\sigma_i}G(z)\\
&=&\overline{\textup{Res}}_{z=\infty}F(z)-\sum_{i=0}^n\overline{\textup{Res}}_{z=\sigma_i}F(z)
\end{eqnarray}
for any $F,G\in\mathcal{A}^{\rm sv}_{X,\mathbb{C}}$ such that $\partial_zF(z)=\partial_{\overline{z}}G(z)=f(z)$.
\end{propos}
\begin{proof} 
First of all we remark that, as mentioned in Section~\ref{sec:ringsv}, the ring~$\mathcal{A}^{\rm sv}_{X,\mathbb{C}}$ is closed under the operation of taking primitives with respect to~$\partial_z$ and~$\partial_{\overline{z}}$, so we can always find such~$F$ and~$G$. Moreover,~$F$ is uniquely determined up to adding a function in~$\overline{\mathcal{O}_{\Sigma,\mathbb{C}}}$, as well as~$G$ up to adding a function in~$\mathcal{O}_{\Sigma,\mathbb{C}}$. We will only show the first equality, as the second follows by repeating exactly the same steps in the anti-holomorphic case. One has 
\[
f(z)d^2z=-\frac{f(z)}{2\pi i}dz\wedge d\overline{z}=d\bigg(\frac{G(z)}{2\pi i}\bigg)dz,
\]
therefore $f(z)d^2z$ is exact on $\mathbb{P}^1_{\mathbb{C}}\setminus (X\cup\{\infty\})$. Let $B_a(r)$ denote the ball centered in $a$ of radius $r$, $S^\pm_a(r)=\partial^\pm B_a(r)$ and $\varepsilon>0$ such that $B_\eta(\varepsilon)\cap B_\theta(\varepsilon)=\emptyset$ for all finite $\eta, \theta\in X$. Then $f(z)d^2z$ is exact on the oriented manifold $V_\varepsilon:=\mathbb{P}^1_{\mathbb{C}}\setminus (\bigcup_{\eta\in X}B_\eta(\varepsilon)\cup B_0(\varepsilon^{-1}))$ with boundaries $S^+_0(\varepsilon^{-1})$, $S^-_\eta(\varepsilon)$, and so by Stokes's theorem we have
\[
\int_{V_{\varepsilon}}f(z)d^2z=-\frac{1}{2\pi i}\bigg(\int_{S^+_0(\varepsilon^{-1})}+\sum_{\eta\in X}\int_{S^-_\eta(\varepsilon)}\bigg)G(z)dz.
\]
Parametrizing $S_\eta(\varepsilon)=\{z=\eta+\varepsilon e^{i\theta}\,:\,0\leq \theta < 2\pi \}$ and integrating term by term the expansion of the integrand given by eqs.~(\ref{FinPtsExp}) and~(\ref{InfExp}) we obtain the first formula in~(\ref{resFormula}). Moreover, this does not depend on the choice of~$G$, because the sum of the residues of a function belonging to~$\mathcal{O}_{\Sigma,\mathbb{C}}$ is zero.
\end{proof}

Proposition~\ref{ResidueThm} gives us a powerful instrument to compute integrals of single-valued hyperlogarithms over the Riemann sphere, and it is one of the ingredients of the proof of the following theorem, which is the main new result of this section.
\begin{thm}\label{LemmaAppD}
Let $X=\{0,1,\sigma_2,\ldots ,\sigma_N\}$, $X_N=X\setminus\{\sigma_N\}$, $m,n\leq N$ and $\{\sigma_{i_r}\}_{r=1}^m, \{\sigma_{j_s}\}_{s=1}^n\subset X$ (with possibly non-empty intersection). Let $f(z)=\sum_uc_u\mathcal{L}_u(z)$ be a finite linear combination of single-valued hyperlogarithms with coefficients $c_u\in\mathcal{S}^{\rm sv}_{X,\mathbb{Q}}$ such that the integral
\begin{equation}\label{applemma}
I:=\int_{\mathbb{P}^1_{\mathbb{C}}}\frac{f(z)\,d^2z}{\prod_{r=1}^m(z-\sigma_{i_r})\prod_{s=1}^n(\overline{z}-\overline{\sigma}_{j_s})}
\end{equation}
is absolutely convergent. Then there exists a finite linear combination $g(\sigma_N)=\sum_{v\in X_N^*}k_v\mathcal{L}_v(\sigma_N)$ with $k_v\in\mathcal{S}^{\rm sv}_{X_N,\mathbb{Q}}$ such that
\begin{equation}
I=\sum_{r=1}^m\sum_{s=1}^nh_r\overline{h}_sg(\sigma_N),
\end{equation}
where
\begin{equation}\label{hrhs}
h_r:=\prod_{\substack{k=1\\ k\neq r}}^m\frac{1}{\sigma_{i_r}-\sigma_{i_k}},\,\,\,\,\,\,\, \overline{h}_s:=\prod_{\substack{k=1\\ k\neq s}}^n\frac{1}{\overline{\sigma}_{j_s}-\overline{\sigma}_{j_k}}.
\end{equation}
\end{thm}
\begin{proof} 
First of all, we recall the partial-fraction identities
\begin{eqnarray}
\prod_{r=1}^m\frac{1}{z-\sigma_{i_r}}=\sum_{r=1}^m\frac{h_r}{z-\sigma_{i_r}},\label{uno!}\\
\prod_{s=1}^n\frac{1}{\overline{z}-\overline{\sigma}_{j_s}}=\sum_{s=1}^n\frac{\overline{h}_s}{\overline{z}-\overline{\sigma}_{j_s}},\label{due!}
\end{eqnarray}
with $h_r, \overline{h}_s$ as in eq.(~\ref{hrhs}). As mentioned in Section~\ref{sec:ringsv}, the ring $\mathcal{A}^{\rm sv}_{X,\mathbb{C}}$ is closed under the operation of taking primitives with respect to $\partial/\partial z$. In the case of the integrand of~(\ref{applemma}), it is easy to see from~(\ref{uno!}) and~(\ref{due!}) that a primitive in $\mathcal{A}^{\rm sv}_{X,\mathbb{C}}$ is given by
\begin{equation}
\sum_{r=1}^m\sum_{s=1}^nh_r\frac{\overline{h}_s}{\overline{z}-\overline{\sigma}_{j_s}}\sum_{u}c_u\mathcal{L}_{\sigma_{i_r}u}(z),
\end{equation}
and so by Proposition~\ref{ResidueThm} we have
\begin{equation}
I=-\sum_{r=1}^m\sum_{s=1}^nh_r\overline{h}_s\sum_{u}c_u\mathcal{L}_{\sigma_{i_r}u}(\sigma_{j_s}).
\end{equation}
Theorem~\ref{ThmLabelArgumentSV}, applied for each~$u$ both to $\mathcal{L}_{\sigma_{i_r}u}(\sigma_{j_s})$ and to its coefficient~$c_u$, concludes the proof.
\end{proof} 

It is possible with a little more work to obtain an effective version of this theorem, which could be implemented on a computer. More precisely, this would follow from an effective version of Theorem~\ref{ThmLabelArgumentSV}, and in particular of Lemma~\ref{CrucialLemmaSV}. Part of this work was recently done by Schnetz (see his Maple procedures~\cite{SchnetzProc}).

\section{The $\alpha'$-expansion of closed string amplitudes}\label{Sec:Final}

For $N\geq 1$ let~$\pmb s$ denote as usual the collection of Mandelstam kinematic invariants $s_{ij}:=\alpha'k_i\cdot k_j$ with $0\leq i<j\leq N+1$, as introduced in Section~\ref{sec:closedamp}. We define a family of integrals\footnote{The usual notation for this family in the literature is rather $J_{\rho,\sigma}(\pmb s)$.}
\begin{equation}\label{SupAmpJ}
M_{\rho,\sigma}(\pmb s)\,:=\,\int_{(\mathbb{P}^1_{\mathbb{C}})^N}\frac{\prod_{1\leq i<j\leq N}|z_j-z_i|^{2s_{ij}}\prod_{i=1}^{N}|z_i|^{2s_{0i}}|z_i-1|^{2s_{iN+1}}\,d^2z_i}{z_{\rho(1)}\,\overline{z}_{\sigma(1)}(1-z_{\rho(N)})(1-\overline{z}_{\sigma(N)})\prod_{i=2}^N(z_{\rho(i)}-z_{\rho(i-1)})(\overline{z}_{\sigma(i)}-\overline{z}_{\sigma(i-1)})},
\end{equation}
indexed by two permutations $\rho,\sigma\in\mathfrak S_{N}$. These integrals have non-empty convergent regions, which we denote by~$\mathcal{C}_{\rho,\sigma}$, and the origin $\pmb s=\textbf{0}$ always belongs to the boundary of~$\mathcal{C}_{\rho,\sigma}$ (see Appendix~\ref{Sec:ConvReg}). 

This is a special subfamily of the closed string partial amplitudes $M_{N+3}(\pmb s,\pmb n,\pmb {\tilde{n}})$ defined by~(\ref{e:IntGeneric}), obtained for very specific choices of the integer tuples $\pmb n,\pmb {\tilde{n}}$. It is known (see e.g.~\cite{Schlotterer:2012ny}), that all $M_{N+3}(\pmb s,\pmb n,\pmb {\tilde{n}})$ which arise from superstring theory can be written in terms of the functions $M_{\rho,\sigma}(\pmb s)$.

It is also known\footnote{See for instance~\cite{Brown:2019wna} for a more precise statement and a rigorous proof.} that, for each $\rho,\sigma$, there are sets $I$ of pairs of indices $(i,j)$ such that the product
\begin{equation}\label{eqalphexpdef}
\prod_{I}\bigg(\sum_{(i,j)\in I}s_{ij}\bigg)\,\times\,M_{\rho,\sigma}(\pmb s)
\end{equation}
has a Taylor series expansion at $\pmb s=0$, which is the \emph{$\alpha'$-expansion} of the closed string integrals $M_{\rho,\sigma}(\pmb s)$. 
 
This section is devoted to prove the following statement, which was conjectured in~\cite{Stieberger:2013wea}\footnote{Other two different proofs of this result, obtained independently, have appeared in~\cite{Schlotterer:2018zce, Brown:2019wna} while we were writing this paper.}.
\begin{thm}\label{MainThmSec7}
For any $N\geq 1$ and any $\rho,\sigma\in\mathfrak{S}_N$ the coefficients of the $\alpha'$-expansion of $M_{\rho,\sigma}(\pmb s)$ belong to $\mathcal{Z}^{\rm sv}$.
\end{thm}

The method of the proof is constructive, it ultimately relies on the machinery developed in Section~\ref{sec:inthyp}, and it allows in principle to determine also the precise form of the first factor of~\eqref{eqalphexpdef}. We begin by proving the statement in two special cases. First we will look at the only 1-dimensional integral of the family, which corresponds to the classical Virasoro amplitude, and which is also the only closed string integral for which the statement was already proven (see Section~\ref{sec:svM4}). In particular, we will explain in details how to isolate and calculate, by analytic considerations, the polar contributions, and we will obtain the explicit asymptotic behaviour at $\pmb s=0$. Then we will look at a 2-dimensional integral, where both the analysis of the polar contributions and that of the Taylor coefficients become more subtle than in dimension one, and resemble those of the general case. Finally, we will prove the general case, but we will skip details, especially on the analysis of the polar contributions, in order to avoid a lenghty discussion of different cases involving very complicated notations; we believe that the familiarity with our methods aquired from the previous special cases should suffice to understand how to work out the missing steps out.

\subsection{Dimension one}\label{Sectionk=1}

Let $N=1$. The symmetric group $\mathfrak{S}_1$ only contains the identity, and so the only 1-dimensional integral in our family is
\begin{equation}\label{4closedStrings}
M_{\text{Id},\text{Id}}(s_{01},s_{12})=\int_{\mathbb{P}^1_{\mathbb{C}}}|z|^{2s_{01}-2}|1-z|^{2s_{02}-2}d^2z.
\end{equation}
This complex analogue of Euler's beta function is essentially the Virasoro bosonic string amplitude~\cite{Virasoro:1969me}, and it is a special case of the integrals $M_4(\pmb s,\pmb n,\pmb{\tilde n})$ from Section~\ref{sec:svM4}, obtained by setting $n_{01}=\tilde n_{01}=n_{12}=\tilde n_{12}=-1$. 

The region of absolute convergence is $\textup{Re}(s_{01})>0$, $\textup{Re}(s_{12})>0$ and $\textup{Re}(s_{01}+s_{12})<1$. In particular, the point $s_{01}=s_{12}=0$ lies at the boundary of this region. It is well known, and it also directly follows from the computations of Section~\ref{sec:svM4}, that 
\begin{align}\label{expansioncomplexbeta}
M_{\text{Id},\text{Id}}(s_{01},s_{12})&=\frac{s_{01}+s_{12}}{s_{01}s_{12}}\,\frac{\Gamma(1+s_{01})\,\Gamma(1+s_{12})\,\Gamma(1-s_{01}-s_{12})}{\Gamma(1-s_{01})\,\Gamma(1-s_{12})\,\Gamma(1+s_{01}+s_{12})} \notag\\
&=\frac{s_{01}+s_{12}}{s_{01}s_{12}}\,\exp\bigg(-\sum_{n\geq 1}\frac{2\zeta(2n+1)}{(2n+1)}\,\Big(s_{01}^{2n+1}+s_{12}^{2n+1}-(s_{01}+s_{12})^{2n+1}\Big)\bigg). 
\end{align}
Recall from Section~\ref{Sec:Hyperlogs} that $2\zeta(2n+1)=\zeta^{\rm sv}(2n+1)=\mathcal{L}_{0^{2n}1}(1)$, and therefore the coefficients of this series expansion belong to $\mathcal{Z}^{\rm sv}$. This gives a proof of the following special case of Theorem~\ref{MainThmSec7}:
\begin{propos}
The product $s_{01}s_{12}M_{\text{Id},\text{Id}}(s_{01},s_{12})$ is holomorphic at $s_{01}=s_{12}=0$, and its Taylor coefficients belong to $\mathcal{Z}^{\rm sv}$.
\end{propos}
However, as already mentioned, we find it very instructive to give an alternative proof, which will be later generalised to demonstrate Theorem~\ref{MainThmSec7}.
\begin{proof}[Alternative proof] 
We write
\begin{equation}
M_{\text{Id},\text{Id}}(s_{01},s_{12})\,=\,\int_{\mathbb{P}^1_{\mathbb{C}}}\frac{(|z|^{2s_{01}}-1)(|1-z|^{2s_{02}}-1)}{|z|^2|1-z|^2}d^2z\,+\,\int_{\mathbb{P}^1_{\mathbb{C}}}\frac{|z|^{2s_{01}}+|1-z|^{2s_{02}}-1}{|z|^2|1-z|^2}d^2z\,,
\end{equation}
and we first look at the first term of the right hand side. This term is absolutely convergent at $s_{01}=s_{12}=0$ and thereby has a Taylor expansion, given by Taylor expanding the integrand and then interchanging summation and integration (this is justified by the absolute convergence). We get
\begin{equation}
\int_{\mathbb{P}^1_{\mathbb{C}}}\frac{(|z|^{2s_{01}}-1)(|1-z|^{2s_{02}}-1)}{|z|^2|1-z|^2}d^2z\,=\,\sum_{p,q\geq 1}s_{01}^ps_{12}^q\int_{\mathbb{P}^1_{\mathbb{C}}}\frac{\mathcal{L}_{0^p}(z)\mathcal{L}_{1^q}(z)}{|z|^2|1-z|^2}d^2z\,,
\end{equation}
where we recall that $\mathcal{L}_{0^p}(z)=(\log|z|^2)^p/p!$ and $\mathcal{L}_{1^q}(z)=(\log|1-z|^2)^q/q!$.

Since single-valued hyperlogarithms satisfy shuffle product identities, we have
\begin{equation}
\frac{\mathcal{L}_{0^p}(z)\mathcal{L}_{1^q}(z)}{|z|^2|1-z|^2}=\sum_{w=0^p\shuffle 1^q}\frac{\mathcal{L}_w(z)}{|z|^2|1-z|^2}=\sum_{w=0^p\shuffle 1^q}\frac{1}{\overline{z}(1-\overline{z})}\bigg(\frac{\mathcal{L}_w(z)}{z}-\frac{\mathcal{L}_w(z)}{z-1}\bigg),
\end{equation}
hence if we set
\begin{equation}
F(z):=\sum_{w=0^p\shuffle 1^q}\frac{\mathcal{L}_{0w}(z)-\mathcal{L}_{1w}(z)}{\overline{z}(1-\overline{z})}
\end{equation}
we find that
\begin{equation}\label{aaaaa}
\partial_z F(z)=\frac{\mathcal{L}_{0^p}(z)\mathcal{L}_{1^q}(z)}{|z|^2|1-z|^2}.
\end{equation}
Therefore by Proposition~\ref{ResidueThm} and by a careful analysis of the asymptotic expansion of~$F$ we obtain
\begin{align}
\int_{\mathbb{P}^1_{\mathbb{C}}}\frac{\mathcal{L}_{0^p}(z)\mathcal{L}_{1^q}(z)}{|z|^2|1-z|^2}d^2z&=\overline{\textup{Res}}_{z=\infty}F(z)-\overline{\textup{Res}}_{z=0}F(z)-\overline{\textup{Res}}_{z=1}F(z) \notag \\
&=\sum_{w=0^p\shuffle 1^q}\mathcal{L}_{0w}(1)-\mathcal{L}_{1w}(1), \notag
\end{align}
which belongs to the algebra~$\mathcal{Z}^{\rm sv}$. This proves the statement for the first term.

We consider now the second term. For $\varepsilon\in \mathbb{R}_+$ we define $U_\varepsilon:=\mathbb{P}^1_{\mathbb{C}}\setminus(B_0(\varepsilon)\cup B_1(\varepsilon)\cup B_0(\varepsilon^{-1}))$, where we denote by $B_x(r)$ the ball centered at~$x$ of radius~$r$. On the one hand, the integrand is an entire function in~$s_{01}$ and~$s_{12}$ for any~$z\in U_\varepsilon$, so it admits a Taylor expansion at the origin. Moreover, its integral over~$U_\varepsilon$ is absolutely convergent for any~$s_{01}$ and~$s_{12}$ and also defines an entire function of~$s_{01}$ and~$s_{12}$, whose Taylor expansion at the origin is simply given, because of absolute convergence, by interchanging summation and integration:
\begin{equation}\label{brbrbr}
\int_{U_\varepsilon}\frac{|z|^{2s_{01}}+|1-z|^{2s_{02}}-1}{|z|^2|1-z|^2}d^2z\,=\,\sum_{\substack{p,q\geq 0\\p\cdot q =0}}s_{01}^ps_{12}^q\int_{U_\varepsilon}\frac{\mathcal{L}_{0^p}(z)\mathcal{L}_{1^q}(z)}{|z|^2|1-z|^2}d^2z.
\end{equation}

On the other hand, for any $s_{01}$ and $s_{12}$ inside the region of absolute convergent we have 
\begin{equation}
\int_{\mathbb{P}^1_{\mathbb{C}}}\frac{|z|^{2s_{01}}+|1-z|^{2s_{02}}-1}{|z|^2|1-z|^2}d^2z\,=\,\lim_{\varepsilon\rightarrow 0}\int_{U_\varepsilon}\frac{|z|^{2s_{01}}+|1-z|^{2s_{02}}-1}{|z|^2|1-z|^2}d^2z,
\end{equation}
hence we find that, in this region,
\begin{equation}\label{LimitBeta}
\int_{\mathbb{P}^1_{\mathbb{C}}}\frac{|z|^{2s_{01}}+|1-z|^{2s_{02}}-1}{|z|^2|1-z|^2}d^2z=\lim_{\varepsilon\rightarrow 0}\bigg(\sum_{\substack{p,q\geq 0\\p\cdot q=0}}s_{01}^ps_{12}^q\int_{U_\varepsilon}\frac{\mathcal{L}_{0^p}(z)\mathcal{L}_{1^q}(z)}{|z|^2|1-z|^2}d^2z\bigg).
\end{equation}

By the Stokes theorem and by eq.~\eqref{aaaaa} we can write the right hand side of eq.~\eqref{brbrbr} as
\begin{equation}
\sum_{\substack{p,q\geq 0\\p\cdot q=0}}s_{01}^ps_{12}^q\Big(\int_{\partial^+B_0(\varepsilon^{-1})}+\int_{\partial^-B_0(\varepsilon)}+\int_{\partial^-B_1(\varepsilon)}\Big)\sum_{w=0^p\shuffle 1^q}\frac{\mathcal{L}_{0w}(z)-\mathcal{L}_{1w}(z)}{\overline{z}(1-\overline{z})}\,\frac{id\overline{z}}{2\pi}.
\end{equation}
Analysing separately the contributions over the three boundary components, and using polar coordinates (see the proof of Proposition~\ref{ResidueThm}), we find that the contribution from $\partial^-B_0(\varepsilon)$ is
\begin{equation}
  -\sum_{p\geq0}
   \left(\frac{(\log\varepsilon^2)^{p+1}}{(p+1)!}\right)\,s_{01}^p\,+\,O(\varepsilon)\, =\,\frac{1}{s_{01}}(1-\varepsilon^{2s_{01}})\,+\,O(\varepsilon),
\end{equation}
the contribution from $\partial^+B_0(\varepsilon^{-1})$ is $O(\varepsilon)$ and the contribution from $\partial^-B_1(\varepsilon)$ is
\begin{align}
  &\sum_{q\geq 1}\mathcal{L}_{01^q}(1)\,s_{12}^q-\sum_{q\geq0}
   \left(\frac{(\log\varepsilon^2)^{q+1}}{(q+1)!}\right)s_{12}^q-\sum_{p\geq 1}\mathcal{L}_{10^p}(1)\,s_{01}^p + O(\varepsilon) \\
   &=\,\sum_{q\geq 1}\mathcal{L}_{01^q}(1)\,s_{12}^q+\frac{1}{s_{12}}(1-\varepsilon^{2s_{12}})-\sum_{p\geq 1}\mathcal{L}_{10^p}(1)\,s_{01}^p + O(\varepsilon).\notag
\end{align}

Taking the limit $\varepsilon\rightarrow 0$ and noting\footnote{Recall that $L_{0^n1}(1)=-\zeta(n+1)$. The first identity follows from the fact that $L_{01^n}(1)=(-1)^{n+1}L_{0^n1}(1)$, which can be seen by substituting $x_i\rightarrow 1-x_i$ in the integrand of $L_{01^n}(z)$, and from the fact that $\zeta^{\rm sv}(2n)=0$. The second identity follows from the more general formula given in~\cite[Example 2.10]{Schnetz:2013hqa}.} that $\mathcal{L}_{01^n}(1)=\zeta^{\rm sv}(n+1)=-\mathcal{L}_{10^n}(1)$ for $n\geq 1$, we conclude that the product $s_{01}s_{12}M_{\text{Id},\text{Id}}(s_{01},s_{12})$ is holomorphic at $s_{01}=s_{12}=0$ and the coefficients of its Taylor expansion, explicitly given by
\begin{equation}
s_{01}+s_{12}\,+\,2\sum_{n\geq 1}\zeta(2n+1)(s_{01}^{2n+1}s_{12}+s_{01}s_{12}^{2n+1})+\sum_{p,q\geq 1}\bigg(\sum_{w=0^p\shuffle 1^q}\mathcal{L}_{0w}(1)-\mathcal{L}_{1w}(1)\bigg)\,s_{01}^{p+1}s_{12}^{q+1},
\end{equation}
belong to the ring $\mathcal{Z}^{\rm sv}$.
\end{proof}
We remark that our proof implies that there exists a rational function, in this case $s_{01}^{-1}+s_{12}^{-1}$, which can be subtracted to $M_{\text{Id},\text{Id}}(s_{01},s_{12})$ to obtain a function which is holomorphic at $s_{01}=s_{02}=0$. This is stronger than the statement of the proposition\footnote{For instance, the case where $s_{01}s_{02}M_{\text{Id},\text{Id}}(s_{01},s_{12})$ only depends on $s_{01}$ and is not rational would not satisfy this stronger condition.}, and it should be possible to prove this stronger kind of statement also in the general case, by refining our proof of Theorem~\ref{MainThmSec7}.

\subsection{Dimension two}\label{Sectionk=2}

Let $N=2$. We have four 2-dimensional integrals $M_{\rho,\sigma}(\pmb s)$ with $\rho,\sigma\in\mathfrak{S}_2$. We content ourselves to study their sum
\begin{equation}\label{FivePts}
\sum_{\rho,\sigma\in\mathfrak{S}_2}M_{\rho,\sigma}(\alpha_1,\alpha_2,\beta_1,\beta_2,\gamma)\,=\,\int_{(\mathbb{P}^1_{\mathbb{C}})^2}|z|^{2\alpha_1-2}|u|^{2\alpha_2-2}|1-z|^{2\beta_1-2}|1-u|^{2\beta_2-2}|z-u|^{2\gamma}d^2zd^2u,
\end{equation}
where we have set $\alpha_i:=s_{0,i}, \beta_i:=s_{i,3}$ and $\gamma:=s_{12}$ to lighten the notation. This is a good prototype of all of the 2-dimensional integrals, whose singularities are slightly simpler.
\begin{propos}
The product 
\begin{equation}
\alpha_1\,\alpha_2\,\beta_2\,\beta_2\,(\alpha_1+\alpha_2+\gamma)\,(\beta_1+\beta_2+\gamma)\,\sum_{\rho,\sigma\in\mathfrak{S}_2}M_{\rho,\sigma}(\alpha_1,\alpha_2,\beta_1,\beta_2,\gamma)
\end{equation}
is holomorphic at $\alpha_1=\alpha_2=\beta_1=\beta_2=\gamma=0$, and its Taylor coefficients belong to $\mathcal{Z}^{\rm sv}$.
\end{propos}
\begin{proof}
We want to study the asymptotic behaviour at the origin, which is situated at the boundary of the region of convergence (see Appendix~\ref{Sec:ConvReg}). Similarly to the 1-dimensional case, we first consider the integral
\begin{equation}\label{bbbbbb}
\int_{(\mathbb{P}^1_{\mathbb{C}})^2}\frac{(|z|^{2\alpha_1}-1)(|u|^{2\alpha_2}-1)(|1-z|^{2\beta_1}-1)(|1-u|^{2\beta_2}-1)|z-u|^{2\gamma}}{|z|^2|u|^2|1-z|^2|1-u|^2}d^2zd^2u,
\end{equation}
which is absolutely convergent in a strictly bigger region of convergence which contains the origin, and then we will take care of the difference between~(\ref{FivePts}) and~\eqref{bbbbbb}, which diverges at the origin.

Writing $\log|z-u|^2=\mathcal{L}_z(u)-\mathcal{L}_0(z)$ and using absolute convergence to interchange summation and integration, we can rewrite~\eqref{bbbbbb} as
\begin{align}
&\sum_{\substack{p_1,p_2,q_1,q_2\geq 1 \\r_1,r_2\geq 0}}\alpha_1^{p_1}\alpha_2^{p_2}\beta_1^{q_1}\beta_2^{q_2}\gamma^{r_1+r_2}(-1)^{r_2}\binom{r_1+r_2}{r_2}\times\notag \\
&\times\int_{(\mathbb{P}^1_{\mathbb{C}})^2}\frac{\mathcal{L}_{0^{p_1+r_2}}(z)\mathcal{L}_{0^{p_2}}(u)\mathcal{L}_{1^{q_1}}(z)\mathcal{L}_{1^{q_2}}(u)\mathcal{L}_{z^{r_1}}(u)}{|z|^2|1-z|^2|u|^2|1-u|^2}d^2zd^2u.
\end{align}
We first compute the integral over $u$, given by
\begin{equation}\label{jpqr}
J_{p_2,q_2,r_1}(z):=\int_{\mathbb{P}^1_{\mathbb{C}}}\frac{\mathcal{L}_{0^{p_2}}(u)\mathcal{L}_{1^{q_2}}(u)\mathcal{L}_{z^{r_1}}(u)}{|u|^2|1-u|^2}d^2u.
\end{equation}
By Proposition~\ref{ResidueThm} we can write $J_{p_2,q_2,r_1}(z)$ as
\begin{equation}\label{jpqr2}
\overline{\textup{Res}}_{u=\infty}F(z,u)-\overline{\textup{Res}}_{u=0}F(z,u)-\overline{\textup{Res}}_{u=1}F(z,u)-\overline{\textup{Res}}_{u=z}F(z,u),
\end{equation}
where
\begin{equation}
F(z,u):=\sum_{w=0^{p_2}\shuffle 1^{q_2}\shuffle z^{r_1}}\big(\mathcal{L}_{1w}(u)-\mathcal{L}_{0w}(u)\big)\bigg(\frac{1}{\overline{u}-1}-\frac{1}{\overline{u}}\bigg)
\end{equation}
is such that $\partial_u F(z,u)$ is equal to the integrand of~(\ref{jpqr}).
The only non-vanishing residues in~(\ref{jpqr2}) are obtained at $u=1$, thus
\begin{equation}
J_{p_2,q_2,r_1}(z)=\sum_{w=0^{p_2}\shuffle 1^{q_2}\shuffle z^{r_1}}\big(\mathcal{L}_{0w}(1)-\mathcal{L}_{1w}(1)\big).
\end{equation}
By Theorem~\ref{ThmLabelArgumentSV}, this is a $\mathcal{Z}^{\rm sv}$-linear combination of single-valued multiple polylogarithms in the variable $z$. Therefore, the same argument used in 1-dimensional case\footnote{We remark that, since $p_2,q_2\geq 1$, both $\lim_{z\rightarrow 0}J_{p_2,q_2,r_1}(z)$ and $\lim_{z\rightarrow 1}J_{p_2,q_2,r_1}(z)$ are finite, which serves as a double-check that the integral~\eqref{intjjj} is absolutely convergent.} applies to
\begin{equation}\label{intjjj}
\int_{\mathbb{P}^1_{\mathbb{C}}}\frac{J_{p_2,q_2,r_1}(z)\mathcal{L}_{0^{p_1+r_2}}(z)\mathcal{L}_{1^{q_1}}(z)}{|z|^2|1-z|^2}d^2z,
\end{equation}
thus proving the desired statement for the integral~\eqref{bbbbbb}.

We are now left with computing the asymptotic expansion of the  difference between~(\ref{FivePts}) and~\eqref{bbbbbb}, that we will call $P_2(\alpha_1,\alpha_2,\beta_1,\beta_2,\gamma)$. Let $\varepsilon\in\mathbb{R}^+$, and let
\begin{equation}
U_{\varepsilon,1}=\{z,u\in\mathbb{P}^1_{\mathbb{C}}:|z|,|1-z|,|u|,|1-u|,|z|-|u|>\varepsilon,\,|z|,|u|<\varepsilon^{-1}\},
\end{equation}
\begin{equation}
U_{\varepsilon,2}=\{z,u\in\mathbb{P}^1_{\mathbb{C}}:|z|,|1-z|,|u|,|1-u|,|u|-|z|>\varepsilon,\,|z|,|u|<\varepsilon^{-1}\}.
\end{equation}
The same argument used for the 1-dimensional case allows us to write $(\ref{FivePts})-\eqref{bbbbbb}$, inside the region of convergence of~(\ref{FivePts}), as
\begin{align}
&P_2(\alpha_1,\alpha_2,\beta_1,\beta_2,\gamma)=\lim_{\varepsilon\rightarrow 0}\sum_{\substack{p_1,p_2,q_1,q_2,r\geq 0\\ p_1\cdot p_2\cdot q_1\cdot q_2=0}}\alpha_1^{p_1}\alpha_2^{p_2}\beta_1^{q_1}\beta_2^{q_2}\frac{\gamma^r}{r!}\times \notag\\
&\times\bigg(\int_{U_{\varepsilon,1}}+\int_{U_{\varepsilon,2}}\bigg)\frac{\mathcal{L}_{0^{p_1}}(z)\mathcal{L}_{0^{p_2}}(u)\mathcal{L}_{1^{q_1}}(z)\mathcal{L}_{1^{q_2}}(u)\big(\log|z-u|^2\big)^r}{|z|^2|1-z|^2|u|^2|1-u|^2}d^2zd^2u.
\end{align}
Writing $\log|z-u|^2=\mathcal{L}_z(u)-\mathcal{L}_0(z)$ on $U_{\varepsilon,1}$ and $\log|z-u|^2=\mathcal{L}_u(z)-\mathcal{L}_0(u)$ on $U_{\varepsilon,2}$ we can rewrite $P_2(\alpha_1,\alpha_2,\beta_1,\beta_2,\gamma)$ as
\begin{align}\label{LimitBeta2}
&\lim_{\varepsilon\rightarrow 0}\sum_{\substack{p_1,p_2,q_1,q_2,r_1,r_2\geq 0\\ p_1\cdot p_2\cdot q_1\cdot q_2=0}}\alpha_1^{p_1}\alpha_2^{p_2}\beta_1^{q_1}\beta_2^{q_2}\gamma^{r_1+r_2}(-1)^{r_2}\binom{r_1+r_2}{r_2}\times\notag \\
&\times\bigg(\int_{U_{\varepsilon,1}}\frac{\mathcal{L}_{0^{p_1+r_2}}(z)\mathcal{L}_{0^{p_2}}(u)\mathcal{L}_{1^{q_1}}(z)\mathcal{L}_{1^{q_2}}(u)\mathcal{L}_{z^{r_1}}(u)}{|z|^2|1-z|^2|u|^2|1-u|^2}d^2zd^2u \notag\\
&+\int_{U_{\varepsilon,2}}\frac{\mathcal{L}_{0^{p_1}}(z)\mathcal{L}_{0^{p_2+r_2}}(u)\mathcal{L}_{1^{q_1}}(z)\mathcal{L}_{1^{q_2}}(u)\mathcal{L}_{u^{r_1}}(z)}{|z|^2|1-z|^2|u|^2|1-u|^2}d^2zd^2u\bigg).
\end{align}
We now distinguish between different cases.

\textbf{The case $q_1=q_2=0$}. First of all, this case is identical to the case $p_1=p_2=0$, which is therefore omitted. We consider the integrals over $U_{\varepsilon,1}$ and $U_{\varepsilon,2}$ separately. Since the situation is symmetric, we just focus on $\int_{U_{\varepsilon,1}}$. This integral (as well as $\int_{U_{\varepsilon,2}}$) is not convergent because of the singularity of the integrand at the origin. Let us consider the change of variables $t=z$, $st=u$, which means that $s=u/z$ and that $d^2z\,d^2u=|t|^2\,d^2t\,d^2s$. Moreover, let us denote by $\tilde{U}_{\varepsilon,1}$ the image of $U_{\varepsilon,1}$ under the change of coordinates. By deforming the shape of $U_{\varepsilon,1}$, we can suppose that $\tilde{U}_{\varepsilon,1}$ is obtained by removing a neighborhood of the origin which is a sphere of radius $\varepsilon$. What we need to compute is
\begin{align}
&\sum_{p_1,p_2,r\geq 0}\frac{\alpha_1^{p_1}\alpha_2^{p_2}\gamma^r}{p_1!\,p_2!\,r!}\int_{\tilde{U}_{\varepsilon,1}}\frac{(\log|t|^2)^{p_1}(\log|s|^2+\log|t|^2)^{p_2}(\log|1-s|^2+\log|t|^2)^r}{|t|^2|s|^2|1-t|^2|1-ts|^2} \notag\\
&=\sum_{p_1,p_2,r\geq 0}\frac{\alpha_1^{p_1}\alpha_2^{p_2}\gamma^r}{p_1!}\sum_{\substack{i+j=p_2\\ k+l=r}}\frac{1}{i!\,j!\,k!\,l!}\int_{\tilde{U}_{\varepsilon,1}}\frac{(\log|t|^2)^{p_1+i+k}(\log|s|^2)^j(\log|1-s|^2)^l}{|t|^2|s|^2|1-t|^2|1-ts|^2} \notag\\
&=\sum_{n,j,l\geq 0}(\alpha_1+\alpha_2+\gamma)^n\alpha_2^j\gamma^l\int_{\tilde{U}_{\varepsilon,1}}\frac{\mathcal{L}_{0^n}(t)\mathcal{L}_{0^j}(s)\mathcal{L}_{1^l}(s)}{|t|^2|s|^2|1-t|^2|1-ts|^2}.
\end{align}
In the last integral we can separate the variables and use the same method seen in the 1-dimensional case to obtain the polar part in the limit $\varepsilon\rightarrow 0$, producing the quadratic-denominator term $\alpha_2^{-1}(\alpha_1+\alpha_2+\gamma)^{-1}$ as well as other polar contributions with a linear denominator given either by~$\alpha_2$ or by $(\alpha_1+\alpha_2+\gamma)$ (note that the coefficients of the linear-denominator contributions are not rationals anymore, but they still belong to~$\mathcal{Z}^{\rm sv}$). It is at this point an easy exercise to see that the remaining contributions around $u=1$ or $z=1$ give rise to power series with coefficients in~$\mathcal{Z}^{\rm sv}$.

\textbf{The case $q_1=p_2=0$}. Again, this case is identical to the case $q_2=p_1=0$, which is therefore omitted. Moreover, this case is simpler than the previous one, because the only problem occurs at $(z,u)=(0,1)$. One can therefore simply consider the integral over the union of~$U_{\varepsilon,1}$ and~$U_{\varepsilon,2}$ and split it locally into a product of integrals like those considered in the 1-dimensional case, obtaining the quadratic-denominator polar contribution $\alpha_1^{-1}\beta_2^{-1}$ as well as linear-denominator polar  contributions (with denominator~$\alpha_1$ or~$\beta_2$) and power series contributions with coefficients in~$\mathcal{Z}^{\rm sv}$.

\textbf{The remaining cases}. The other possible cases with just one~$p_i$ or~$q_i$ vanishing are analogous but simpler; they give linear-denominator polar contributions as well as power series contributions, both with coefficients in $\mathcal{Z}^{\rm sv}$, and the denominators are always given by one of the factors of the quadratic contributions. 

\textbf{Summary}. The product of $P_2(\alpha_1,\alpha_2,\beta_1,\beta_2,\gamma)$ by the expression $\alpha_1\alpha_2\beta_2\beta_2(\alpha_1+\alpha_2+\gamma)(\beta_1+\beta_2+\gamma)$ is a function which is holomorphic at the origin, whose Taylor coefficients belong to~$\mathcal{Z}^{\rm sv}$. By the previous analysis of the Taylor expansion of~\eqref{bbbbbb}, the same is true also for~\eqref{FivePts}.
\end{proof}

\subsection{The general case}\label{SectionGen}

\begin{proof}[Sketch of the proof of Theorem~\ref{MainThmSec7}] We need to prove that, for any $N\geq 1$ and any $\rho, \sigma\in\mathfrak{S}_N$, there are sets $I$ of distinct pairs of indices $(i,j)$ with $0\leq i<j\leq N+1$ such that the product
\begin{equation}
\prod_I\bigg(\sum_{(i,j)\in I}s_{ij}\bigg)\times M_{\rho,\sigma}(\pmb s)
\end{equation}
is holomorphic at $\pmb s=0$, and its Taylor coefficients belong to $\mathcal{Z}^{\rm sv}$.

The main challenge of generalising the constructive analysis of the polar contributions seen in the 2-dimensional case, is to find an acceptable notation. Otherwise, it is just a lengthy distinction of different cases, each of which can be proven exactly as in the $N=2$ case to be given, close to the origin, by a quotient between a holomorphic function with~$\mathcal{Z}^{\rm sv}$-coefficients, and a product of at most~$N$ linear terms of the kind $\sum_I s_{ij}$. The relation between the combinatorics of these denominators, and that of the singularity of the integrand, is nicely described in~\cite{Brown:2019wna}, and it can be explicitly worked out also from our method. For this reason, we only focus on the integrals
\begin{equation}\label{SupAmpJtilde}
\int_{(\mathbb{P}^1_{\mathbb{C}})^N}\frac{\prod_{1\leq i<j\leq N}\big(|z_j-z_i|^{2s_{ij}}-1\big)\prod_{i=1}^{N}\big(|z_i|^{2s_{0i}}-1\big)\big(|z_i-1|^{2s_{iN+1}}-1\big)\,d^2z_i}{z_{\rho(1)}\,\overline{z}_{\sigma(1)}(1-z_{\rho(N)})(1-\overline{z}_{\sigma(N)})\prod_{i=2}^k(z_{\rho(i)}-z_{\rho(i-1)})(\overline{z}_{\sigma(i)}-\overline{z}_{\sigma(i-1)})},
\end{equation}
which are absolutely convergent and holomorphic at the origin, and prove that their Taylor coefficients belong to $\mathcal{Z}^{\rm sv}$ (proving the same statement for the ``polar terms'' is simpler). More precisely, after Taylor-expanding the integrand and interchanging summation with integration, we want to demonstrate that for $m_i, n_i, l_{i,j}\geq 1$ and for arbitrary $\rho,\sigma\in\mathfrak{S}_N$ the (absolutely convergent) integrals
\begin{equation}\label{Last}
\int_{(\mathbb{P}^1_{\mathbb{C}})^N}\frac{\prod_{1\leq i<j\leq N}\mathcal{L}_{z_i^{l_{i,j}}}(z_j)\prod_{i=1}^N\mathcal{L}_{0^{m_i}}(z_i)\mathcal{L}_{1^{n_i}}(z_i)\,d^2z_i}{z_{\rho(1)}\,\overline{z}_{\sigma(1)}(1-z_{\rho(k)})(1-\overline{z}_{\sigma(k)})\prod_{i=2}^N(z_{\rho(i)}-z_{\rho(i-1)})(\overline{z}_{\sigma(i)}-\overline{z}_{\sigma(i-1)})}
\end{equation}
belong to $\mathcal{Z}^{\rm sv}$.

To do this, the idea is to iteratively integrate~(\ref{Last}) one variable at a time, and use Theorem~\ref{LemmaAppD} at each step. This can be done, because of the following remarks:
\begin{itemize}
\item[(i)] The special cyclic structure of the denominator of~(\ref{Last}) implies that after each integration, even though $h_r$ and $\overline{h}_s$ from the statement of Theorem~\ref{LemmaAppD} introduce new factors, we always get a denominator which is the product of distinct \emph{linear} holomorphic factors with distinct \emph{linear} anti-holomorphic factors, as required by the assumptions of Theorem~\ref{LemmaAppD} (we leave this as an exercise to the reader).
\item[(ii)] The fact that $m_i, n_i, l_{i,j}$ are bigger than~$1$, together with the previous remark, implies that at each step we get a numerator~$f(z)$ of the integrand such that the integral converges absolutely\footnote{This and the previous remark can be seen also as a double-check that the integrals~(\ref{Last}) are indeed absolutely convergent: suppose for instance that at some integration step we could get a factor $(z_i-z_j)^2(\overline{z}_i-\overline{z}_j)$, then the next integral in $z_i$ would be divergent, and therefore~(\ref{Last}) must be divergent too.}, as required by the assumptions of Theorem~\ref{LemmaAppD}. 
\end{itemize}
It is clear that, after performing~$N$ integrations using Theorem~\ref{LemmaAppD}, we land on a number belonging to $\mathcal{Z}^{\rm sv}=\mathcal{S}^{\rm sv}_{\{0,1\},\mathbb{Q}}$, as claimed.
\end{proof}

\section*{Acknowledgments}
We would like to thank Francis Brown, Hao Fu, Hubert Saleur, Oliver
Schnetz for discussions. We are especially grateful to Cl\'ement Dupont, Erik Panzer and Oliver
Schlotterer for useful comments and remarks on this work.
  The research of P. Vanhove has received funding from the ANR grant
  ``Amplitudes'' ANR-17- CE31-0001-01,  the ANR grant ``SMAGP''
ANR-20-CE40-0026-01, and is partially supported by
  Laboratory of Mirror Symmetry NRU HSE, RF Government grant,
  ag. N$^\circ$ 14.641.31.0001. The research of F. Zerbini was supported by a French public grant as part of the Investissement d'avenir project,
reference ANR-11-LABX-0056-LMH, LabEx LMH, by the People Programme (Marie Curie Actions) of the European Union's Seventh Framework Programme
(FP7/2007-2013) under REA grant agreement n. PCOFUND-GA-2013-609102, through the PRESTIGE programme coordinated by Campus France, and by the LabEx IRMIA. F. Zerbini would like to thank the Galileo Galilei Institute for Theoretical Physics for the
hospitality and the INFN for partial support during the completion of part of this work.

\appendix

\section{Regions of absolute convergence of the integrals considered}\label{Sec:ConvReg}
\begin{propos}\phantomsection\label{ConvRegion}
For each subset of indices $I=\{i_1,\ldots ,i_h\}\subset\{1,\ldots N\}$, with $i_1<i_2<\cdots <i_h$, and for $(\pmb a,\tilde{\pmb a},\pmb b,\tilde{\pmb b},\pmb c,\tilde{\pmb c},\pmb d,\tilde{\pmb d})\in\mathbb{C}^{N(N+5)}$ as in Section~\ref{sec:closedampcorrfun}, we define
\begin{equation}
U_{I,0}:=\bigg\{(\pmb a,\tilde{\pmb a},\pmb b,\tilde{\pmb b},\pmb c,\tilde{\pmb c},\pmb d,\tilde{\pmb d})\,:\,\textup{Re}\bigg(\sum_{s=1}^h\frac{a_{i_s}+\tilde a_{i_s}}{2}+\sum_{1\leq s<r\leq h}\frac{d_{i_s,i_r}+\tilde d_{i_s,i_r}}{2}\bigg)>-h\bigg\},
\end{equation}
\begin{equation}
U_{I,1}:=\bigg\{(\pmb a,\tilde{\pmb a},\pmb b,\tilde{\pmb b},\pmb c,\tilde{\pmb c},\pmb d,\tilde{\pmb d})\,:\,\textup{Re}\bigg(\sum_{s=1}^h\frac{b_{i_s}+\tilde b_{i_s}}{2}+\sum_{1\leq s<r\leq h}\frac{d_{i_s,i_r}+\tilde d_{i_s,i_r}}{2}\bigg)>-h\bigg\},
\end{equation}
\begin{equation}
U_{I,\eta}:=\bigg\{(\pmb a,\tilde{\pmb a},\pmb b,\tilde{\pmb b},\pmb c,\tilde{\pmb c},\pmb d,\tilde{\pmb d})\,:\,\textup{Re}\bigg(\sum_{s=1}^h\frac{c_{i_s}+\tilde c_{i_s}}{2}+\sum_{1\leq s<r\leq h}\frac{d_{i_s,i_r}+\tilde d_{i_s,i_r}}{2}\bigg)>-h\bigg\}.
\end{equation}
Morever, if $|I|=h\geq 2$ we define
\begin{equation}
D_I:=\bigg\{(\pmb a,\tilde{\pmb a},\pmb b,\tilde{\pmb b},\pmb c,\tilde{\pmb c},\pmb d,\tilde{\pmb d})\,:\,\textup{Re}\bigg(\sum_{1\leq s<r\leq h}\frac{d_{i_s,i_r}+\tilde d_{i_s,i_r}}{2}\bigg)>1-h\bigg\}.
\end{equation}
Finally, let
\begin{equation}
U_{\infty}:=\bigg\{(\pmb a,\tilde{\pmb a},\pmb b,\tilde{\pmb b},\pmb c,\tilde{\pmb c},\pmb d,\tilde{\pmb d})\,:\textup{Re}\bigg(\sum_{i=1}^N\frac{a_{i}+\tilde a_{i}+b_{i}+\tilde b_{i}+c_{i}+\tilde c_{i}}{2}+\sum_{1\leq i<j\leq N}\frac{d_{i,j}+\tilde d_{i,j}}{2}\bigg)<-N\bigg\}.
\end{equation}
The region of absolute convergence $\mathcal{C}_N\subset\mathbb{C}^{N(N+5)}$ of the correlation function $\mathcal{G}_{N}(\eta)$~\eqref{e:Gdefgeneral} is, for any fixed $\eta\neq 0,1$, given by the intersection of all the domains $U_{I,0}, U_{I,1}, U_{I,\eta}, D_I, U_{\infty}$. Moreover, the region of common absolute convergence $\mathcal{D}_N\subset \mathbb{C}^{N(N+5)/2}$ of all the Aomoto-Gel'fand functions~\eqref{AGgendef1var}, i.e. the region of absolute convergence of the integral
\begin{equation}\label{realInt}
\int_{\mathbb{R}^N}\prod_{i=1}^N\,|z_i|^{a_i}\,|z_i-1|^{b_i}\,|z_i-\eta|^{c_i}\,\prod_{1\leq i<j\leq N}|z_i-z_j|^{d_{ij}}\,\prod_{i=1}^N\,dz_i\,,
\end{equation}
is also given by the intersection of all the domains $U_{I,0}, U_{I,1}, U_{I,\eta}, D_I, U_{\infty}$ restricted  to $\mathbb{C}^{N(N+5)/2}$ via $\tilde{\pmb a}:=\pmb a$, $\tilde{\pmb b}:=\pmb b$, $\tilde{\pmb c}:=\pmb c$ and $\tilde{\pmb d}:=\pmb d$.
\end{propos}
\begin{proof}[Idea of the proof]
Proving this result is essentially a long exercise in elementary analysis. In fact, there are different ways of computing this convergence region, which give seemingly different but equivalent conditions on the parameters. The method that we suggest is to divide the domain of integration into all possible regions $0\leq |z_{i_k}|\leq\cdots\leq |z_{i_1}|$ and operate the change of variables $u_j=z_{i_{j}}/z_{i_{j-1}}$ (setting $z_{i_{0}}=1$). From this, passing to polar coordinates, one easily works out the domains~$U_{I,0}$ and~$U_{\infty}$, where the integral is convergent near some $u_j=0$ or near $u_1=\infty$, respectively. The regions~$U_{I,1}$ and~$U_{I,\eta}$, which take into account the singularies $z_i=1$ and $z_i=\eta$, respectively, are then obtained by substituting $(\pmb a,\tilde{\pmb a})\leftrightarrow (\pmb b_i,\tilde{\pmb b_i})$ and $(\pmb a,\tilde{\pmb a})\leftrightarrow (\pmb c,\tilde{\pmb c})$, respectively. Finally, one must take care of integrals over $|u_i|\in[0,1]$ of terms $\prod_{i<j}|1-u_{i+1}\cdots u_j|^{d_{ij}+\tilde{d}_{ij}}$, which account for singularities along the diagonals but away from~$0$ or~$\infty$. This can be done by substituting $u_i=1-\rho_i e^{i\theta_i}$ and analysing the asymptotics as $\rho_i\rightarrow 0$, and it leads to considering the regions $D_I$. The region of convergence of the real integrals~\eqref{realInt} turns out to be the same as that of the complex ones with $\tilde{\pmb a}:=\pmb a$, $\tilde{\pmb b}:=\pmb b$, $\tilde{\pmb c}:=\pmb c$ and $\tilde{\pmb d}:=\pmb d$. This is easily seen immediately after passing to polar coordinates.
\end{proof}

This result can be specialised (either by setting $\eta=0$ or $\eta=1$, or by setting ${\pmb c}={\pmb 0}$) with the correct parameter identifications to obtain the convergence region for the closed string building blocks~\eqref{e:IntGeneric}. For instance, considering the tachyonic integral~\eqref{tachyonicInt} for $N=1$ and $N=2$ we find that
\begin{equation}
\mathcal{C}^{tachyon}_1\,=\,\{(a_1,b_1)\,:\,\textup{Re}(a_1),\,\textup{Re}(b_1)>-1,\,\,\textup{Re}(a_1+b_1)<-1\},
\end{equation}
\begin{align}
\mathcal{C}^{tachyon}_2\,=\,&\{(a_1,a_2,b_1,b_2,c_{1,2})\,:\,\textup{Re}(a_1),\,\textup{Re}(a_2),\,\textup{Re}(b_1),\,\textup{Re}(b_2),\,\textup{Re}(c_{1,2})>-1 \notag \\
&\textup{Re}(a_1+b_1+c_{1,2}),\,\textup{Re}(a_2+b_2+c_{1,2})>-2,\notag\\
&\textup{Re}(a_1+a_2+b_1+b_2+c_{1,2})<-2\}
\end{align}

We remark that the point $(\pmb a,\tilde{\pmb a},\pmb b,\tilde{\pmb b},\pmb c,\tilde{\pmb c},\pmb d,\tilde{\pmb d})$ with $\pmb a=\tilde{\pmb a}=\pmb b=\tilde{\pmb b}=\pmb c=\tilde{\pmb c}=(-1,\ldots ,-1)$ and $\pmb d=\tilde{\pmb d}=(0,\ldots ,0)$ belongs to the boundary of $\mathcal{C}_N$, and that for $\mbox{Re}(\varepsilon)>0$ small enough all the points $(\pmb a,\tilde{\pmb a},\pmb b,\tilde{\pmb b},\pmb c,\tilde{\pmb c},\pmb d,\tilde{\pmb d})$ with $\pmb a=\tilde{\pmb a}=\pmb b=\tilde{\pmb b}=\pmb c=\tilde{\pmb c}=(-1+\varepsilon,\ldots ,-1+\varepsilon)$ and $\pmb d=\tilde{\pmb d}=(0,\ldots ,0)$ are contained in $\mathcal{C}_N$, which in particular is always non-empty.

\newpage
\bibliographystyle{ieeetr}

\begin{thebibliography}{10}

 \bibitem[Aom]{Aomoto} K.~Aomoto, ``Gauss-Manin connection of integrals of
   difference products.'' J. Math. Soc. Jpn {\bf 39}, 191-208 (1987)

\bibitem[ACJS]{Azevedo:2018dgo}
  T.~Azevedo, M.~Chiodaroli, H.~Johansson and O.~Schlotterer,
  ``Heterotic and bosonic string amplitudes via field theory'',
  JHEP {\bf 1810} (2018) 012
  [arXiv:1803.05452 [hep-th]].
  
\bibitem[BPP]{Banks:2018rul}
P.~Banks, E.~Panzer and B.~Pym,
``Multiple zeta values in deformation quantization'',
Invent. Math. \textbf{222}, 79-159 (2020)
[arXiv:1812.11649 [math.QA]].

\bibitem[BBDSV]{BjerrumBohr:2010hn}
  N.~E.~J.~Bjerrum-Bohr, P.~H.~Damgaard, T.~Sondergaard and P.~Vanhove,
  ``The Momentum Kernel of Gauge and Gravity Theories'',
  JHEP {\bf 1101} (2011) 001
  [arXiv:1010.3933 [hep-th]].

\bibitem[BBDV]{BjerrumBohr:2009rd}
  N.~E.~J.~Bjerrum-Bohr, P.~H.~Damgaard and P.~Vanhove,
  ``Minimal Basis for Gauge Theory Amplitudes'',
  Phys.\ Rev.\ Lett.\  {\bf 103} (2009) 161602
  [arXiv:0907.1425 [hep-th]].


\bibitem[BGVZG]{Bocardo-Gaspar:2019pzk}
M.~Bocardo-Gaspar, W.~Veys and W.~A.~Z\'u\~niga-Galindo,
``Meromorphic Continuation of Koba-Nielsen String Amplitudes'',
JHEP \textbf{09}, 138 (2020)
[arXiv:1905.10879 [math-ph]].

\bibitem[BK]{Broedel:2019gba}
J.~Broedel and A.~Kaderli,
``Amplitude recursions with an extra marked point'',
[arXiv:1912.09927 [hep-th]].

\bibitem[BMMS]{Broedel:2014vla}
  J.~Broedel, C.~R.~Mafra, N.~Matthes and O.~Schlotterer,
  ``Elliptic multiple zeta values and one-loop superstring amplitudes'',
  JHEP {\bf 1507} (2015) 112
  [arXiv:1412.5535 [hep-th]].

\bibitem[BMS]{Broedel:2015hia}
J.~Broedel, N.~Matthes and O.~Schlotterer,
``Relations Between Elliptic Multiple Zeta Values and a Special Derivation Algebra,''
J. Phys. A \textbf{49} (2016) no.15, 155203
[arXiv:1507.02254 [hep-th]].

\bibitem[BMRS]{Broedel:2017jdo}
J.~Broedel, N.~Matthes, G.~Richter and O.~Schlotterer,
``Twisted Elliptic Multiple Zeta Values and Non-Planar One-Loop Open-String Amplitudes,''
J. Phys. A \textbf{51} (2018) no.28, 285401
[arXiv:1704.03449 [hep-th]].
 
  
\bibitem[BSST]{Broedel:2013aza}
  J.~Broedel, O.~Schlotterer, S.~Stieberger and T.~Terasoma,
  ``All order $\alpha^{\prime}$-expansion of superstring trees from the Drinfeld associator'',
  Phys.\ Rev.\ D {\bf 89} (2014) no.6,  066014
  [arXiv:1304.7304 [hep-th]].

\bibitem[BSZ]{Broedel:2018izr}
J.~Broedel, O.~Schlotterer and F.~Zerbini,
``From Elliptic Multiple Zeta Values to Modular Graph Functions: Open and Closed Strings at One Loop,''
JHEP \textbf{01} (2019), 155
[arXiv:1803.00527 [hep-th]].

\bibitem[Bro04a]{BrownNote}
F.~Brown, ``{Single-valued hyperlogarithms and unipotent differential
  equations}.'' {preprint on webpage at {\tt
  www.ihes.fr/\~{}brown/RHpaper5.pdf}}, 2004.

\bibitem[Bro04b]{BrownSVMPL}
F.~Brown, ``Single-valued multiple polylogarithms in one variable'', {\em C.R.
  Acad. Sci. Paris}, vol.~338, pp.~527--532, 2004.

  \bibitem[Bro09]{Brown:2009qja}
  F.~Brown,
  ``Multiple zeta values and periods of moduli spaces $\mathfrak{M}_{0,n}(\mathbb{R})$'',
  Annales Sci.\ Ecole Norm.\ Sup.\  {\bf 42} (2009) 371
  [math/0606419 [math.AG]].

\bibitem[Bro13]{Brown:2013gia}
  F.~Brown,
  ``Single-valued Motivic Periods and Multiple Zeta Values'',
  SIGMA {\bf 2} (2014) e25
  [arXiv:1309.5309 [math.NT]].

\bibitem[Bro17a]{Brown:2017qwo}
  F.~Brown,
  ``A class of non-holomorphic modular forms I'',
  arXiv:1707.01230 [math.NT].

\bibitem[Bro17b]{BrownNewClassII}
F.~Brown, ``A class of non-holomorphic modular forms {II} : equivariant
  iterated eisenstein integrals'', {arXiv:1708.03354 [math.NT]}, 2017.

\bibitem[BD18]{Brown:2018omk}
F.~Brown and C.~Dupont,
``Single-valued integration and double copy'',
[arXiv:1810.07682 [math.NT]].

\bibitem[BD19a]{Brown:2019jng}
F.~Brown and C.~Dupont,
``Lauricella hypergeometric functions, unipotent fundamental groups of the punctured Riemann sphere, and their motivic coactions'',
[arXiv:1907.06603 [math.AG]].

\bibitem[BD19b]{Brown:2019wna}
F.~Brown and C.~Dupont,
``Single-valued integration and superstring amplitudes in genus zero'',
[arXiv:1910.01107 [math.NT]].
 
\bibitem[CMT]{Casali:2019ihm}
E.~Casali, S.~Mizera and P.~Tourkine,
``Monodromy relations from twisted homology'',
JHEP \textbf{12} (2019), 087
[arXiv:1910.08514 [hep-th]].

\bibitem[D5MPV]{DelDuca:2016lad}
  V.~Del Duca, S.~Druc, J.~Drummond, C.~Duhr, F.~Dulat, R.~Marzucca, G.~Papathanasiou and B.~Verbeek,
  ``Multi-Regge kinematics and the moduli space of Riemann spheres with marked points'',
  JHEP {\bf 1608} (2016) 152
  [arXiv:1606.08807 [hep-th]].

\bibitem[DG]{DHoker:2019xef}
E.~D'Hoker and M.~B.~Green,
``Absence of Irreducible Multiple Zeta-Values in Melon Modular Graph Functions,''
Commun. Num. Theor. Phys. \textbf{14} (2020) no.2, 315-324
[arXiv:1904.06603 [hep-th]].


 
\bibitem[DGGV]{DHoker:2015wxz}
  E.~D'Hoker, M.~B.~Green, \"O~G\"urdogan and P.~Vanhove,
  ``Modular Graph Functions'',
  Commun.\ Num.\ Theor.\ Phys.\  {\bf 11} (2017) 165
    [arXiv:1512.06779 [hep-th]].
  
    \bibitem[DP]{DHoker:1988pdl}
  E.~D'Hoker and D.~H.~Phong,
  ``The Geometry of String Perturbation Theory'',
  Rev.\ Mod.\ Phys.\  {\bf 60} (1988) 917.
   
\bibitem[DMS]{DiFrancesco:1997nk}
  P.~Di Francesco, P.~Mathieu and D.~Senechal,
  ``Conformal Field Theory'', Graduate Texts in Contemporary Physics, New York: Springer-Verlag,
  1997.
  
  \bibitem[DF84]{Dotsenko:1984nm}
  V.~S.~Dotsenko and V.~A.~Fateev,
  ``Conformal Algebra and Multipoint Correlation Functions in Two-Dimensional Statistical Models'',
  Nucl.\ Phys.\ B {\bf 240} (1984) 312.

\bibitem[DF85]{Dotsenko:1984ad}
  V.~S.~Dotsenko and V.~A.~Fateev,
  ``Four Point Correlation Functions and the Operator Algebra in the Two-Dimensional Conformal Invariant Theories with the Central Charge c < 1'',
  Nucl.\ Phys.\ B {\bf 251} (1985) 691.

\bibitem[FFST]{Fan:2017uqy}
  W.~Fan, A.~Fotopoulos, S.~Stieberger and T.~R.~Taylor,
  ``SV-map between Type I and Heterotic Sigma Models'',
  Nucl.\ Phys.\ B {\bf 930} (2018) 195
  [arXiv:1711.05821 [hep-th]].

\bibitem[Gel]{GelfandAlone}  I.M.~Gel'fand
   ``General theory of hypergeometric functions'',
   Soviet Math. Dokl.  {\bf 33}, 573-577 (1986)


\bibitem[Gin]{Ginsparg:1988ui}
  P.~H.~Ginsparg,
  ``Applied Conformal Field Theory'',
  hep-th/9108028.
  

\bibitem[GL]{GonzalesLorca}
J.~Gonzales-Lorca, ``S\'erie de {D}rinfel'd, monodromie et alg\`ebres de
  {H}ecke'', {\em PhD thesis, ENS}, 1998.
 
\bibitem[GSW]{Green:1987sp}
  M.~B.~Green, J.~H.~Schwarz and E.~Witten,
  ``Superstring Theory. Vol. 1: Introduction'',
  Cambridge, Uk: Univ. Pr. (1987) 

\bibitem[GS]{Gross:1986mw}
  D.~J.~Gross and J.~H.~Sloan,
  ``The Quartic Effective Action for the Heterotic String'',
  Nucl.\ Phys.\ B {\bf 291} (1987) 41.

\bibitem[KLT]{Kawai:1985xq}
  H.~Kawai, D.~C.~Lewellen and S.~H.~H.~Tye,
  ``A Relation Between Tree Amplitudes of Closed and Open Strings'',
  Nucl.\ Phys.\ B {\bf 269} (1986) 1.



\bibitem[MSS]{Mafra:2011nv}
  C.~R.~Mafra, O.~Schlotterer and S.~Stieberger,
  ``Complete N-Point Superstring Disk Amplitude I. Pure Spinor Computation'',
  Nucl.\ Phys.\ B {\bf 873} (2013) 419
  [arXiv:1106.2645 [hep-th]].
  

  
\bibitem[Miz16]{Mizera:2016jhj}
  S.~Mizera,
  ``Inverse of the String Theory KLT Kernel'',
  JHEP {\bf 1706} (2017) 084
  [arXiv:1610.04230 [hep-th]].


\bibitem[Miz17]{Mizera:2017cqs}
  S.~Mizera,
  ``Combinatorics and Topology of Kawai-Lewellen-Tye Relations'',
  JHEP {\bf 1708} (2017) 097
  [arXiv:1706.08527 [hep-th]].

\bibitem[Miz19]{Mizera:2019gea}
S.~Mizera,
``Aspects of Scattering Amplitudes and Moduli Space Localization'',
Springer theses
[arXiv:1906.02099 [hep-th]].
 
 \bibitem[Pan14]{Panzer:2014caa}
  E.~Panzer,
  ``Algorithms for the symbolic integration of hyperlogarithms with applications to Feynman integrals'',
  Comput.\ Phys.\ Commun.\  {\bf 188} (2015) 148
  [arXiv:1403.3385 [hep-th]].

 \bibitem[Pan15]{Panzer:2015ida}
  E.~Panzer,
  ``Feynman integrals and hyperlogarithms'',
  \newblock PhD thesis,
  Humboldt-Universit\"at, 2015
  arXiv:1506.07243 [math-ph].


\bibitem[Pol98a]{Polchinski:1998rq}
  J.~Polchinski,
  ``String theory. Vol. 1: An introduction to the bosonic string'',
Cambridge University Press (2007-12-19)

\bibitem[Pol98b]{Polchinski:1998rr}
  J.~Polchinski,
  ``String theory. Vol. 2: Superstring theory and beyond'',
 Cambridge University Press (2007-12-19)

\bibitem[SS18]{Schlotterer:2018zce}
O.~Schlotterer and O.~Schnetz,
``Closed Strings as Single-Valued Open Strings: a Genus-Zero Derivation,''
J. Phys. A \textbf{52} (2019) no.4, 045401
[arXiv:1808.00713 [hep-th]].

\bibitem[SS12]{Schlotterer:2012ny}
  O.~Schlotterer and S.~Stieberger,
  ``Motivic Multiple Zeta Values and Superstring Amplitudes'',
  J.\ Phys.\ A {\bf 46} (2013) 475401
  [arXiv:1205.1516 [hep-th]].
 
\bibitem[Sch13]{Schnetz:2013hqa}
  O.~Schnetz,
  ``Graphical functions and single-valued multiple polylogarithms'',
  Commun.\ Num.\ Theor.\ Phys.\  {\bf 08} (2014) 589
  [arXiv:1302.6445 [math.NT]].

  
\bibitem[Sch18]{SchnetzProc}
O.~Schnetz,
``Hyperlog Procedures'',
https://www.math.fau.de/person/oliver-schnetz/, (2018).


\bibitem[Stie09]{Stieberger:2009hq}
  S.~Stieberger,
  ``Open \& Closed vs. Pure Open String Disk Amplitudes'',
  [arXiv:0907.2211 [hep-th]].
 
\bibitem[Stie13]{Stieberger:2013wea}
  S.~Stieberger,
  ``Closed superstring amplitudes, single-valued multiple zeta values and the Deligne associator'',
  J.\ Phys.\ A {\bf 47} (2014) 155401
  [arXiv:1310.3259 [hep-th]].

\bibitem[ST14]{Stieberger:2014hba}
  S.~Stieberger and T.~R.~Taylor,
  ``Closed String Amplitudes as Single-Valued Open String Amplitudes'',
  Nucl.\ Phys.\ B {\bf 881} (2014) 269
  [arXiv:1401.1218 [hep-th]].

\bibitem[VZ]{Vanhove:2020qtt}
P.~Vanhove and F.~Zerbini,
``Building Blocks of Closed and Open String Amplitudes,''
Contribution the proceedings of "Mathemamplitudes", Padova December 2019
[arXiv:2007.08981 [hep-th]].

\bibitem[VGZ]{Gelfand}  
V.A.~Vasil'ev, I.M.~Gel'fand, A.V.~Zelevinskii,
   ``General hypergeometric functions on complex Grassmannians'',
   Funct. Anal. Prilozhen.  {\bf 21}, 19-31 (1987)
   
\bibitem[Vir]{Virasoro:1969me}
  M.~A.~Virasoro,
  ``Alternative constructions of crossing-symmetric amplitudes with regge behavior'',
  Phys.\ Rev.\  {\bf 177} (1969) 2309.

\bibitem[Zer15]{Zerbini:2015rss}
  F.~Zerbini,
  ``Single-valued multiple zeta values in genus 1 superstring amplitudes'',
  Commun.\ Num.\ Theor.\ Phys.\  {\bf 10} (2016) 703
   [arXiv:1512.05689 [hep-th]].

\bibitem[Zer18]{Zerbini:2018sox}
  F.~Zerbini,
  ``Elliptic multiple zeta values, modular graph functions and genus 1 superstring scattering amplitudes'',
 Diss., Univ. Bonn 2017, Bonn, 2018.
  arXiv:1804.07989 [math-ph].


\bibitem[ZZ]{Zagier:2019eus}
D.~Zagier and F.~Zerbini,
``Genus-zero and genus-one string amplitudes and special multiple zeta values,''
Commun. Num. Theor. Phys. \textbf{14} (2020) no.2, 413-452
[arXiv:1906.12339 [math.NT]].

 



\end{thebibliography}
\markboth{\textsc{Bibliography}}{\textsc{Bibliography}}

\end{document}